\documentclass[12pt]{iopart}

\usepackage{graphicx} 
\usepackage{dcolumn}
\usepackage{bm}
\usepackage{epsfig}
\usepackage{pdflscape}
\usepackage[colorlinks=true,citecolor=blue,filecolor=blue,linkcolor=blue,urlcolor=blue,pdftex]{hyperref}
\usepackage{rotating}
\usepackage{longtable}

\usepackage{fancyhdr}
\begin{document}

\title[Dark matter direct-detection experiments]{Dark matter direct-detection experiments}

\author{Teresa Marrod\'an Undagoitia and Ludwig Rauch}
\address{Max-Planck-Institut f\"ur Kernphysik, Saupfercheckweg 1, 69117 Heidelberg, Germany}

\ead{marrodan@mpi-hd.mpg.de, rauch@mpi-hd.mpg.de}
\begin{abstract}

In the past decades, several detector technologies have been developed with the quest to directly detect dark matter interactions and to test one of the most important unsolved questions in
modern physics. The sensitivity of these experiments has improved with a tremendous speed due to a constant development of the detectors and analysis methods, proving uniquely suited devices to solve the dark matter puzzle, as all other discovery strategies can only indirectly infer its existence. Despite the overwhelming evidence for dark matter from cosmological indications at small and large scales, a clear evidence for a particle explaining these observations remains absent. This review summarises the status of direct dark matter searches, focussing on the detector technologies used to directly detect a dark matter particle  producing recoil energies in the keV energy scale. The phenomenological signal expectations, main background sources, statistical treatment of data and calibration strategies are discussed.

\end{abstract}


\maketitle

\tableofcontents



\section{Introduction}
\label{sec:Introduction}

Overwhelming evidence of gravitational interactions between baryonic and a new form of non-luminous matter can be observed on cosmological as well as astronomical scales.
Its nature, however, remains uncertain.  It is commonly assumed that elementary particles could be the constituents of this 'dark' matter.  Such new particles, that could account for dark matter, appear in various theories beyond the standard model of particle physics. A variety of experiments have been developed over the past decades, aiming to detect these massive particles via their scattering in a detector medium.  Measuring this process would provide information on the dark-matter particle mass and its interaction probability with ordinary matter. The identification of the nature of dark matter would answer one of the most important open questions in physics and would help to better understand the Universe and its evolution. The main goal of this article is to review current and future direct-detection experimental efforts.

This article is organised in the following way. In section\,\ref{sec:DMpuzzle},  the different phenomena indicating  the existence of dark matter and possible explanations or candidates emphasising particle solutions are presented. If, indeed, particles are the answer to the dark matter puzzle, there are three main possibilities for a verification: to produce them at particle accelerators, to look for products of e.g. their self-annihilations at locations with a high dark matter density, or to directly measure their scattering off a detector's target material. This article is dedicated to direct detection searches  for massive particles producing recoil energies in the keV energy scale. The production of dark matter particles at accelerators and searches for indirect signals are discussed only  briefly. As the local density and velocity distributions of dark matter are relevant for the interpretation of the experimental results, the main characteristics of the Milky Way halo are presented in section\,\ref{sec:DMpuzzle}. Next, in section\,\ref{sec:intro_prin_dir_det}, the principles of direct detection of WIMPs including the expected signal signatures are explained.  Assumptions on particle-  and nuclear physics aspects which  are necessary for the derivation of the results are summarised, and possible interpretations of the results are given.
In section\,\ref{Sec:BG}, a general overview of background sources in direct-detection experiments is given considering different types of radiation and sources both internal and external contributions to the target material. 
In section\,\ref{sec:Gene_results}, the basic detector technologies are introduced along with their capability to distinguish between signal and background events. Furthermore, statistical methods and the general result of an experiment  are discussed. 
Afterwards, in section\,\ref{Sec:Calibr},  the required calibrations to determine the energy scale, energy threshold as well as signal and background regions are detailed.
In the main part of this review, section\,\ref{Sec:Tech_Res},  the working principles of different direct detection technologies and the current experimental status are reviewed. Finally, in section\,\ref{Sec:SumAndProsp}, the experimental results are summarised, and the prospects for the next years are discussed.


\section{The dark matter puzzle}
\label{sec:DMpuzzle}

A wealth of observational data from gravitational effects at very different length scales supports the existence of an unknown component in our Universe. After a brief review of these observations ranging from cosmological to Milky Way-sized galaxies, various explanations and elementary-particle candidates are discussed in the following. At the end of the section, possible methods to detect particle dark matter are presented.

\subsection{Dark matter indications from Cosmology and Astronomy}\label{sec:DMindications}

Temperature anisotropies in the cosmic microwave background (CMB), precisely measured by WMAP\,\cite{Hinshaw:2012aka} and more recently by the Planck satellite\,\cite{Ade:2013sjv}, give access to the Universe when it was about 400\,000 years old.
 The power spectrum of temperature fluctuations can be evaluated by a six parameter model which contains, among others, the baryonic matter, dark matter and dark energy contents of the Universe.   This cosmological standard model, which  fits the data with high significance, is denoted $\Lambda$CDM ($\Lambda$ cold dark matter) indicating that dark matter with a small random velocity is a fundamental ingredient. The $\Lambda$ refers to the cosmological constant necessary to explain the current accelerated expansion of the Universe\,\cite{Riess:1998cb}.
Oscillations of the baryon-photon fluid in the gravitational potential dominated by cold dark matter density perturbations give rise to the characteristic oscillation pattern in the CMB power spectrum (acoustic peaks).
From the relative height of these acoustic peaks, the amount of baryonic matter can be estimated, which allows to calculate the total dark matter density in the Universe.
Present estimates\,\cite{Planck:2015xua} show a flat Universe with $\Omega_{\textrm{\scriptsize{DM}}}=0.265$, $\Omega_{b}=0.049$ and $\Omega_{\Lambda}= 0.686$ representing the densities of dark matter, baryonic matter and dark energy, respectively.

In the standard scenario, the anisotro\-pies of the CMB originate from  quantum fluctuations during inflation. 
In order to understand the formation of matter distributions from the time of recombination to the present state, N-body simulations of dark matter particles have been carried out\,\cite{Springel:2006vs}. 
These simulations\,\cite{Springel:2005nw}\cite{Schaye:2015xx}\cite{Vogelsberger:2014dza} propagate particles using super computers aiming to describe the  structure growth, producing a cosmic web ranging from $\sim 10$\,kpc objects to the largest scales. 
Meanwhile, this type of simulations reproduce very accurately the measurements made by galaxy surveys\,\cite{Colless:2003wz}\cite{Anderson:2013zyy}\cite{Sanchez:2009xx}. Measurements of the Lyman$- \alpha$ forest\,\cite{Cen:1994da}\cite{Lee:2014mea} and weak lensing\,\cite{VanWaerbeke:2004af}\cite{Bartelmann:1999yn} confirm the cosmic structure considering not only galaxies and gas clouds but also non-luminous and non-baryonic matter. Large scale simulations, which consider only dark matter, have been used to confirm theories of large scale structure formation which serve as seeds for galaxy and cluster formation. Recently, gas and stars have been included into the simulations and it is shown that they can significantly alter the distribution of the dark matter component on small scales\,\cite{Pontzen:2014lma}.

A further hint for the existence of dark matter arises from gravitational lensing measurements\,\cite{Bartelmann:1999yn}. This effect discussed by Albert Einstein\,\cite{Einstein1936} in 1936 and later by Zwicky\,\cite{Zwicky:1937zzb}  occurs when a massive object is in the line of sight between the observer at the Earth and the object under study. The light-rays are deflected through their path due to the gravitational field resulting, for example, in multiple images or a deformation of the observable's image (strong and weak lensing, respectively). The degree of deformation can be used to reconstruct the gravitational potential of the object that deflects the light along the line of sight. From various observations it has been found that the reconstructed mass using this method is greater than the luminous matter, resulting in very large mass to light ratios (from a few to hundreds). 
Gravitational lensing has also been applied in galaxy-cluster collisions to reconstruct the mass distributions in such events where mass to light ratios of $>$\,200 are measured.  In some examples\,\cite{Clowe:2006eq}\cite{Bradac:2008eu}\cite{Dawson:2011kf} and in an extensive study of 72 cluster collisions\,\cite{Harvey:2015hha}, the reconstructed gravitational centers appear clearly separated from the main constituent of the ordinary matter, i.e. the gas clouds which collide and  produce detectable X-rays. This can be interpreted as being due to dark matter haloes that continue their trajectories independently of the collision. An upper limit to the self-interaction 
cross-section for dark matter can be derived from these observations\,\cite{Kahlhoefer:2013dca}. 

Indications for non-luminous matter appear in our Universe also at smaller scales. Historically, the first indications for dark matter arose from astronomical observations. In order to explain measurements of the dynamics 
of stars in our Galaxy, the word "dark matter" was already used by Kapteyn\,\cite{Kapteyn:1922} in 1922 but it was not the correct physical explanation of the observed phenomenon. The first evidence of dark matter
in the present understanding was the measurement  of unexpectedly high velocities of nebulae in the Coma cluster which brought Fritz Zwicky\,\cite{Zwicky1933} to the idea that a large amount of dark matter could be the 
explanation for the unexpected high velocities. 
In 1978,  Vera C. Rubin {\it et al.}\,\cite{Rubin1978}  found that rotation velocities of stars in galaxies stay approximately constant with increasing distance to their galactic center. This observation was in contradiction with the expectation, as objects outside the visible mass distribution should have velocities $v \propto 1/\sqrt{r}$ following Newtonian dynamics. 
A uniformly-distributed halo of dark matter could explain both the velocities in clusters and the rotation velocities of objects far from the luminous matter in galaxies (e.g.\,\cite{Richards:2015gla}).

\subsection{The nature of dark matter: possible explanations and candidates}\label{intro:DMcandidates}


A plausible solution to describe some of the astronomical measurements mentioned in section\,\ref{sec:DMindications} is a modification of gravitation laws to accommodate the observations. Such modified Newtonian dynamic models like MOND\,\cite{Milgrom:1983ca}  or its relativistic extension TeVeS\,\cite{Bekenstein:2004ne} can, for instance, successfully describe rotational velocities measured in galaxies. 
However, MOND fails or needs unrealistic parameters to fit observations on larger scales such us structure formation or the CMB structure and violates fundamental laws such as momentum conservation and
the cosmological principle\,\cite{Felten1984}. While TeVeS can solve some of the conceptual problems of MOND, the required parameters seem to generate an unstable Universe\,\cite{Seifert:2007fr} or fails to 
simultaneously fit lensing and rotation curves\,\cite{Mavromatos2009}. 

Massive astrophysical compact halo objects (MACHOs) have also been considered as a possible explanation for the large mass to light ratios detected in the astronomical observations described in the previous section. These objects could be neutron stars, black holes, brown dwarfs or unassociated planets that would emit very little to no radiation. Searches for such objects using gravitational microlensing\,\cite{Paczynski:1985jf} towards the Large Magellanic Cloud have been performed\,\cite{Alcock:2000ph}. Extrapolations to the Galactic dark matter halo showed that MACHOs can make up about 20\% of the dark matter in our galaxy and that a model with MACHOs accounting
entirely for the dark-matter halo is ruled out at 95\% confidence level\,\cite{Alcock:2000ph}. 
The baryonic nature of dark matter is actually also ruled out by Big-Bang nucleosynthesis (BBN). 
The abundance of light elements predicted by BBN depends on the baryon density and, in fact, measurements constrain the baryon density to a value around $\Omega_{b}=0.04$\,\cite{Iocco:2008va} close to the value derived from CMB. 
An example of baryonic dark matter, which is not affected by the BBN and CMB constraints mentioned above, are primordial black holes\,\cite{Carr:2016drx}\cite{Carr:2009jm}. Even though a large part of the viable parameter space is already excluded, for some black hole masses this explanation is still possible.

A more common ansatz is to assume that dark matter is made out of massive neutral particles featuring a weak self-interaction. From the known particles in the standard model, only the neutrino could be considered. Due to its relativistic velocity in the early Universe, the neutrino would constitute a hot dark matter candidate. Cosmological simulations have shown, however, that a Universe dominated by neutrinos would not be in agreement with the observed clustering scale of galaxies\,\cite{White:1984yj}. 
Furthermore, due to the fermionic character of neutrinos, their occupation number is constrained by the Fermi-Boltzmann distribution thus, they can not account for the observed dark-matter density in halos\,\cite{Tremaine1979}.
Sterile neutrinos are hypothetical particles which were originally introduced to explain the smallness of the neutrino masses\,\cite{Abazajian:2012ys}. Additionally, they provide a viable dark matter candidate.
Depending on their production mechanism, they would constitute cold (non relativistic at all times) or a warm (relativistic only in an early epoch) dark matter candidate\,\cite{Kusenko:2009up}\cite{Boyarsky:2009ix}.
Possible masses, which are not yet constrained by X-ray measurements or the analysis of dwarf spheroidal galaxies, range from 1\,keV to  tens of keV. Given this very low mass, and the low interaction strength, the existence of sterile neutrinos is not tested by direct detection experiments. An indication could, for example, arise from the X-ray measurement of the sterile neutrino decay via the radiative channel $N\to \nu\gamma$\,\cite{Abazajian:2001}.

Models beyond the standard model of particle physics suggest the existence of new particles which could account for the dark matter. If such hypothetical particles would be stable, neutral and have a mass from below GeV/$c^2$ to several TeV/$c^2$, they could be the weakly interacting massive particle (WIMP). The standard production mechanism for WIMPs assumes that in the early Universe these particles were in equilibrium with the thermal plasma\,\cite{Gelmini:2010zh}. As the Universe expanded, the temperature of the plasma became lower than the WIMP mass resulting in the decoupling from the plasma.  At this freeze-out temperature, when the WIMP annihilation rate was smaller than the Hubble expansion rate, the dark matter relic density was reached. The cross-section necessary to observe the current dark matter density is of the order of the weak interaction scale. It appears as a great coincidence that a particle interacting via the weak force would produce the right relic abundance and, therefore, the WIMP is a theoretically well motivated dark matter candidate.  This hypothesis is being thoroughly tested experimentally with no unambiguous signal appearing.  If the absence of signals remains in the upcoming generation of experiments, the WIMP paradigm might be challenged\,\cite{Duerr:2016tmh}\cite{Arcadi:2017kky}.

Supersymmetry models\,\cite{Jungman:1995df} are proposed as extensions of the standard model of particle physics to solve the  hierarchy problem as well as the unification of weak, strong and electromagnetic interactions.   In this model, a whole new set of particles are postulated such that for each particle in the standard model there is a supersymmetric partner. Each particle differs from its partner by 1/2 in spin and, consequently, bosons are related to fermions and vice versa. 
The neutralino, the lightest neutral particle which appears as a superposition of the partners of the standard model bosons, constitutes an example of a new particle fulfilling the properties of a WIMP. The typical masses predicted for the neutralino range from few GeV/$c^2$ to several TeV/$c^2$. 
A WIMP candidate appears also in models with extra-dimensions. In such models N spatial dimensions are added to the $(3+1)$ space time classical ones. They appeared already around 1920 to unify electromagnetism with gravity. The lightest stable particle is called 'lightest Kaluza particle' and constitutes also a good WIMP candidate\,\cite{Kaluza:1921}\cite{Klein:1926tv}.

Among the non-WIMP candidates, 'superheavy dark matter' or 'WIMPzillas' are postulated to explain the origin of ultra high-energy cosmic rays\,\cite{Kuzmin:1998uv}. At energies close to $10^{20}$\,eV, cosmic protons can interact with the cosmic microwave background and, thus, their mean free path is reduced resulting in a suppressed measured flux\,\cite{Greisen:1966jv}\cite{Zatsepin:1966jv}. Experimental results include, however, the detection of a few events above the expected cut-off, motivating a superheavy dark matter candidate. Decays of these non-thermally-produced\,\cite{Chung:1998zb} superheavy particles with masses of $(10^{12} - 10^{16})$\,GeV/$c^2$ could account for the observations, being at the same time responsible for the dark matter in the Universe.

Finally, a very well motivated particle and dark matter candidate is the axion. In the standard model of particle physics, there is no fundamental reason why QCD should conserve P and CP. However, from the experimental bound on the neutron electric dipole moment\,\cite{Baker:2006ts}, very small values of  P and CP violation can be derived. 
In order to solve this so-called 'strong CP-problem'\,\cite{Raffelt:2006cw}, a new symmetry was postulated\,\cite{Peccei:1977hh}  
in 1977. When this symmetry is spontaneously broken, a massive particle, the axion, appears. The axion mass and the coupling strength to ordinary matter are inversely proportional to the breaking scale which was originally associated to the electroweak scale. This original axion model is ruled out by laboratory experiments\,\cite{Weinberg:1977ma}. Cosmological and astrophysical results provide as well very strong bounds on the axion hypothesis\,\cite{Raffelt:2006cw}. There exist, however, further 'invisible' axion models in which the breaking scale is a free parameter, KSVZ\,\cite{Kim:1979if}\cite{Shifman:1979if} and DFSZ\,\cite{Zhitnitsky:1980tq}\cite{Dine:1981rt}, and still provide a solution to the CP-problem.
Invisible axions or axion-like particles, would have been produced non-thermally in the early Universe by mechanisms like the vacuum realignment\,\cite{Sikivie:1983ip}\cite{Abbott:1982af} for example, giving the right dark matter abundance. The resulting free streaming length would be small and, therefore, these axions are a "cold" candidate. For certain parameters, axions could account for the complete missing matter\,\cite{Visinelli:2009zm}.

Sterile neutrinos, WIMPs, superheavy particles and axions are not the only particle candidates proposed. The candidates mentioned above arise from models that were proposed originally with a different motivation and not to explain dark matter. The fact that the models are motivated by different unresolved observations strengthen the relevance of the predicted dark matter candidate. A more comprehensive review on dark matter candidates can be found for example in~\cite{Bertone:1900zza}. This article focuses on the direct detection of WIMPs and just some brief information on searches for particles that would induce an electronic recoil (e.g. axion-like particles) will be given in the following.

\subsection{Searches for dark matter particles}

\begin{figure}[h]
  \begin{center}
   \includegraphics[angle=0,width=0.65\textwidth]{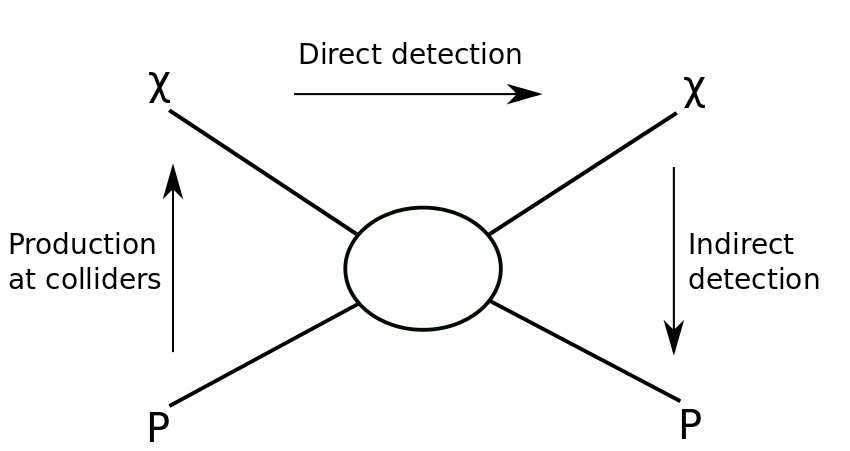}
   \caption[]{Schematic showing the possible dark matter detection channels.\label{Fig:Chann_Det}}
  \end{center}	
\end{figure}

The particle dark matter hypothesis can be tested via three processes: the production at particle accelerators, indirectly by searching for signals from annihilation products, or directly via scattering on target nuclei. Figure\,\ref{Fig:Chann_Det} shows a schematic representation of the possible dark matter couplings to a particle, P, of ordinary matter.  
While the annihilation of dark matter particles (downwards direction) could give pairs of standard model particles, the collision of electrons or protons at colliders could produce pairs of dark matter particles.
In this section the production and indirect detection methods as well as the current status of searches are briefly summarised. The subsequent sections and main part of this review are then devoted to the direct detection of dark matter, $\chi \textrm{P} \to \chi \textrm{P}$ (horizontal direction in figure\,\ref{Fig:Chann_Det}).

Since the start of the Large Hadron Collider (LHC) at CERN in 2008, the CMS\,\cite{Chatrchyan:2008aa} and ATLAS\,\cite{Aad:2008zzm} experiments have searched for new particles in proton-proton collisions at a center-of-mass energy of 7\,TeV. Besides the discovery of the Higgs particle\,\cite{Chatrchyan:2012ufa}\cite{Aad:2012tfa}, CMS and ATLAS have studied a number of new particle signatures by scanning the parameter space of different supersymmetric and extra-dimensions models. The presence of a dark matter particle would only be inferred by observing events with missing transferred momentum and energy. Therefore, events with, e.g., an energetic jet and an imbalanced momentum transfer are selected for analysis. Reactions of the type  
\begin{equation}\label{eq:dm_prod}
pp\to\chi\overline{\chi} + x
\end{equation}
 are probed, being $x$ a hadronic jet, a photon or a leptonically decaying $Z$ or $W$ boson. The results obtained so far are consistent with the standard model expectations (see for example~\cite{Sirunyan:2017hci}\cite{Aaboud:2016tnv}\cite{Sirunyan:2016iap}) but further searches will be performed in the next years. The derived bounds can be translated into limits on the cross-section for a given particle mass. Bounds arising from accelerator searches are most constraining below $\sim  5$\,GeV and $\sim$ a few hundreds of GeV for spin-independent and spin-dependent (proton coupling, see section\,\ref{CrosSec_NuclPhys}) interactions, respectively.
However, a direct comparison of these experimental results to other detection methods is, in general, model dependent (see the discussion in section\,\ref{Sec:SumAndProsp}).

Dark matter particles can gravitationally accumulate in astrophysical objects such as stars, galaxies or our Sun.
 The most favoured sources to search for indirect signals are the galactic centre and halo, close galaxy clusters or dwarf galaxies also called dwarf spheroidals. The latter are very popular locations due to their large measured mass to light ratio and their small background.
Due to the increased dark-matter density, an enhanced self-annihilation,  scattering or decay into standard model particles could produce a measurable particle flux (see~\cite{Strigari:2013iaa} for a detailed discussion). The measurement of this secondary particles is a further detection mechanism usually denoted as 'indirect detection'. Examples of possible annihilation channels are 
\begin{equation}\label{eq:dm_ind1}
\chi\overline{\chi} \to \gamma \gamma, \gamma Z, \gamma H ~~ \textrm{or}
\end{equation}
\begin{equation}\label{eq:dm_ind2}
\chi\overline{\chi} \to q \overline{q}, W^- W^+, ZZ
\end{equation}
some of the products decay further into $e^- e^+$, $p\overline{p}$, $\gamma$-rays and neutrinos.  A second mechanism to generate charged (anti-) particles, photons
or neutrinos from dark matter is given by its decay. In contrast to self-annihilation processes, where the production rate shows a quadratic dependence of the dark matter density, decaying dark matter scales only linearly (e.g.\,\cite{Ibarra:2013cra}).
In addition, dark matter particles might be gravitationally captured inside the Sun due to the elastic scattering with its nuclei. The annihilation of captured dark-matter particles can produce neutrinos which
can propagate out of the Sun and might be detectable with Earth-based neutrino telescopes. Note that the total number of captured particles is less affected by uncertainties of the dark matter halo since this process lasts for billions of years and dark matter density variations are averaged out\,\cite{Ibarra:2013cra}.

Produced charged particles are deflected in the interstellar magnetic fields loosing the information on their origin. Due to their charge neutrality, $\gamma$-rays and neutrinos point, instead, to the 
source where they were produced.  While neutrinos travel unaffected from the production source, $\gamma$-rays can be affected by absorption in the interstellar medium.  

Imaging atmospheric Cherenkov telescopes for TeV $\gamma$-ray detection can look specifically in the direction of objects where a large amount of dark matter is expected. Either a $\gamma$-flux in dwarf galaxies or galaxy clusters, or mono-energetic line signatures are searched for. So far no significant signal from dark matter annihilations has been observed, and upper limits are derived by the MAGIC\,\cite{Aleksic:2013xea}\cite{Ahnen:2016qkx}, HESS\,\cite{Abramowski:2014tra}\cite{Abdallah:2016ygi} and VERITAS\,\cite{Pfrommer:2012mm}\cite{Archambault:2017wyh} telescopes. 
Indirect searches can be also performed by satellite-based instruments capable of detecting low-energy $\gamma$-rays (approx. 20\,MeV\,--\,300\,GeV) like Fermi-LAT\,\cite{Conrad:2015bsa}. Although some gamma-ray features identified in the Fermi data are intriguing (for example~\cite{Fermi-LAT:2014sfa}\cite{Ackermann:2013uma}\cite{Hooper:2010mq}),  the Fermi collaboration have performed several searches\,\cite{Ackermann:2015tah}\cite{Ackermann:2015lka} in which no evidence for a dark-matter signal is found. One of the strongest and most robust constraints can be derived by the Fermi-LAT observation of dwarf spheroidal satellite galaxies of the Milky Way 
as those are some of the most dark matter dominated objects known\,\cite{Ackermann:2015zua}\cite{Fermi-LAT:2016uux}. Consequently, conservative limits on the annihilation cross-section of dark matter particles ranging from a few GeV to a few tens of TeV are derived. 
In the energy region of $(0.1-10)$\,keV, X-ray satellites as XMM-Newton and Chandra provide data to search for indirect dark matter signals. In 2014, an unexpected line at 3.5\,keV was found in the data recorded by  both satellites\,\cite{Bulbul:2014sua}\cite{Boyarsky:2014jta}. This signal can be interpreted by a decay of dark matter candidates, for instance, from sterile neutrinos or axions\,\cite{Abazajian:2014gza}\cite{Ishida:2014dlp}\cite{Higaki:2014zua}\cite{Jaeckel:2014qea}. Other astrophysical explanations have been, however, proposed and thus, the origin of the signal remains controversial (see e.g.~\cite{Jeltema:2014qfa}\cite{Carlson:2014lla}\cite{Bulbul:2014ala}\cite{Shah:2016efh}).
 Large neutrino detectors like Ice Cube, ANTARES or Super-Kamiokande are able to search for dark matter annihilations into neutrinos. No evidence for such a signal has been observed, resulting in constrains on the cross-section\,\cite{Aartsen:2016zhm}\cite{Adrian-Martinez:2016gti}\cite{Choi:2015ara}. 
Finally, also charged particles like protons, antiprotons, electrons and positrons can be detected by satellites. Measurements on the steadily increasing positron fraction from 10 to $\sim$\,250\,GeV by Pamela\,\cite{Adriani:2008zr} and AMS\,\cite{Aguilar:2013qda} among others rise discussions on its possible dark matter origin. However, given that such a spectrum could be also described by astrophysical objects like pulsars (rapidly rotating neutron stars) or by 
the secondary production of $e^{+}$ by the collision of cosmic rays with interstellar matter\,\cite{Blum:2013zsa}, this cannot be considered as a clear indication of a dark matter signal.


\section{Principles of  WIMP direct detection}
\label{sec:intro_prin_dir_det}

Large efforts have been  pursued to develop experiments which are able to directly test the particle nature of dark matter. The aim is to identify nuclear recoils produced by the collisions between the new particles 
and a detector's target nuclei.
The elastic scattering of WIMPs with masses of  $(10-1000)$\,GeV/$c^2$ would produce nuclear recoils in the range of $(1-100)$\,keV\,\cite{Lewin:1995rx}.  To unambiguously identify such low-energy interactions, a detailed knowledge on the signal signatures, the particle physics aspects and nuclear physics modelling is mandatory. Furthermore, for the calculation of event rates in direct detection experiments, the dark matter density and the halo velocity distribution in the Milky Way are required. This section is devoted to review all these aspects focussing on WIMP dark matter, whereas non-WIMP candidates are briefly discussed in section\,\ref{sec:OtherInterpr}.

\subsection{Experimental signatures of dark matter}\label{sec:signatures}

The signature of dark matter in a direct detection experiment consists of a recoil spectrum of single scattering events. Given the low interaction strength expected for the dark matter particle, the probability of 
multiple collisions within a detector is negligible. In case of a WIMP, a nuclear recoil is expected\,\cite{Goodman:1984dc}. 
The differential recoil spectrum resulting from dark matter interactions can be written, following~\cite{Lewin:1995rx}, as:
\begin{equation}\label{eq:diff_rate1}
\frac{dR}{dE} (E,t) = \frac{\rho_0}{m_{\chi} \cdot m_{A}}   \cdot \int v \cdot f({\bf v},t) \cdot \frac{d\sigma}{dE}(E,v)~\textrm{d}^3v,
\end{equation}
 where $m_{\chi}$ is the dark matter mass and $\frac{d\sigma}{dE}(E,v)$ its differential cross-section. The WIMP cross-section $\sigma$ and $m_{\chi}$ are the two observables of a dark matter experiment.  The dark matter  velocity $v$ is defined in the rest frame of the detector and $m_{A}$ is the nucleus mass.  Equation\,\ref{eq:diff_rate1} 
shows explicitly the astrophysical parameters,  the local dark matter density $\rho_0$ and  $f({\bf v},t)$, which accounts for the WIMP velocity distribution in the detector reference frame. This
velocity distribution is time dependent due to the revolution of the Earth around the Sun. Based on equation\,\ref{eq:diff_rate1}, detection strategies can exploit the energy, time or direction dependencies of the signal.

The most common approach in direct detection experiments is the attempt to measure the energy dependence of dark matter interactions. 
According to\,\cite{Lewin:1995rx}, equation\,\ref{eq:diff_rate1} can be approximated by
\begin{equation}\label{eq:diff_rate_simpl}
\frac{dR}{dE} (E) \approx \left (\frac{dR}{dE} \right)_0 F^2(E) \exp\left(-\frac{E}{E_c} \right),
\end{equation}
where $\left (\frac{dR}{dE} \right)_0 $ denotes the event rate at zero momentum transfer and $E_c$ is a constant parameterizing a characteristic energy 
scale which depends on the dark matter mass and target nucleus\,\cite{Lewin:1995rx}. Hence, the signal is dominated at low recoil energies by the exponential function.  $F^2(E)$ is the form-factor correction which will be described in more detail in section\,\ref{CrosSec_NuclPhys}.

Another possible dark matter signature is the so-called 'annual modulation'. As a consequence of the Earth rotation around the Sun, the speed of the dark matter particles in the Milky Way halo relative to 
the Earth is largest around June 2nd and smallest in December. Consequently, the amount of particles able to produce nuclear recoils above the detectors' energy threshold is also largest in June\,\cite{Drukier:1986tm}. 
As the amplitude of the variation is expected to be small, the temporal variation of the differential event rate can be written, following\,\cite{Freese:2012xd}, as 
\begin{equation}\label{eq:diff_rate2}
\frac{dR}{dE} (E,t) \approx S_0(E) + S_m(E) \cdot \cos \Bigg( \frac{2\pi (t-t_0)}{T}\Bigg),
\end{equation}
where $t_0$ is the phase which is expected at about 150~days and $T$ is the expected period of one year. The time-averaged event rate is denoted by $S_0$, whereas the modulation amplitude is given by $S_m$.  
A rate modulation would, in principle, enhance the ability to discriminate against background and help to confirm a dark matter detection.

Directionality is another dark-matter signature which can be employed for detection as the direction of the nuclear recoils resulting from WIMP interactions has a strong angular dependence\,\cite{Spergel:1987kx}. This dependence can be seen in the differential rate equation when it is explicitly written as a function of the angle $\gamma$, defined by the 
direction of the nuclear recoil relative to the mean direction of the solar motion
\begin{equation}\label{eq:diff_angDep}
\frac{dR} {dE ~ d \cos \gamma}  \propto \exp \Bigg[\frac{-[(v_E + v_{\odot})\cos \gamma - v_{min}]^2}{v_c^2}\Bigg].
\end{equation}
In equation\,\ref{eq:diff_angDep}, $v_E $ represents the Earth's motion, $v_{\odot}$ the velocity of the Sun around the galactic centre, $v_{min}$ the minimum WIMP velocity that can produce a nuclear recoil of an energy $E$ and  $v_c$  the halo circular velocity $v_c = \sqrt{3/2}v_{\odot}$. The integrated rate of events scattering in the forward direction will, therefore, exceed the rate for backwards scattering events by 
an order of magnitude\,\cite{Spergel:1987kx}. An oscillation of the mean direction of recoils over a sidereal day is also expected due to the rotation of the Earth and if the detector is placed at an appropriate latitude.
This directional signature allows to discriminate potential backgrounds\,\cite{SnowdenIfft:1999hz}.  A detector able to determine the direction of the WIMP-induced nuclear recoil would provide a powerful tool to confirm the measurement of dark matter particles. Such directional searches are summarized in section\,\ref{sec:DirectionalDet}.

\subsection{Cross-sections and nuclear physics aspects}\label{CrosSec_NuclPhys}

To interpret the data of dark matter experiments, further assumptions on the specific particle-physics model as well as on the involved nuclear-physics processes have to be made.
This section summarises the most common interactions between dark matter particles and the target nucleons.
  
For WIMP interactions that are independent of spin, it is assumed that neutrons and protons contribute equally to the scattering process (isospin conservation). For sufficiently low momentum transfer $q$, 
the scattering amplitude of each nucleon adds in phase and results in a coherent process.  
For spin-dependent interactions, only unpaired nucleons contribute to the scattering. Therefore, only nuclei with an odd number of protons or neutrons are sensitive to these interactions. In this case, the cross-section is related to the quark spin content of the nucleon with components from both proton and neutron couplings.

When the momentum transfer is such that the particle wavelength is no longer large compared to the nuclear radius, the cross-section decreases with increasing $q$. The form factor $F$ accounts for this 
effect and the cross-section can be expressed as: $\sigma \propto \sigma_0 \cdot F^2$, where $\sigma_0$ is the cross-section at zero momentum transfer. In general, the differential WIMP-nucleus cross 
section, $d\sigma / dE$ shown in equation\,\ref{eq:diff_rate1}, can be written as the sum of a spin-independent (SI) contribution and a spin-dependent (SD) one,
\begin{equation}\label{eq:diff}
\frac{d\sigma} {dE}  = \frac{m_A}{2\mu _A ^2 v^2 } \cdot (\sigma_0^{\scriptsize{\textrm{SI}}} \cdot F_{\scriptsize{\textrm{SI}}}^2(E) +\sigma_0^{\scriptsize{\textrm{SD}}} \cdot F_{\scriptsize{\textrm{SD}}}^2(E) ).
\end{equation}
The WIMP-nucleus reduced mass is described by $\mu _A$.
For spin independent interactions, the cross-section at zero momentum transfer can be expressed as
\begin{equation}\label{eq:3SI}
\sigma_0^{\scriptsize{\textrm{SI}}} = \sigma _p \cdot \frac{\mu _{A} ^2}{\mu _p ^2} \cdot [Z \cdot f^p + (A-Z)\cdot f^n]^2
\end{equation}
where $f^{p,n}$ are the contributions of protons and neutrons to the total coupling strength, respectively, and $\mu _p$ is the WIMP-nucleon reduced mass. Usually, $f^{p} = f^{n}$ is  assumed and the dependence of the cross-section with the number of nucleons $A$ takes an $A^2$ form. The impact of $f^{p} \neq f^{n}$ (isospin-violating dark matter) on experimental results is discussed in\,\cite{Yaguna:2016bga}.
The form factor for SI interactions is calculated assuming the distribution of scattering centres to be the same as the charge distribution derived from electron scattering experiments\,\cite{Lewin:1995rx}. Commonly, the Helm parameterisation\,\cite{Helm:1956zz} is used to describe the form factor. Recent shell-model calculations\,\cite{Vietze:2014vsa} show that the derived structure factors are in good agreement with the classical parameterisation.  

\begin{figure}[h]
  \begin{center}
   \includegraphics[angle=0,width=0.49\textwidth]{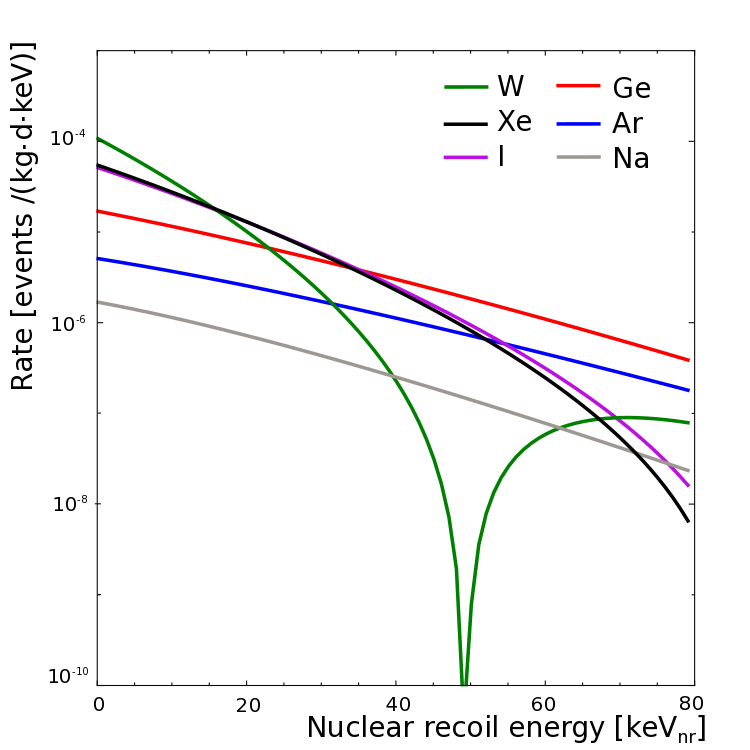}
    \includegraphics[angle=0,width=0.49\textwidth]{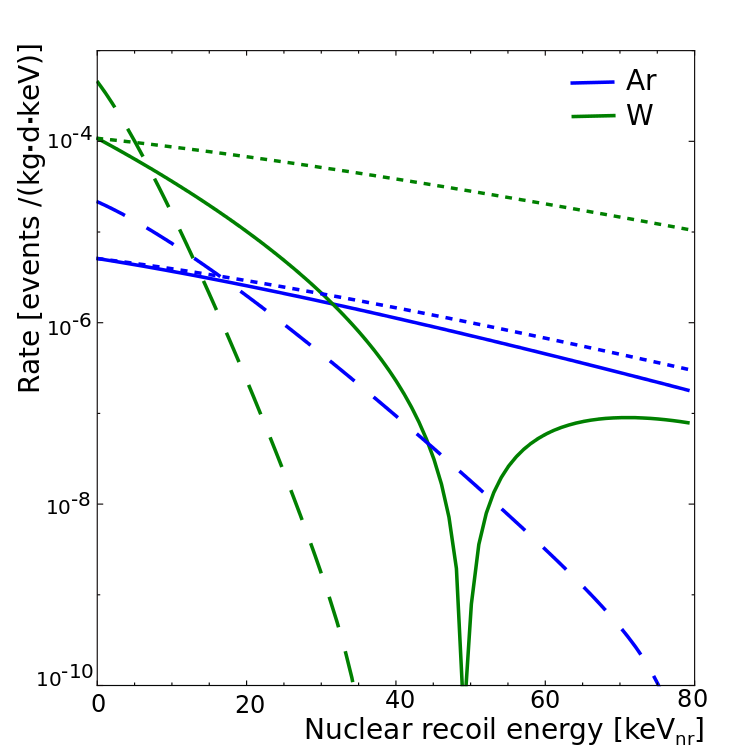}
   \caption[]{(Left) Differential event rate for the direct detection of a 100\,GeV/$c^2$ WIMP with a cross-section of $10^{-45}$\,cm$^{2}$ in experiments using tungsten (green), xenon (black), iodine 
(magenta), germanium (red), argon (blue) and sodium (grey) as target materials. (Right) The event rate is shown for a heavy
and a light target as indicated in green (tungsten) and blue (argon), respectively, showing the effect of neglecting the form factor correction (dotted line) and the effect of a lower WIMP mass  of 25\,GeV$c^2$ (dashed line). \label{Fig:DD_diffRate}}
  \end{center}	
\end{figure}

To visualise the effect of the target isotope and the form-factor correction,  figure\,\ref{Fig:DD_diffRate} (left) shows the event rate given in number of events per keV, day and kg (equation\,\ref{eq:diff_rate1}) for spin-independent interactions in different target materials: tungsten in green, xenon in black, iodine 
in magenta, germanium in red, argon in blue and sodium in grey. A WIMP mass of 100\,GeV/$c^2$ and a cross-section of $10^{-45}$\,cm$^{2}$ are assumed for the calculation.
In these curves both the $A^2$ dependence of the cross-section and the form factor correction affect the shape of the energy spectrum. Heavier elements profit from the $A^2$ enhancement with a higher event rate at low deposited
energies but the coherence loss due to the form factor suppresses the event rate especially at higher recoil energies. Therefore, for lighter targets a low energy threshold is of less relevance than for the heavier ones.
Figure\,\ref{Fig:DD_diffRate} (right) shows separately the WIMP mass and the form factor effect on the differential event rate without considering the nuclear recoil acceptance and the energy threshold of the detector. 
Solid lines show the expected rates for a 100\,GeV/$c^2$ WIMP as in the left figure for a heavy and a light target as indicated in green (tungsten) and blue (argon), respectively.
In comparison to the heavy WIMP mass the rates 
for a 25\,GeV/$c^2$ dark matter particle (dashed line) drop steeper as the momentum transfer is smaller. 
The form factor correction for a heavy target is more important than for light targets. This can be seen by the dotted lines representing rates  for a 100\,GeV/$c^2$ WIMP, calculated without the form factor correction.     

For spin-dependent interactions, the form factor is written in terms of the spin structure function whose terms are determined from nuclear shell model 
calculations\,\cite{Ressell:1997kx}\cite{Toivanen:2009zza}. 
A common practice is to express the cross-section for the interaction with protons and with neutrons
\begin{equation}\label{eq:3SD}
\sigma ^{\scriptsize{\textrm{SD}}} _0 = \frac{32}{\pi} \mu_A ^2 \cdot G_{F}^2\cdot \left [a_p \cdot \langle S^p \rangle + a_n\cdot \langle S^n \rangle \right ]^2 \cdot \frac{J+1}{J}.
\end{equation}
where $G_{F}^2$ is the Fermi coupling constant, $J$ the total nuclear spin and $a_{p,n}$ the effective proton (neutron) couplings. The expectation value of the nuclear spin content due to the proton (neutron) 
group is denoted by $ \langle S^{p,n} \rangle $.
New calculations performed in\,\cite{Klos:2013rwa} use chiral effective-field theory currents to determine the couplings of WIMPs to nucleons up to the leading two-nucleon currents. This method 
yields to an improved agreement between the calculated and measured energy spectra of the considered nuclei as well as the ordering of the nuclear levels (e.g.\,\cite{Menendez:2012tm}). These calculations have been used to calculate the couplings for the most relevant isotopes in direct detection experiments: $^{129,131}$Xe, $^{127}$I, $^{73}$Ge, $^{19}$F, $^{23}$Na, $^{27}$Al and $^{29}$Si.

In the context of a non-relativistic effective field theory (EFT) for WIMP-like interactions, a more detailed formulation of possible couplings from dark matter to baryons has 
been proposed\,\cite{Fitzpatrick:2012ib}\cite{Anand:2013yka}\cite{Fitzpatrick:2012ix} and is applied by some experiments\,\cite{Schneck:2015eqa}.
Instead of the classical two (spin-independent and -dependent) couplings, six possible nuclear response-functions are assumed which are described by 14 different operators.
In this model, the nucleus is not treated as a point-like particle, instead, its composite nature is reflected. Thus, the spin response function is split in transverse and longitudinal components and
new response functions arise from the intrinsic velocities of the nucleons. Note that the form factor F, as introduced above, tries to account for the finite spatial extend of the nuclear 
charge and spin densities. This correction, however, is only approximate. The EFT operators are constructed by four three-vectors $i\frac{\vec{q}}{m_N}, \vec{v}^{\perp},\vec{S}_N,\vec{S}_{\chi}$ which 
describe the momentum transfer $q$ scaled with the nucleon mass $m _N$, the WIMP-nucleon relative velocity $\vec{v}^{\perp}$, the spin of the nucleus $\vec{S}_N$ and the possible spin of the dark 
matter particle $\vec{S}_{\chi}$, respectively. 
The standard spin independent (equation\,\ref{eq:3SI}) and spin-dependent interactions (equation\,\ref{eq:3SD}) are described by operators $\mathcal{O}_1$ and $\mathcal{O}_4$ with $1$ being the identity matrix
\begin{equation}\label{eq:O14}
\mathcal{O}_1 = 1_{\chi} 1_N\,,\,\mathcal{O}_4 =  \vec{S_{\chi} } \cdot \vec{S_N} .
\end{equation}
The spin dependent interactions are, furthermore, decomposed into two longitudinal and a transversal spin components, as in general interactions do not couple to all spin projections symmetrically.
New operators arise also by a direct velocity dependence. The impact of the detailed EFT approach on the dark matter limits in comparison to the conventional SI\,/\,SD interaction has been calculated in\,\cite{Schneck:2015eqa}
and shows that, in some cases, the compatibility of results among experiments using different targets is significantly affected. 
Furthermore, destructive interference effects among operators can weaken standard direct detection exclusion limits by up to one order of magnitude in the coupling constants\,\cite{Catena:2015uua}.
This approach not only generalises the traditional SI and SD parameter space but also allows to constrain, in a easier way, dark matter models due to the variety of constrained operators.

\subsection{Other interpretations}\label{sec:OtherInterpr}

The previous section describes a model where dark-matter particles scatter off the target nucleus producing nuclear recoils, however, various other models exist. This section briefly summarises a selection of  alternative dark matter interactions for which experiments have derived results. 

An extension of the standard elastic scattering off nuclei is an inelastic scattering off the WIMP, which was motivated to solve discrepancies among experimental results\,\cite{TuckerSmith:2001hy}.
In this approach, WIMPs are assumed to only scatter off nuclei by simultaneously getting excited to a higher state with an energy $\delta$ above the ground state. The elastic scattering would be, in this case, highly suppressed or even forbidden. The energy spectrum is suppressed at low energies due to the velocity threshold for the inelastic scattering process. Experimental constraints on this model have been shown e.g. in\,\cite{Akimov:2010vk}\,\cite{Arrenberg:2011zz}\,\cite{Aprile:2011ts}.
Another possibility is the inelastic WIMP-nucleus scattering in which the target nucleus is left in a low-lying nuclear excited state\,\cite{Ellis:1988nb}. The signal would have a signature of a nuclear recoil followed by a $\gamma$-ray from the prompt de-excitation of the nucleus. A simultaneous measurement of the elastic and the inelastic signals would allow to distinguish between spin-independent and -dependent interactions in a single detector. The prospects for search in the next years using xenon detectors are studied in\,\cite{McCabe:2015eia}. The xenon inelastic structure functions have been calculated\,\cite{Baudis:2013bba} and are used\,\cite{Uchida:2014cnn} to derive exclusion limits for this process.

In contrast to interactions with nucleons, various models allow a dark matter scattering off electrons. 
For instance, sub-GeV dark matter particles could produce detectable ionisation signals\,\cite{Essig:2011nj} and, indeed, limits have been derived for such candidates\,\cite{Essig:2012yx}. Furthermore, if new forms of couplings are introduced to mediate the dark matter - electron interactions, further models become viable. By assuming an axial-vector coupling\,\cite{Kopp:2009et}, the dark matter-lepton interactions dominate at tree level and can not be probed by dark matter - baryon scattering. 
Furthermore, models such as kinematic-mixed mirror dark matter\,\cite{Foot:2013uxa} or luminous dark matter\,\cite{Feldstein:2010su} also predict interactions with atomic electrons.   

New couplings are, as well, introduced to mediate interactions of axion-like particles (ALPs) with electrons via the axioelectric (also photoelectric-like) or Primakov processes\,\cite{Bernabei:2005ca}\,(see section\,\ref{intro:DMcandidates}). These processes, invoked  by sufficiently massive particles in direct detection experiments,  involve only the emission of electrons and X-rays and can therefore not be separated from the experimental electronic recoil background. Nevertheless, bounds on these models have been derived from data of various experiments\,\cite{Abe:2012ut}\cite{Ahmed:2009ht}\cite{Armengaud:2013rta}\cite{Aprile:2014eoa}.
The same interactions are assumed for bosonic super-weakly interacting massive dark matter candidates\,\cite{Pospelov:2008jk} but their electronic recoil energy scale is in general higher and limits are derived in\,\cite{Abe:2014zcd}.

\subsection{Distribution of dark matter in the Milky Way}\label{sec:DMdistrib}

The dark matter density in the Milky way at the position of the Earth and its velocity distribution are astrophysical input parameters, needed to interpret the results of direct detection experiments. 
In this section, the parameters of the standard halo model typically used to derive the properties of dark matter interactions, their uncertainties and the differences in modelling the dark-matter halo itself are summarised.

It is common to assume a local dark matter density of 0.3\,GeV/cm$^3$ which results from mass modelling of the Milky Way, using parameters in agreement with observational data\,\cite{Green:2011bv}. However, depending on 
the profile model used for the halo, a density range from $(0.2-0.6)$\,GeV/cm$^3$ can be derived (see~\cite{Read:2014qva} for a review on this topic). 

The dark matter velocity profile is commonly described by an isotropic Maxwell-Boltzmann distribution
\begin{equation}\label{eq:vel_distr}
f(\textbf{v})=  \frac{1}{\sqrt{2\pi\sigma}}\cdot \exp\bigg(-\frac{|\textbf{v}|^2}{2\sigma^2}\bigg)
\end{equation}
which is truncated at velocities exceeding the escape velocity. Here, the dispersion velocity $\sigma$ is related to the circular velocity via $\sigma =\sqrt{3/2}\,v_c$. A standard value of $v_c =220$\,km/ is used for 
the local circular speed. This value results from an average of values found in different analyses\,\cite{Kerr:1986hz}. More recent studies using additional data and/or different
methods, find velocities ranging from $(200\pm20)$\,km/s to $(279\pm33)$\,km/s\,\cite{Green:2011bv}. Finally, the escape velocity 
defines a cut-off in the description of the standard halo profile. The commonly used value of 544\,km/s is the likelihood median 
calculated using data from the RAVE survey\,\cite{Smith:2006ym}. The 90\% confidence interval contains velocities from 498\,km/s to 608\,km/s. 
These large ranges of possible values for the dark matter density, circular  speed and escape velocity illustrate that the uncertainties in the halo modelling are significant\,\cite{Green:2017odb}. The GAIA 
satellite\,\footnote{http://sci.esa.int/gaia}, in orbit since January 2014, has been designed to measure about a billion stars in our Galaxy and throughout the Local Group. These unprecedented positional 
and radial velocity measurements will reduce the uncertainties on the local halo model of the Milky Way.

Not only the parameters of the dark matter halo show uncertainties but also modelling the halo itself inherits strong assumptions. A sharp truncation of the assumed Maxwell-Boltzmann distribution at the escape
velocity has to be unphysical, which motivated the idea of King models (e.g.\,\cite{King:1966}\cite{Chaudhury:2010hj}) trying to account naturally for the finite size of the dark matter halo. 
Its also possible that the velocity distribution is anisotropic giving rise to triaxial models, allowing different velocities in each dimension of the velocity vector (e.g.\,\cite{Evans:2000}\cite{Vogelsberger:2008qb}). If the dark matter halo is not virialized, it could give rise to local inhomogeneities e.g. subhalos, tidal streams or unbound dark matter particles with velocities exceeding the escape velocity. It is worth mentioning that the effect of these assumptions on the astrophysical parameters and dark-matter halo distributions on the results of different experiments is reduced by choosing the common values
as introduced above. However, the effects can be also energy dependent, thus altering the detector response for diverse target materials.  Therefore, other analysis methods are necessary to resolve these ambiguities (see section\,\ref{sec:Gene_results}).  
 
The dark matter density profile can only be indirectly observed (e.g. rotation velocities of stars), therefore, numerical simulations have been performed in order to understand the structure of halos. These simulations contained, traditionally, only dark matter\,\cite{Navarro:1995iw}\cite{Springel:2008cc}\cite{Stadel:2008pn}\cite{Diemand:2008in} and showed triaxial velocity distributions\,\cite{Vogelsberger:2008qb}.
The resulting haloes feature, however, cusped profiles with steeper density variations towards the centre of the halo, while observations favoured flatter cored-profiles. Moreover, the simulations predict a large amount of substructure, i.e. large number of subhaloes, in contradiction with the few haloes present in the Milky Way.
These issues, currently under investigation, might challenge the validity of the $\Lambda$CDM model and different possible solutions are discussed. 
One solution could be related to the nature of dark matter or its properties\,\cite{Weinberg:2013aya}. A warm dark matter candidate with a larger free-steaming length could, for instance, modify the halo density profile 
resulting in the observed cored-type profiles and suppressing the formation of small structure.
Another possibility is to consider candidates with weak interaction with matter but strong 
self-interaction\,\cite{Spergel:1999mh}. The elastic scattering of these particles in the dense central region could modify the energy and momentum distribution resulting in cored dark matter profiles.
Probably, the solution could be related to the absence of baryonic matter in the simulations. The effect of baryons to the halo mass distribution is observed, for instance, in the recent Illustris-1 simulation\,\cite{Vogelsberger:2014dza} which considers the coevolution of both dark and visible matter in the Universe.
Furthermore, sudden mass outflows can alter substantially the central structure of haloes\,\cite{Navarro:1996bv}. Dark matter simulations including also baryons\,\cite{Pontzen:2014lma} show how gas outflows can change the 
distribution of gas and stars. For sufficiently fast outflows, the dark matter distribution can be also affected explaining hereby the low central-halo densities. 

Nevertheless even with large simulations containing baryons, uncertainties in the dark matter halo remain and, thus, direct detection experiments generally use the common assumption of an isotropic Maxwell-Boltzmann distribution using values for astrophysical parameters as introduced above. In section\,\ref{sec:GenericResults}, a method to display results in an astrophysical independent representation is described.

\section{Background sources and reduction techniques}
\label{Sec:BG}

In order to identify unambiguously interactions from dark matter particles, ultra-low background experimental conditions are required.  This section summarises the various background contributions for a direct dark matter experiment. It includes external radiation by $\gamma$-rays, neutrons and neutrinos which is common for all experiments and internal backgrounds for solid-state and for liquid detectors.  The main strategies to suppress these backgrounds through shielding, material selection, and reduction in data analysis are also discussed.

\subsection{Environmental gamma-ray radiation}

The dominant radiation from gamma-decays originates from the decays in the natural uranium and thorium chains, as well as from decays of common isotopes e.g. $^{40}$K, $^{60}$Co and $^{137}$Cs present in the surrounding materials.  The uranium ($^{238}$U) and thorium ($^{232}$Th) chains, have a series of alpha and beta decays accompanied by the emission of several gamma rays with energies from tens of keV up to 2.6\,MeV (highest $\gamma$-energy from the thorium chain).  
The interactions of $\gamma$-rays with matter include the photoelectric effect, Compton scattering and $e^-\,e^+$ pair production\,\cite{Leo1987}. While the photoelectric effect has the highest cross-section at energies up to few hundred keV, the cross-section for pair production dominates above several MeV. For the energies in between, the Compton scattering is the most probable process. All these reactions result in the emission of an electron (or electron and positron for the pair production) which can deposit its energy in the target medium. Such energy depositions can be at energies of a few keV affecting the sensitivity of the experiments because this is the energy region of interest for dark matter searches.

Gamma radiation close to the sensitive volume of the detector can be reduced by selecting materials with low radioactive traces. Gamma-spectrometry using high-purity germanium detectors is a common and powerful technique to screen and select radio-pure materials.  Other techniques such as mass spectrometry or neutron activation analysis are also used for this purpose\,\cite{Heusser:2005hz}.
The unavoidable gamma activity from natural radioactivity outside the experimental setup can be shielded by surrounding the detector by a material with a high atomic number and a high density, i.e. good stopping power, and low internal contamination.  Lead is a common material used for this purpose. Large water tanks are also employed as they provide a homogeneous shielding as well as the background requirements. 
To reduce the gamma-ray activity from radon in the air, the inner part of the detector shield is either flushed with clean nitrogen or the radon is reduced using a radon trap facility\,\cite{Armengaud:2013vci}.

Analysis tools can be used to further reduce the rate of background interactions. Given the low probability of dark matter particles to interact, the removal of multiple simultaneous hits in the target volume can be, for instance, used for background-event suppression. This includes tagging time-coincident hits in different crystals or identifying multiple scatters in homogeneous detectors. For detectors with sensitivity to the position of the interaction, an innermost volume can be selected for the analysis (fiducial volume). As the penetration range of radiation has an exponential dependence on the distance, most interactions take place close to the surface and background is effectively suppressed. Finally, detectors able to distinguish electronic recoils from nuclear recoils (see section\,\ref{sec:detectors}) can reduce the background by exploiting the corresponding separation parameter.

\subsection{Cosmogenic and radiogenic neutron radiation}\label{sec:NeutronBG}

Neutrons can interact with nuclei in the detector target via elastic scattering producing nuclear recoils.  This is a dangerous background because the type of signal is identical to the one of the WIMPs. Note that there is also inelastic scattering where the nuclear recoil is typically accompanied by a gamma emission which can be used to tag these events.
Cosmogenic neutrons are produced due to spallation reactions of muons on nuclei in the experimental setup or surrounding rock. These neutrons can have energies up to several GeV\,\cite{Mei:2005gm} and are moderated by the detector surrounding materials resulting in MeV energies which can produce nuclear recoils in the energy regime relevant for dark matter searches. In addition, neutrons are emitted in $(\alpha,n)$- and spontaneous fission reactions from natural radioactivity (called radiogenic neutrons\,\cite{Westerdale:2017kml}). These neutrons have lower energies of around a few MeV.

\begin{figure}[h]
  \begin{center}
   \includegraphics[angle=0,width=0.55\textwidth]{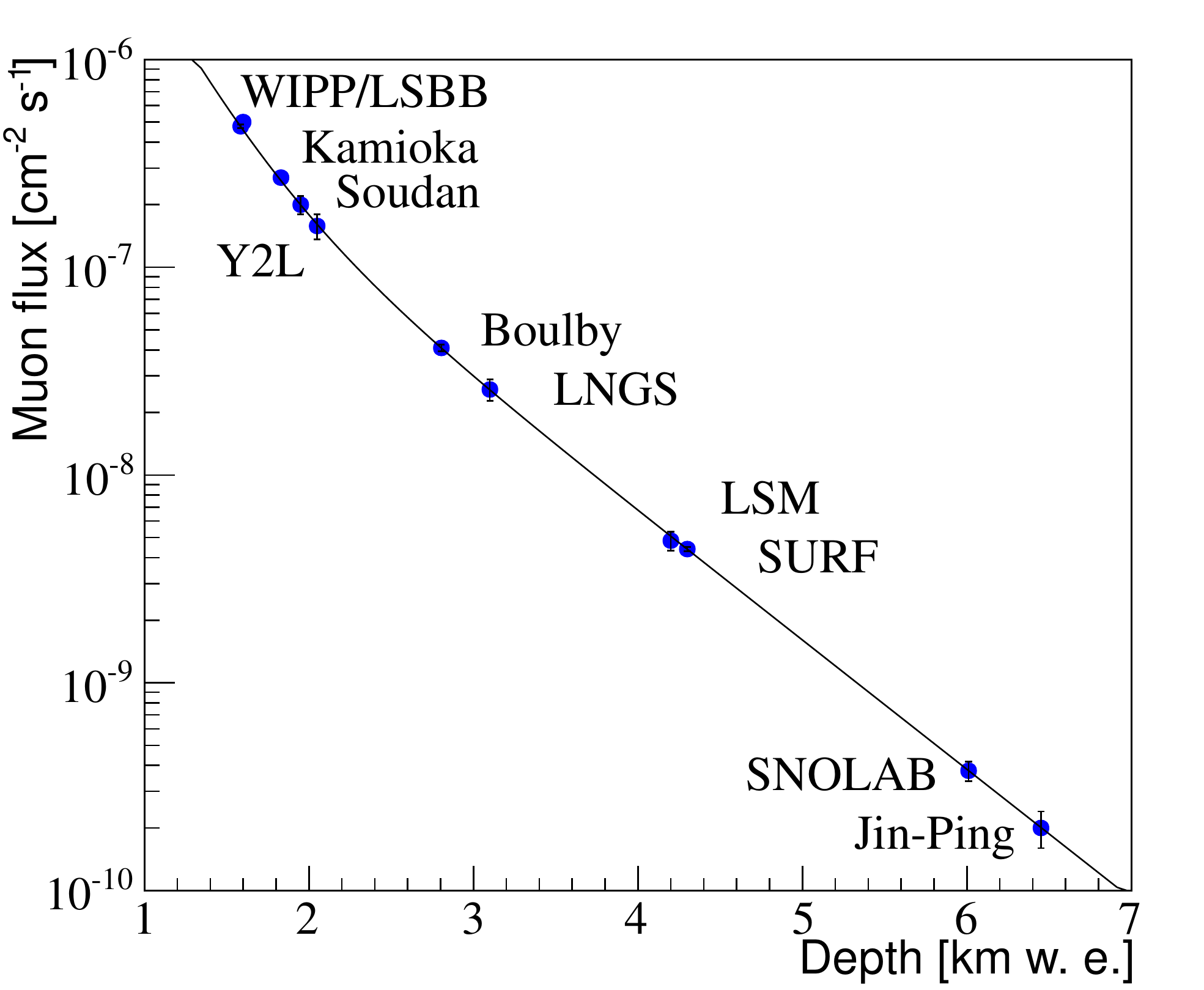}
   \caption[]{Muon flux as function of depth in kilometres water equivalent (km w. e.) for various underground laboratories hosting dark matter experiments. The effective depth is calculated using the parametrisation curve (thin line) from\,\cite{Mei:2005gm}. \label{MuonFluxShielding}}
  \end{center}	
\end{figure}

Dark matter experiments are typically placed at underground laboratories in order to minimise the number of produced muon-induced neutrons. The deeper the location of the experiment, the lower the muon flux. Figure\,\ref{MuonFluxShielding} shows  the muon flux as a function of depth for different laboratories hosting dark matter experiments. 
 The effective depth is calculated using the parametrisation from\,\cite{Mei:2005gm} which is represented by the black line in the figure.
The muon flux for each underground location is taken from the corresponding reference of the list below.
\begin{itemize}
 \item Waste Isolation Pilot Plant (WIPP)\,\cite{Esch:2004zj} in USA
 \item Laboratoire Souterrain \`{a} Bras Bruit (LSBB)\,\cite{Waysand:1999zp} in France
 \item  Kamioka observatory\,\cite{Mei:2005gm} in Japan 
 \item Soudan Underground Laboratory\,\cite{Mei:2005gm} in USA 
 \item  Yang Yang Underground Lab (Y2L)\,\cite{Kim:2012yv} in Corea
 \item Boulby Underground Laboratory\,\cite{Mei:2005gm}  in UK
 \item Laboratori Nazionali del Gran Sasso (LNGS)\,\cite{Mei:2005gm}  in Italy
 \item Laboratoire Souterrain de Modane (LSM)\,\cite{Berger:1989hs} in France
 \item Sanford Underground Research Facility (SURF)\,\cite{Mei:2005gm} in USA
 \item SNOLAB\,\cite{Mei:2005gm} in Canada
  \item Jin-Ping laboratory\,\cite{Yu-Cheng2013} in China 
\end{itemize}

The flux of radiogenic neutrons can be reduced via material selection. Detector materials with low uranium and thorium content give lower $\alpha$- and spontaneous fission rates. 
In addition, detector shielding can be used to reduce the external neutron flux further. Often water or polyethylene layers are installed around the detector setup to moderate the neutrons effectively\,\cite{Aprile:2011dd}. Active vetoes are designed to record interactions of muons.  The data acquired in the inner detector simultaneously to the muon event is discarded in order to reduce the muon-induced neutron background. Plastic scintillator plates are, for example, used for this purpose\,\cite{Armengaud:2013vci}\cite{Akerib:2004in}. This can be improved further by the use of water Cherenkov detectors\,\cite{Akerib:2012ys}\cite{Aprile:2014zvw} as they provide a higher muon tagging efficiency (full coverage), are efficient in stopping neutrons and, for sufficiently large thickness, the external gamma activity is also reduced. To tag directly the interactions of neutrons, shielding using liquid scintillators can be used\,\cite{Bossa:2014cfa}.

Finally, the analysis techniques described in the previous section can also be applied to reduce the neutron background. The multiple scattering tagging is, for instance, particularly effective with growing size of targets. The fiducial volume selection can also be used, however, it has a smaller effect in the reduction of background for neutrons than for gamma interactions because of the larger mean free path of neutrons.

\subsection{Neutrino background}
 
 With increasing target masses approaching hundreds of kilograms to tons, direct dark-matter detectors with sensitivity to keV energies start being sensitive to neutrino interactions. Neutrinos will become, therefore, a significant background contributing both to electronic and nuclear-recoils. Solar neutrinos can scatter elastically with electrons in the target via charged and neutral current interactions for $\nu_e$ and only neutral current for the other neutrino flavours\,\cite{Cabrera:1984rr}. Due to their larger fluxes, pp- and $^7$Be-neutrinos would be the first neutrinos which  could be detected. The resulting signal is a recoiling electron in contrast to the nuclear recoil resulting from WIMP interactions. Therefore, neutrino-electron scattering is an important background mainly for experiments which are not able to distinguish between nuclear and electronic recoils (see section\,\ref{sec:detectors}). Here, we consider neutrino-induced reactions as background but the measurement is interesting on itself as it can confirm the recent pp-neutrino measurement by the Borexino experiment\,\cite{Bellini:2014uqa}, testing in real time the main energy production mechanism inside the Sun.
 
 Neutrinos can also undergo coherent neutrino-nucleus elastic scattering producing nuclear recoils with energies up to few keV\,\cite{Freedman:1973yd}. Although this process has not been measured yet, it is expected to be accessible in the experiments planed to run in the next couple of years. Dark matter detectors could be, hereby, the first to measure this process. Coherent scattering of solar neutrinos would limit the sensitivity of dark matter experiment for low WIMP masses (few GeV) for cross-sections around $\sim10^{-45}$\,cm$^{2}$. For higher WIMP masses, the coherent scattering of atmospheric neutrinos would limit dark matter searches at $\sim 10^{-49}$\,cm$^{2}$\,\cite{Strigari:2009bq}\cite{Gutlein:2010tq}\cite{Ruppin:2014bra} (see also figure\,\ref{fig:SensitEvol}). In case of a positive signal at these cross-sections, in principle, the modulation of the signal along the year could be considered in order to distinguish WIMPs from neutrinos. While the WIMP rate should peak around June 2\,nd (see section\,\ref{sec:signatures}), the rate of solar neutrinos should peak around January 3\,rd due to the larger solid angle during the perihelion.  The rate of atmospheric neutrinos also peaks around January\,\cite{Aartsen:2013jla} due to the changes in atmospheric density resulting from seasonal temperature variations.

\subsection{Internal and surface backgrounds} \label{InternalBG}

In contrast to the external background which are common to all types of detectors, internal backgrounds differ depending on the target state. Therefore, internal backgrounds for crystal and liquid-targets are discussed separately.

Crystalline detectors as germanium or scintillators are grown from high purity powders or melts. During the growth process remaining impurities are effectively rejected as their ionic radius does not necessarily match the space in the crystalline grid. In this way, the crystal growing process itself reduces internal contaminations, for instance with radium, uranium  or thorium\,\cite{Munster:2014mga}\cite{Danevich201144}\cite{Shields:2015wka}\cite{Bernabei:2014kaa}.
Important for these detectors is the surface contamination with radon decay products. Either $\alpha$-, $\beta$-decays or the nuclear recoils associated to the latter can enter the crystal depositing part of its energy.  
The incomplete collection of signal carriers results in events that appear close to the region of interest, where nuclear recoils from WIMP interactions are expected. To identify events happening close to the surface, new detector designs have been developed over the last years. For example, in germanium detectors interleaved electrodes can be placed on the detector surface in order to collect an additional signal identifying the position of the event\,\cite{Agnese:2013ixa}\cite{Broniatowski:2009qu}. In scintillating crystals, an effective reduction of surface-alpha events have been achieved by a new design with a fully scintillating surface\,\cite{Strauss:2014hog}. More details will be discussed in chapter\,\ref{Sec:Tech_Res}.

Furthermore, cosmic activation of the target or detector surrounding materials during the time before the detector is placed underground needs to be considered.  One of the most important processes in the production of long-lived isotopes is the spallation of nuclei by high energy protons and neutrons. As the absorption of protons in the atmosphere is very efficient, 
neutrons dominate the activation at the Earth's surface for energies below GeV\,\cite{Cebrian:2010zz}. Exposure time, height above sea level and latitude affect the yield of isotopes, therefore, by minimising the time at surface and avoiding transportation via airplane, the isotope creation can be reduced. Since these precautions can not always be taken, tools or studies targeted to quantify the background due to cosmogenic activation are required (see for example\,\cite{Martoff:1992oxa}\cite{Cebrian:2010zz}\cite{Back:2007kk}\cite{Amare:2014bea}). 

For noble gases, a contribution to the internal background originates from cosmogenic-activated radioactive isotopes contained in the target nuclei.
For argon, $^{39}$Ar with an endpoint energy at 565\,keV has a large contribution as it is produced from cosmic-ray activation at a level of 1\,Bq/kg in natural argon. In order to reduce it, argon from underground sources is extracted. It has been shown that in this way, the activity is reduced by a factor of 1\,400\,\cite{Agnes:2015ftt}. In xenon, cosmic activation produces also radioactive isotopes, all rather short-lived. $^{127}$Xe has the longest lifetime with 36\,d which is still short enough to decay within the start of the experiment\,\cite{Akerib:2014rda}. Xenon also contains a double beta decaying isotope, $^{136}$Xe, however its lifetime is so large, $2.2\times10^{21}$\,y\,\cite{Albert:2013gpz}, that it doesn't contribute to the background for detectors up to few tons mass. If necessary, this isotope can be removed relatively easy by centrifugation.
In addition, decays from the contamination of the target with krypton and the radon emanation from the detector materials contribute to the internal background. 
The $\beta$-decaying isotope $^{85}$Kr is produced in nuclear fission and it is released to the atmosphere by nuclear-fuel reprocessing plants and in tests of nuclear weapons. 
Krypton can be removed from xenon either by cryogenic distillation\,\cite{Abe:2008py} or using chromatographic separation\,\cite{Akerib:2012ys}. Both methods have been proven to work at the XMASS/XENON and LUX experiments, respectively.
Besides the reduction of krypton in the target, techniques to determine the remaining krypton contamination are necessary in order to precisely quantify  its contribution the remaining contamination. Recently, detections in the ppq (parts per quadrillion) regime of natural Kr in Xe have been achieved\,\cite{Lindemann:2013kna}. Another possible method is the use of an atom-trap trace analysis system\,\cite{Aprile:2013aza}.
Radon is emanated from all detector materials containing traces of uranium or thorium. Once radon is produced in these decay chains, it slowly diffuses throughout the material and can be then dissolved in the liquid target. An approach to reduce radon is to use materials with low radon emanation\,\cite{Zuzel2009}\cite{Battat:2014oqa}. Furthermore, methods to continuously  remove the emanated radon are being investigated\,\cite{Martens:2012sva}\cite{SLindemann2013}\cite{Bruenner:2016ziq}.  

For both solids and liquids, the surface deserves special attention. For example, radium accumulated at the surfaces of the target or in the materials in contact with the liquid can contribute to the background i.e. surface background and radon emanation. Surface treatment with acid cleaning and electropolishing have been proven to be effective in removing radioactive contaminants at the surfaces\,\cite{Heusser:1995wd}.

\section{Result of a direct detection experiment}
\label{sec:Gene_results}

This chapter gives a generic description of a dark matter experimental result starting with the signal production in the target media. The statistical treatment of the measured events is discussed as well as the representation of the derived results.

\subsection{Detector signals} \label{sec:detectors} 

\begin{figure}[b]
  \begin{center}
   \includegraphics[angle=0,width=0.9\textwidth]{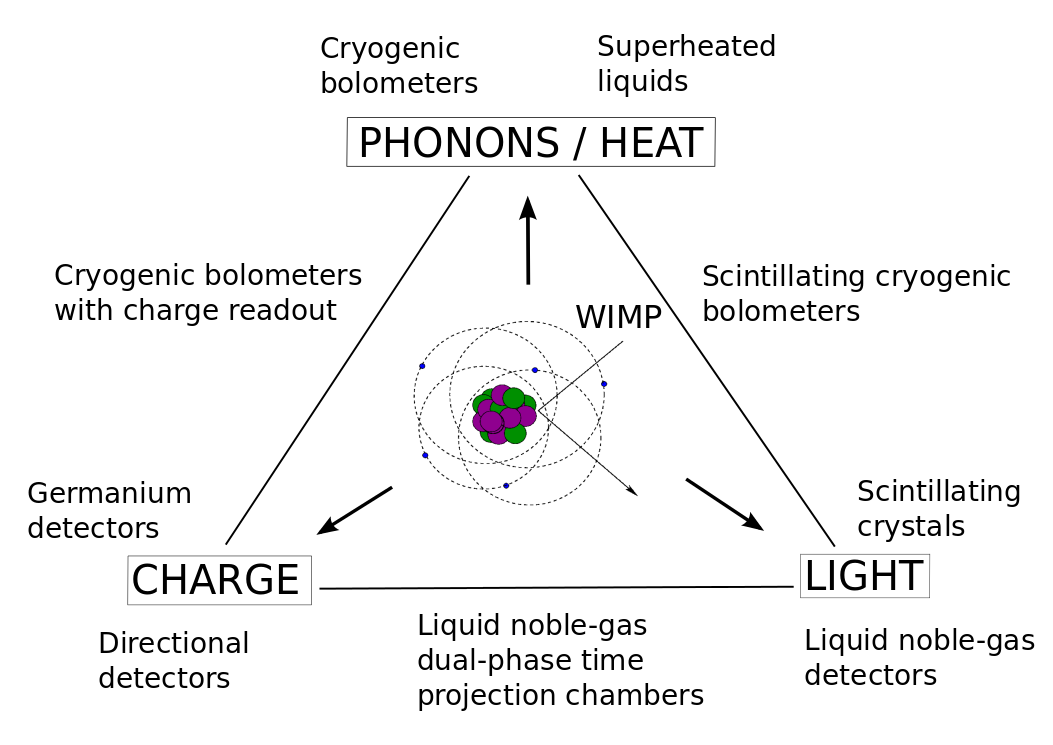}
   \caption[]{Schematic of possible signals that can be measured in direct detection experiments depending on the technology in use. \label{Fig:DetecDM_exp}}
  \end{center}	
\end{figure}

The elastic scattering of a dark matter particle off a target medium induces for the case of the WIMP an energy transfer to  nuclei which can be observed through three different signals, depending on the detector technology in use. These can be 
the production of heat (phonons in a crystal), 
an excitation of the target nucleus which de-excites releasing scintillation photons or by the direct ionisation of the target atoms. Detection strategies focus either on one of the three, or on the combination of two of these signals. Although, in principle, all three signals could be recorded, such an experiment does not exist to date. Figure\,\ref{Fig:DetecDM_exp} shows a scheme of the possible observables, as well as the most common detector technologies. 
A combination of two detection channels turns out to be powerful, since the response of media to an interaction is not only proportional to the 
deposited energy but depends on the type of particle that deposits the energy. More precisely, the relative size of the two signals depends on the type of particle. This enables the discrimination of e.g. nuclear recoils (neutrons, WIMPs) from electronic recoils (e.g. photon interactions, beta decays) which is an important method to reduce the background of the 
experiment. 
To measure the ionisation signal either germanium detectors or gases (low pressure, for directional searches) are employed while scintillation can be recorded for crystals and for noble-gas liquids. To detect heat, the
phonons produced in crystals are collected using cryogenic bolometers at mK temperatures.  The heat signal is also responsible for nucleation processes in experiments using superheated fluids.
Detectors which explore the discrimination power by measuring two signals are positioned in figure\,\ref{Fig:DetecDM_exp} between the corresponding signals: scintillating bolometers for phonon and light detection, germanium or silicon crystals to measure phonon and charge, and double phase (gas-liquid) 
noble-gas detectors for charge and light read-out. It is to mention that discrimination can also be achieved by exploiting other features in the response of the medium. For instance, the pulse shape of the signal depends on the particle type in liquid noble gas scintillators.  Detailed information on the various detector technologies used in direct dark-mater searches is given in section\,\ref{Sec:Tech_Res}.

\subsection{Statistical treatment of data}\label{sec:StatTreatm}

In direct detection experiments, various statistical methods are used to derive upper limits on the WIMP-nucleus cross-section as a function of the dark matter mass or to claim a detection of dark matter.
Over the last years, a number of experiments have recorded events above the expected background and based on those, signal contours in cross-section with nucleons versus dark-matter mass have been derived\,\cite{Bernabei:2010mq}\cite{Angloher:2011uu}\cite{Agnese:2013rvf}\cite{Aalseth:2014eft}. Some of those results have been, later on, disfavoured by the same authors based on new data from upgraded detectors.
In this potential 'discovery' situation, a correct application of statistical methods is essential to avoid a misidentification of up- or downward fluctuations of the background. Common to 
all experiments is not only that the expected signal consists of only a few events per year but also an unavoidable presence of background (see section\,\ref{Sec:BG} for a throughly explanation). Hence, a statistical analysis
has to consider both, the Poisson distribution of the signal events and a correct treatment of systematic uncertainties of the detector response. A detailed description of methods can be found in~\cite{Conrad:2014nna}.

\begin{figure}[h]
  \begin{center}
   \includegraphics[angle=0,width=1.0\textwidth]{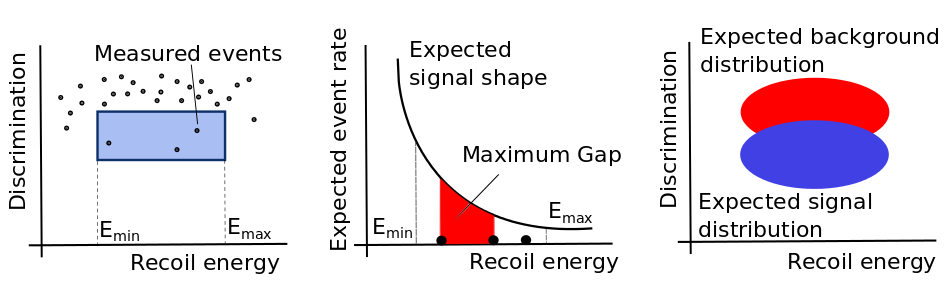}
   \caption[]{The left figure is an illustration of the statistical analysis method  where the result is reduced to a simple counting experiment in the Poisson regime in the presence of 
background (e.g. Feldman and Cousins). Yellin's method improved towards the simple box-based analysis by further considering the signal shape to derive limits (middle figure). If the probability density functions for the 
background and signal distribution are known, limits as well as a discovery can be calculated with a maximum likelihood analysis (right). For more information see text.\label{Fig:Statist_meth}}
  \end{center}	
\end{figure}

For detectors featuring a separation between different types of particles (discrimination), an intuitive approach is to select a signal region where the ratio of signal to expected background is high. This is indicated by the blue rectangle in the left panel of figure\,\ref{Fig:Statist_meth}.  
Due to the generally low number of expected events, their expectation value is described by a Poisson distribution. 
If the knowledge of the background distribution is available, e.g. using calibration data or a Monte Carlo simulation, a background prediction for the
signal region can be estimated. 
An exclusion limit (one sided confidence interval) or an interval representing the uncertainties on a possible signal (two sided confidence interval) can be computed based on a likelihood ratio of Poisson distributions developed by Feldman and Cousins\,\cite{Feldman:1997qc}.
This method gives the correct coverage, i.e. a quantification of how often the interval contains the true value of interest, and is able to decide between one and two sided confidence intervals. However, neither 
the knowledge of the background and signal probability density 
functions can be exploited, nor an uncertainty in the background prediction can be addressed. It is worth mentioning that other methods exist which include systematic uncertainties in the confidence interval 
construction\,\cite{Conrad:2002kn}\cite{Rolke:2004mj}.
The Feldman and Cousins method is used, for example, for the derivation of results of the PandaX (2014)\,\cite{Xiao:2014xyn}, ZEPLIN-III (2012)\,\cite{Akimov:2011tj} and SIMPLE (2012)\,\cite{Felizardo:2011uw} experiments.

If no background prediction is available, possible or its uncertainty is too large, Yellin's method\,\cite{Yellin:2002xd} can be used. In this method, the absence or low density of events in certain energy intervals (maximum gap or optimal interval, respectively) is used to calculate the probability of not measuring dark matter events in that interval. The middle panel of figure\,\ref{Fig:Statist_meth} shows how the maximum gap (red area) is defined given the measured data (black points) and the knowledge on the signal shape (black curve).
By not assuming a specific background model, the method is robust against unexpected background 
events. A one sided interval (upper limit) in the  presence of unknown background can be calculated leading always to conservative results which take into account the signal shape. By construction this method leads to one sided confidence intervals and, therefore, no signal discovery is possible. 
Current published upper limits derived by this method are e.g. SuperCDMS (2014)\,\cite{Agnese:2014aze}, 
CRESST-II (2014)\,\cite{Angloher:2014myn} and PICO (2015)\,\cite{Amole:2015lsj}.

Exploiting the full knowledge of the signal as well as the background distributions, allows to use a maximum likelihood estimation which typically results in stronger exclusion limits or to a higher significance of a signal (right panel in figure \ref{Fig:Statist_meth}). 
Furthermore, nuisance parameters can be treated in the context of a profile likelihood analysis in order to account for systematic uncertainties.
These parameters are not of immediate interest for the analysis of the signal model and, therefore, their uncertainties can be profiled out\,\cite{Cowan:2010js}\cite{Aprile:2011hx}. This method not only allows to penalise the result due to a limited knowledge of the detector response but also enables a natural transition between one and 
two sided confidence intervals. For setting exclusion limits, the maximum or profile likelihood analysis might be combined with a method developed in\,\cite{Junk:1999kv}\cite{Read:2002hq} to reduce the 
impact of a statistical downward fluctuation of the background on the result.
The results of XENON (2016)\,\cite{Aprile:2016swn}, LUX (2016)\,\cite{Akerib:2016vxi} and CDMS II Ge (2015)\,\cite{Agnese:2014xye} use this method to derive upper limits while in CDMS II Si (2013) \cite{Agnese:2013rvf} it was used to calculate the significance of the measured events above the expected background. 

Experimental results can be computed either by a frequentist or 
bayesian interpretation of the data. The former is extensively employed for direct detection experiments as reviewed above. The latter is, so far, less common in the analysis of dark matter experiments. The bayesian interpretation of likelihood and probability differs from the frequentist approach. In contrast to state the frequency 
of possible outcomes of an experiment by confidence intervals, bayesian credible intervals
allows a statement about the degree of belief of the tested hypothesis and is based on the Bayes' theorem. Thus, the computation of a probability for a theoretical model to be true based on the observed data 
is only possible with Bayesian statistics. In addition, it is necessary (or possible) to assign a priori information in form of a prior which might bias
the result if not chosen appropriately. Systematic uncertainties can, similarly to the profile likelihood method, be considered in the Bayesian framework which are later marginalised by 
Monte Carlo Markov-chains (MCMC) e.g. the Metropolis-Hastings algorithm\,\cite{Metropolis_1953}\cite{Hastings_1970}. A more detailed review of a Bayesian analysis of direct detection experiments can be found in~\cite{Arina:2013jma}.

\subsection{Generic result of a direct detection experiment}\label{sec:GenericResults}

The data of a dark matter experiment is an event rate consisting usually of only few counts and featuring a certain spectral shape (see section\,\ref{sec:signatures}). These results are commonly displayed in a parameter space of
the dark matter-nucleon cross-section and the dark matter mass. To derive these physical properties, astrophysical values for the dark matter density and its velocity distribution have to be assumed (see section\,\ref{sec:DMdistrib}). 
   
\begin{figure}[h]
  \begin{center}
   \includegraphics[angle=0,width=0.49\textwidth]{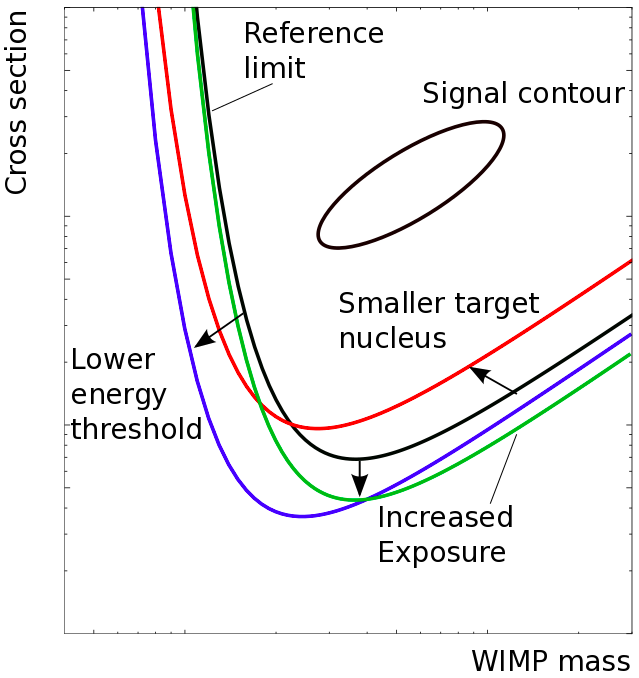}
   \includegraphics[angle=0,width=0.48\textwidth]{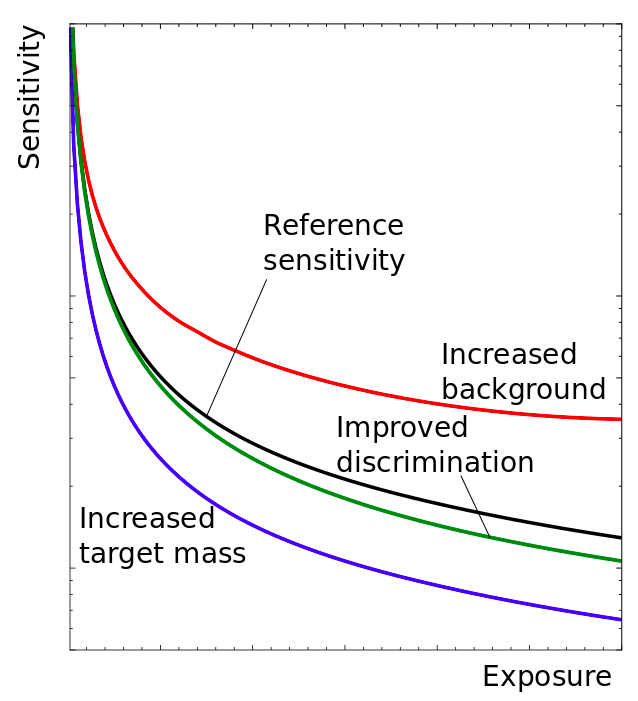}
   \caption[]{Left: Illustration of a result from a direct dark-matter detector derived as a cross-section with matter as function of the WIMP mass. The black line shows a limit and signal for reference, while the coloured limits illustrate the variation of an upper limit due to changes in the detector design or properties. Right: Evolution of the sensitivity versus the exposure. For more information see text. \label{fig:DD_Results}}
  \end{center}	
\end{figure}

The most common way to display direct detection results is based on a differential rate with spin independent and isospin-conserving interactions  
\begin{equation} \label{eq:diff_rate3}
\frac{dR}{dE} (E,t) = \frac{\rho_0}{2 \mu_{A}^2 \cdot m_{\chi}}  \cdot \sigma _0 \cdot A^2 \cdot F^2  \int _{v_{min}}^{v_{esc}}\frac{f({\bf v},t)}{v}~\textrm{d}^3v,
\end{equation}
with $v_{esc}$ the escape velocity (see section\,\ref{sec:DMdistrib}) and the minimal velocity defined as
\begin{equation} \label{eq:vmin}
v_{min} = \sqrt{\frac{m_A \cdot E_{thr}}{2\mu_{A}^2}} .
\end{equation}
The parameter $E_{thr}$ describes the energy threshold of the detector and $ \mu_{\rm{A}}$ is the reduced mass of the WIMP-nucleus system. 
The left plot in figure \ref{fig:DD_Results} shows a generic limit (open black curve) on the dark matter cross-section with respect to the dark matter mass which can be calculated with equation\,\ref{eq:diff_rate3}. 
At low WIMP masses the sensitivity is reduced mainly due to the low-energy threshold of the detector, whereas the minimum of the exclusion curve 
is given by the kinematics of the scattering process which depends on the target nucleus. 
At larger WIMP masses, the event rate is overall suppressed by 1/$m_{\chi}$. Given that the local dark-matter density is a constant of 0.3\,GeV/cm$^3$ (section\,\ref{sec:DMdistrib}), the heavier the individual particles, the less particles are available for scattering. In addition, the form factor reduces the rate for interactions with a large momentum transfer (section\,\ref{CrosSec_NuclPhys}).
The overall sensitivity of the experiment is dominated by the product of the size of the target and duration of the measurement, also called exposure, as well as the ability to avoid or reduce background events. 
A possible detection would be displayed as a contour region (closed black line) representing a certain confidence level. The coloured lines indicate qualitatively the influence of varying certain detector parameters. 
The exposure can be increased by either longer measurements or by an increased target mass.
An increase in exposure enhances the ability to measure lower cross-sections (green line). Note, however, that typically the background scales up with larger target masses and reduces the sensitivity if it is not simultaneously suppressed with improved techniques.
By using lighter target nuclei (red line), not only the kinematics of the elastic scattering is modified but also $v_{min}$ is reduced, resulting in a shift of the maximum sensitivity to lighter WIMP masses. The event rate is proportional to $A^2$ and thus, a smaller value of $A$ reduces the overall sensitivity. 
Lowering the energy threshold of the detector (blue line) allows not only to extend the sensitivity to lighter WIMP masses but
also reduces the value of $v_{min}$ and, hence, allows to test smaller cross-sections.  

The right plot in figure \ref{fig:DD_Results} illustrates the evolution of the sensitivity to the cross-section with respect to the exposure. For a given detector mass, the increase in exposure is caused by the accumulation of measuring time. The black line shows a reference curve. 
A non-discriminating detector, or a discriminating detector with an order of magnitude higher background, reaches its maximum sensitivity sooner (red line), and longer measurements do not improve the sensitivity. Assuming a constant background while enlarging the target mass, the sensitivity still increases with time (blue) and is not yet statistical limited. As mentioned before, to keep a low background level requires a higher purity of the detector and target material. An improved discrimination between background and signal events improves the acceptance to signal events and can also lead to a higher sensitivity (green line). 

\begin{figure}[h]
  \begin{center}
   \includegraphics[angle=0,width=0.49\textwidth]{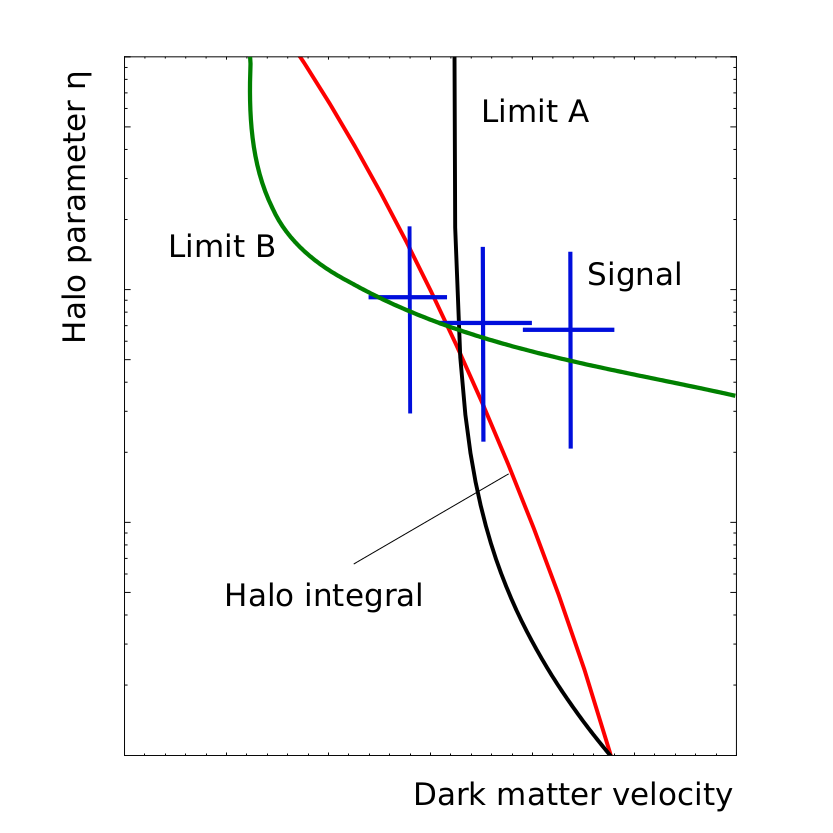}
   \caption[]{Illustration of the result of a direct detection experiment in the parameter space $\eta$ (see text) which is free of astrophysical assumptions and allows a direct comparison of different experiments. The blue markers indicate a dark matter
signal, whereas the black and green line indicate exclusion limits. The red line shows the velocity integral for a fixed choice of halo parameters. \label{fig:DD_vmin}}
  \end{center}	
\end{figure}

The choice of a different dark-matter halo model, $f({\bf v},t)$, affects the comparability of results from experiments using different target materials and technologies, as they might probe different  dark-matter velocity intervals\,\cite{Belli:1999nz}. Therefore, an alternative representation of the data which integrates out astrophysical uncertainties has been proposed\,\cite{Fox:2010bz}\cite{Fox:2010bu}. 
By displaying results in a parameter space which is halo independent, direct comparisons of detectors are feasible. This is possible by defining a parameter $\eta$ containing all astrophysical assumptions 
\begin{equation}
\eta = \frac{\rho_0 \cdot \sigma _0} {m_{\chi}}  \cdot \int _{v_{min}} \frac{f({\bf v},t)}{v} \textrm{d}^3v .
\end{equation}
Using the monotonicity of the velocity integral\,\cite{Fox:2010bz}, $\eta$ can be approximated to be independent of the detector response function and, hence, being common to all experiments. 
Figure \ref{fig:DD_vmin} shows examples of exclusion limits (black and green lines) and signals (blue crosses) for a fixed dark-matter mass, using $\eta$ and the dark matter velocity as free parameters. 
The dark-matter velocity is defined as the minimum speed that a WIMP requires in order to deposit a certain nuclear-recoil energy in the detector.
The red line represents a typical halo model and all velocities below the line describe its feasible physical parameters.
Experiments can probe the minimal velocity region above their exclusion limits up to the halo model (red line). Due to the use of different target elements and energy thresholds, these regions differ for each experiment. Hence, the limits from A and B cannot be compared to each other since they probe different velocity intervals of the halo model. Accordingly, only one and two sided confidence intervals within the same velocity interval are robust against the uncertainties of astrophysical parameters. This parameter space has been used in the last couple of years to display results of direct detection experiments (see as examples\,\cite{McCabe:2010zh}\cite{Frandsen:2011gi}\cite{DelNobile:2014sja}\cite{Blennow:2015gta}).

\section{Detector calibration}
\label{Sec:Calibr}

Results from dark matter detectors contain typically a low number of signal-like interactions for which their recoil energies are measured. In order to understand and interpret this data, the energy scale for the recoiling nucleus has to be characterised. This is particularly important to determine the energy threshold and to possibly constrain the WIMP mass. In addition, the signal and background regions  have to be calibrated in the parameter space relevant for the analysis.
For this purpose, regular calibrations using, for instance, radioactive sources are carried out. This section summarises the strategies and main features of calibration for different detector technologies focussing on the most competitive detector types.

\subsection{Calibration of the recoil-energies}\label{sec:cal_recoil_energy}

Depending on the detector technology (see section\,\ref{Sec:Tech_Res}), scintillation photons, phonons in a crystal and/or charge signal from ionisation can be measured. For a given energy deposition after a particle interaction, the corresponding recoil energy can be calculated applying a conversion function which contains quenching effects, i.e. the losses of signal due to various mechanisms as function of recoil energy. 
This function is, in general, different for electronic recoils and for nuclear recoils as the quenching mechanisms depend on the energy and nature of the interacting particle. In order to emphasise which function has been used for the conversion, energy units are expressed in electronic-recoil equivalent keV$_{ee}$ or nuclear-recoil equivalent keV$_{nr}$. 
Experiments recording phonons in a crystal lattice collect the full recoil energy in the form of phonons\,\cite{Shutt:1992qh}\cite{Alessandrello:1997ca}\cite{Simon:2002cw} and therefore, no signal quenching is usually considered. However, inhomogeneities inside the crystal (e.g. crystal defects) can, in principle, lead to phonon quenching.

\begin{figure}[h]
  \begin{center}
    \includegraphics[angle=-90,width=0.55\textwidth]{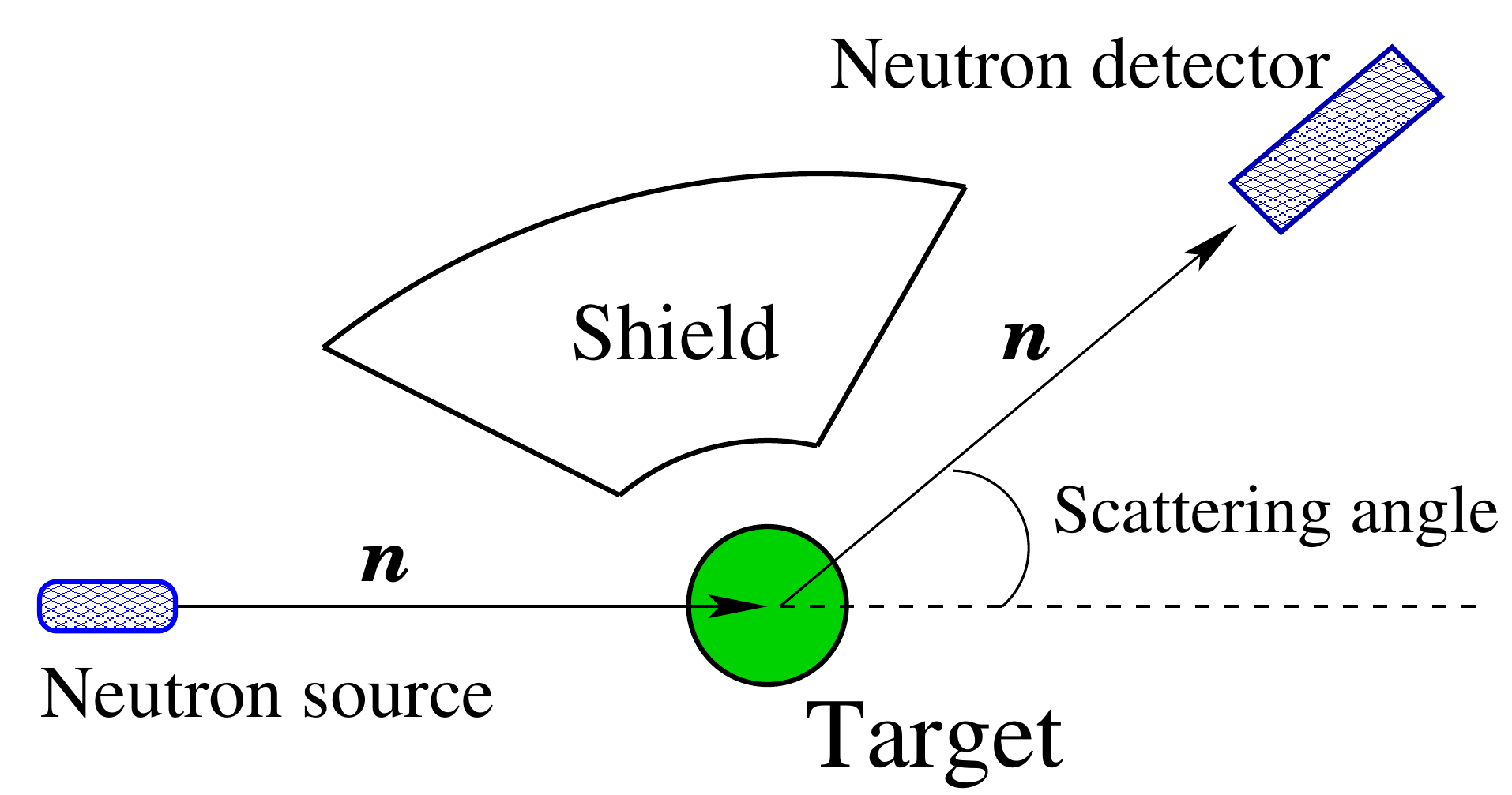}
   \caption[]{Scheme of a neutron scattering experiment using a neutron source, a target material under study and a coincidence neutron detector to determine the response of the target to mono-energetic nuclear recoils.\label{Fig:Scatt_Exp}}
  \end{center}	
\end{figure}

WIMPs are assumed to produce nuclear recoils and, hence,  an energy scale to convert the measured quanta to nuclear-recoil equivalent energy (keV$_{nr}$) is required. This scale is determined either with direct measurements using neutron scattering experiments or by comparing the nuclear recoil spectra of calibration neutron-sources to Monte Carlo (MC) generated ones. The first method is widely used as it is, in general, more robust due to the fewer assumptions involved compared to MC methods. In direct experiments (see scheme in figure\,\ref{Fig:Scatt_Exp}),  mono-energetic neutrons are scattered once in the medium and once in a detector operated in coincidence. 
By choosing a scattering angle and selecting single interactions, the kinematics are determined and a mono-energetic nuclear recoil can be selected. 
Varying the scattering angle provides the energy dependence of the signal yield. The results are usually normalised to the yield of a known electronic energy deposition. This method tests directly the kinematics of the WIMP interaction since single-scattering neutron events are selected similarly to the expected WIMP scattering process.

\begin{figure}[h]
  \begin{center}
     \includegraphics[angle=0,width=0.49\textwidth]{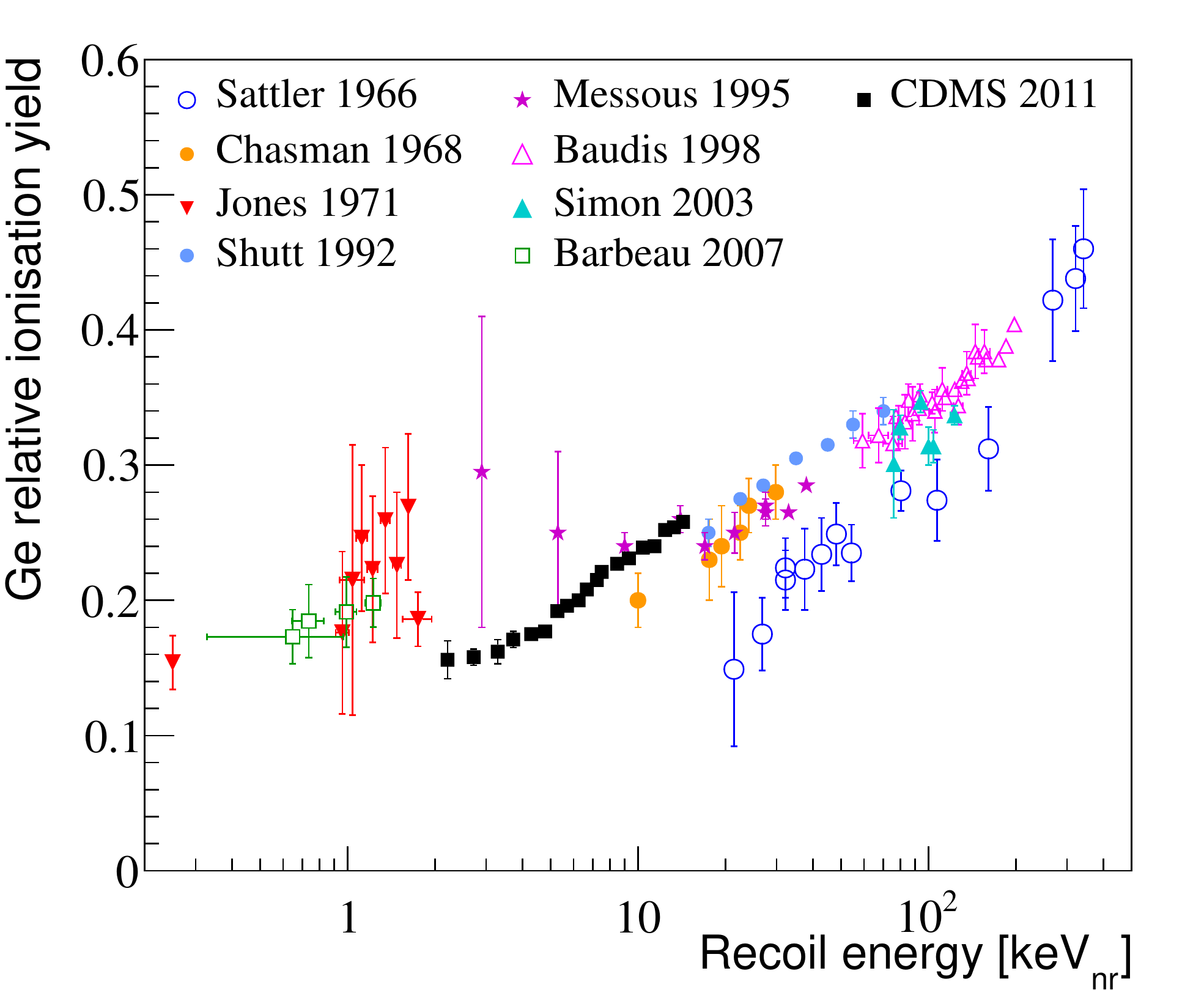}
     \includegraphics[angle=0,width=0.49\textwidth]{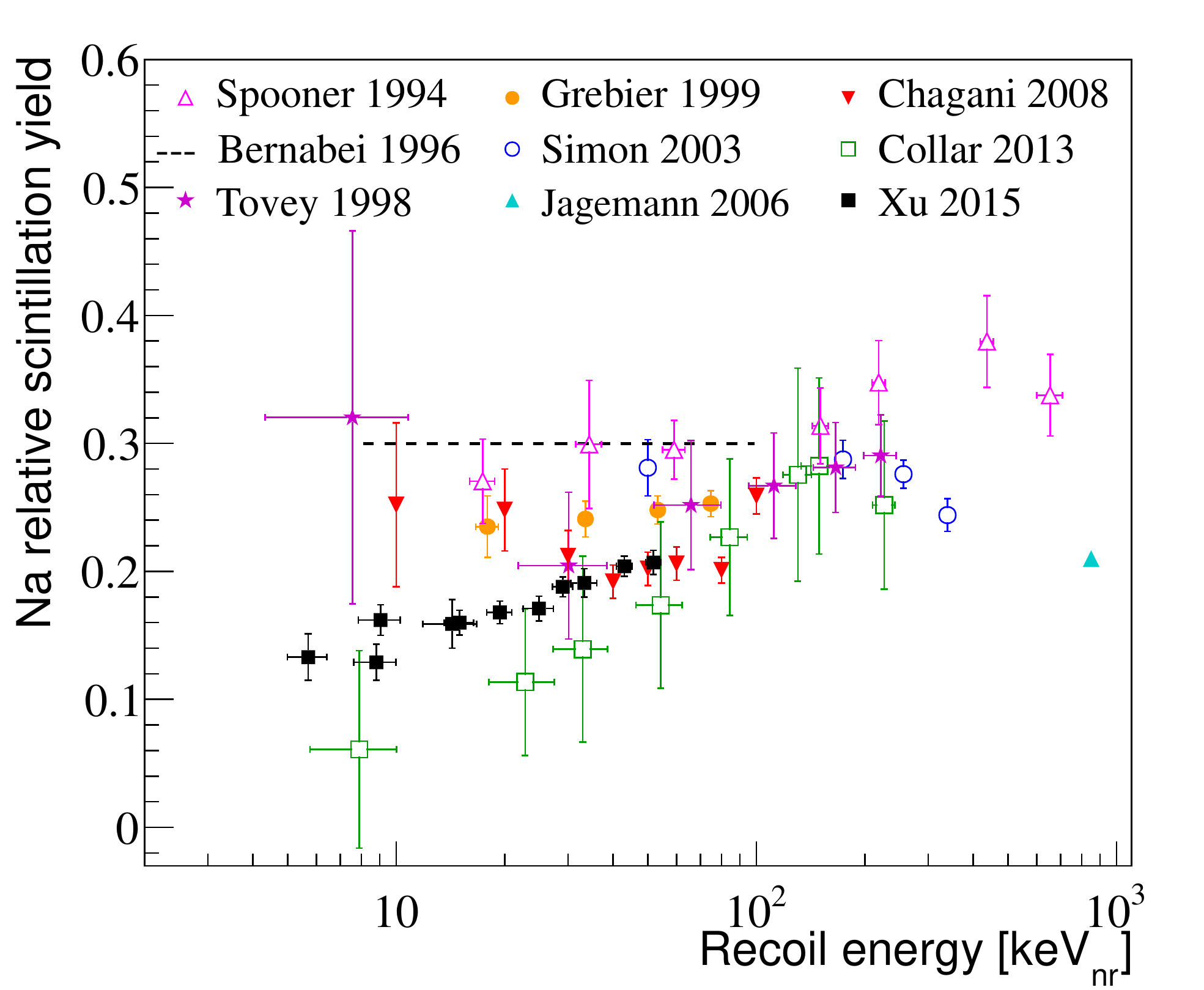}
    \includegraphics[angle=0,width=0.49\textwidth]{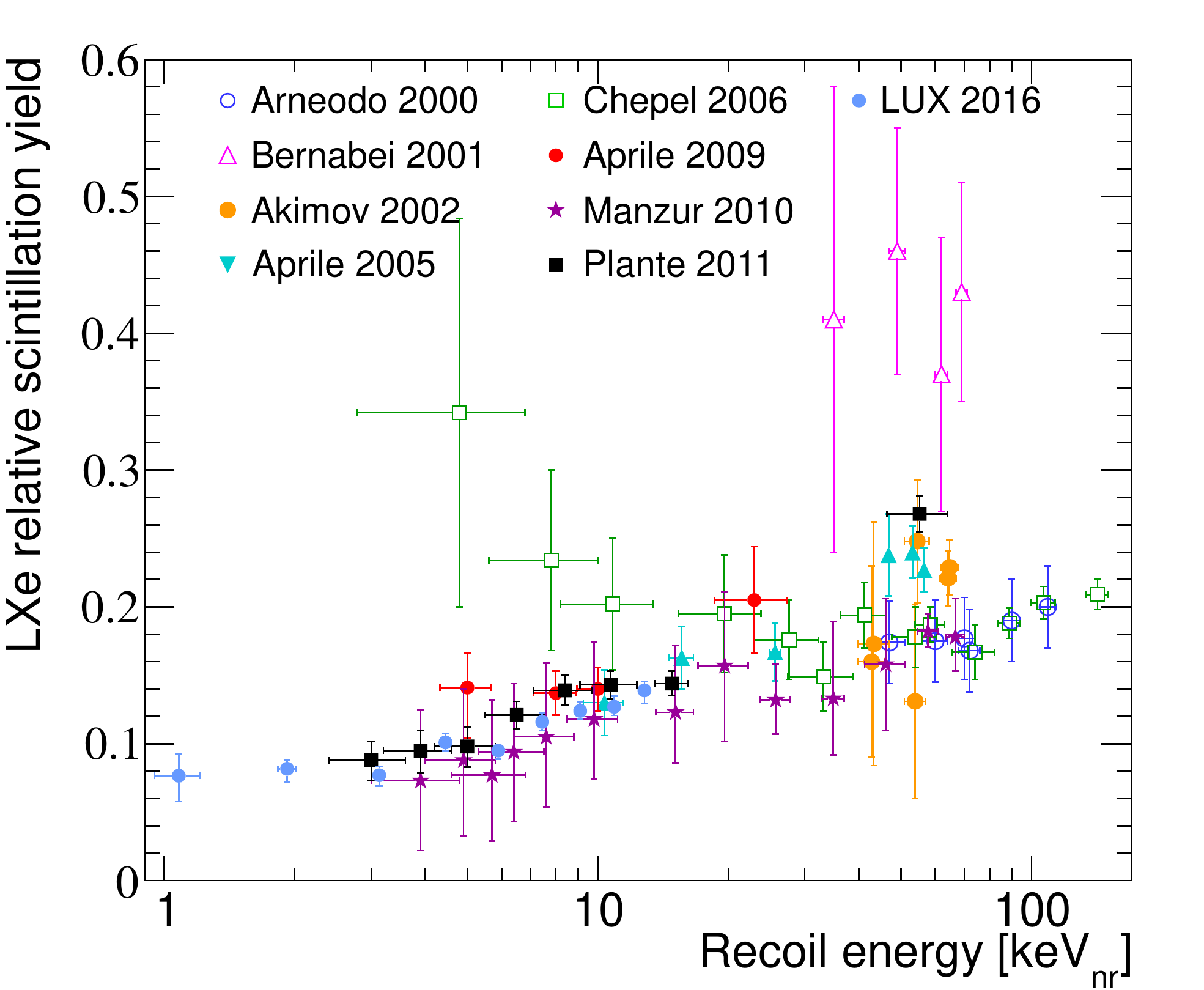}
    \includegraphics[angle=0,width=0.49\textwidth]{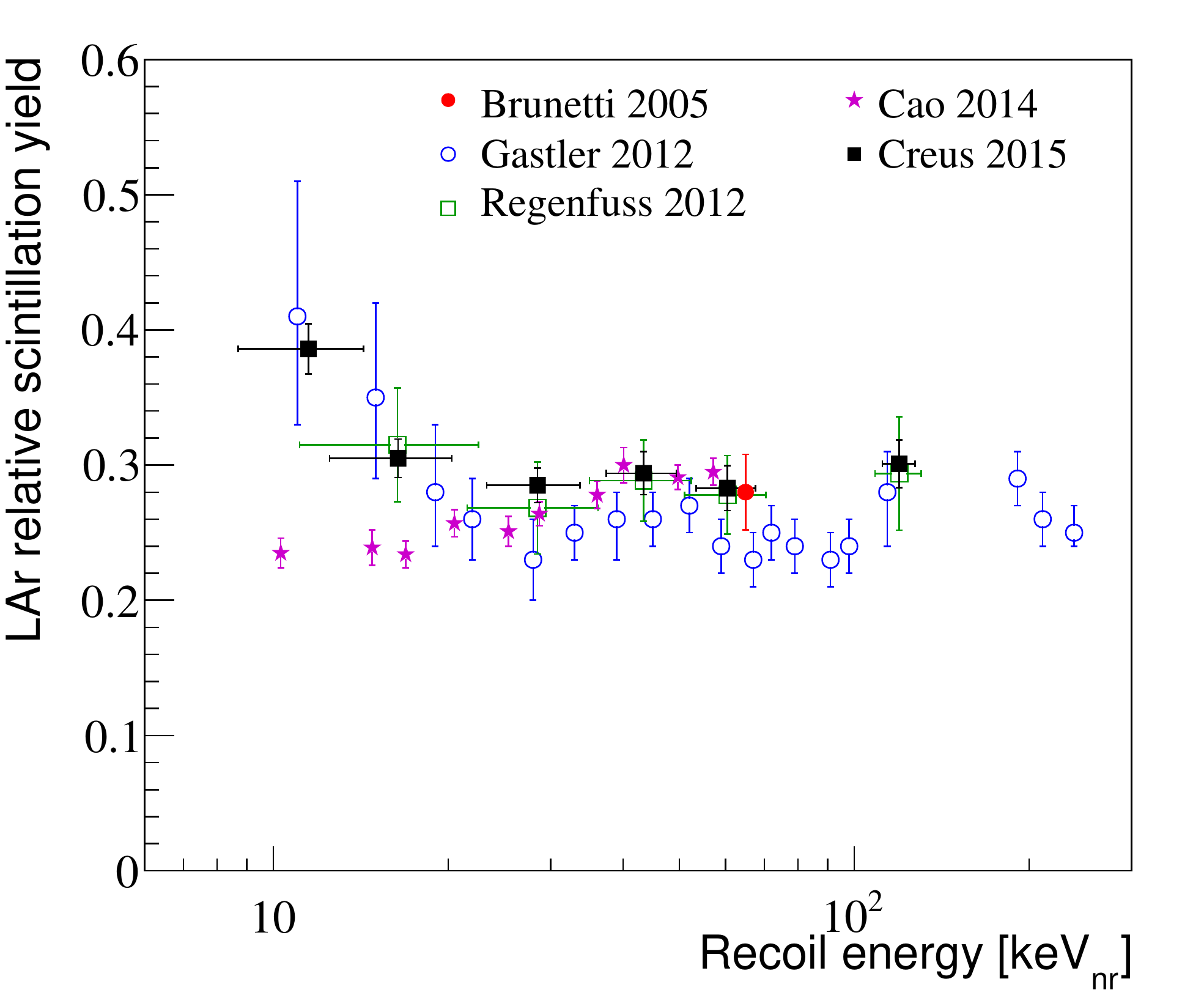}
   \caption[]{Measurements of signal quenching for various detector materials: ionisation efficiency in germanium (top left, from\,\cite{Barker:2012ek}\cite{Barbeau:2007qi} and references therein), scintillation efficiency for Na (top right, from\,\cite{Xu:2015wha} and references therein), for LXe (bottom left, from\,\cite{Plante:2011hw}\cite{Akerib:2016mzi} and references therein) and for argon (bottom right, from\,\cite{Regenfus:2012kh}\cite{Creus:2015fqa} and references therein).\label{Fig:Quenching}}
  \end{center}	
\end{figure}

Figure\,\ref{Fig:Quenching} shows the nuclear-recoil energy dependence of the signal yield for various media at energies relevant for dark matter searches. The top left figure shows the ionisation efficiency for germanium using the data from~\cite{Barker:2012ek}\cite{Barbeau:2007qi} and references therein. A recent measurement employing an $^{88}$Y/Be photo-neutron source confirms the ionisation yield shown in this figure down to approximately 0.5\,keV.
The light yield quenching is shown for sodium in a NaI crystal (top right), for liquid xenon (bottom left) and liquid argon (bottom right). The data is from~\cite{Xu:2015wha}, \cite{Plante:2011hw} and \cite{Regenfus:2012kh}\cite{Creus:2015fqa} including references therein for Na, LXe and LAr, respectively.  It can be seen that for sodium and xenon several measurements exist with some of them exploiting the quenching down to a couple of keV$_{nr}$ energies. 
In case of LXe, most results are compatible with each other. However,
one measurement (green squares in figure\,\ref{Fig:Quenching}) shows an increase of scintillation efficiency with decreasing energy. It is speculated\,\cite{Manalaysay:2010mb} that this might be caused by an
incomplete consideration of the detection efficiency. The most recent experimental
results show, contrarily, a decrease of the scintillation yield at decreasing energies.
The data points reaching the lowest energies correspond to an in-situ measurement with the LUX detector\,\cite{Akerib:2015rjg} performed using a DD neutron-generator\,\cite{Akerib:2016mzi}.

For argon, there exist two sets of recent measurements by~\,\cite{Cao:2014gns} and \cite{Creus:2015fqa} both with relatively small errors. These two results are in contradiction to each other within several sigma and further investigations will be necessary to understand the scintillation behaviour at low recoil energies. For LAr and LXe detectors operated with an electric field (see section\,\ref{LiquidNobleGas}), the field quenching has to be considered. For argon, dedicated measurements have shown a significant dependence of the light yield on the field, up to 32\% for energies between 11 and 50\,keV$_{nr}$\,\cite{Alexander:2013aaq}.
Note that this energy scale is used to derive the energy threshold of the experiment in keV$_{nr}$. As the WIMP recoil spectrum has an exponential shape (see section\,\ref{eq:diff_rate_simpl}), uncertainties in the threshold result in large variation of the expected number of events, especially at low WIMP masses.  Therefore, usually the energy threshold of an experiment should be in a region for which quenching data points exist. In addition to the measurements shown in figure\,\ref{Fig:Quenching}, dedicated quenching measurements have been performed for silicon\,\cite{Izraelevitch:2017gfi} which are for instance of relevance to the DAMIC experiment\,\cite{Barreto:2011zu}  (see section\,\ref{sec:NovelDetectors}).

A second method to derive the energy scale of a detection medium is to compare the spectrum of a neutron calibration-source to a Monte Carlo generated one. This method had been used for example for LXe detectors\,\cite{Horn:2011wz}\cite{Aprile:2013teh} giving results which are compatible with direct measurements. 
A complementary approach to describe the signal yield as a function of recoil energy is the theoretical modelling of the underlying processes, however, the accuracy of such descriptions has to be tested with data.  Often the model proposed by Lindhard\,\cite{Lidnhard:1963} is adopted but several other models exist. Some of them try to describe the electronic and nuclear stopping power at low energies as in\,\cite{Barker:2012ek} for ionisation in germanium. This model is verified by a dedicated data-MC comparison of neutron scattering off germanium\,\cite{Barker:2013nua}.
In~\cite{Mei:2007jn}\cite{Bezrukov:2010qa} and \cite{Szydagis2013}, the scintillation and ionisation for liquid noble-gas detector are modelled.  The former includes scintillation quenching from the Birks\,\cite{Birks1964} model while the latter two incorporate measured data into the description (see also the recent study\,\cite{Wang:2016obw}).

Note that for superheated liquid detectors, the energy calibration differs from the experiments mentioned above. The nuclear recoil scale is calculated using the 'hot spike' model of bubble nucleation\,\cite{Seitz} and it is verified by experimental data.
For example, the three consecutive $\alpha$-decays from $^{222}$Rn can be used to test the model\,\cite{Behnke:2012ys}. By varying the pressure and temperature of the detector volume, the energy threshold for a bubble nucleation is chosen. For different threshold energies, number of events is measured without the determination of the individual recoil energies (see also section\,\ref{sec:SuperheatedLiquids}).

In this section, the energy scale relevant for nuclear recoils produced by a WIMP-like dark matter candidate has been discussed. However, for candidates interacting with electrons instead with the nucleus, the corresponding electronic-recoil energy has to be applied. This scale is measured using mono-energetic signals from photo-/full- absorption of gamma rays, or by Compton effect coincidence experiments (see for example~\cite{Baudis:2013cca}) similar to the neutron-scattering mentioned above.

\subsection{Determination of signal and background regions}

Most experiments searching for WIMP interactions in a target material use either the combination of two signals (phonon, light or charge) or the pulse-shape of the signal to distinguish between the main background from electronic recoils by $\gamma$ and $\beta$-decays from the nuclear recoil signal. 
The signal and background regions are typically defined via dedicated calibration campaigns in between the science data taking. The distribution of nuclear recoils can be studied selecting interactions of neutron sources as $^{241}$AmBe or $^{252}$Cf. It is important to acquire enough nuclear-recoil statistics to have a precise determination of the signal region. In addition, the signal acceptance has to be quantified since this quantity enters directly into the sensitivity of the experiment. 

The modelling of the background composition of each experiment is required to calibrate the various components adequately.
As most of the background arises from electronic recoils from $\gamma$-interactions in the target, the background region can be determined by exposing the detector to gamma sources at different positions. Commonly, radioactive sources like $^{133}$Ba, $^{137}$Cs, $^{60}$Co or $^{232}$Th are used. For liquid noble-gas detectors also internal background contributes.  Dissolved internal sources can be used to characterise background distributions and indeed, a tritiated source has been used by the LUX experiment\,\cite{Akerib:2015wdi} and a $^{220}$Rn source is used in XENON to characterise the electronic recoil band\,\cite{Aprile:2016pmc}. 
For solid-state detectors also surface events need to be characterised (see section\,\ref{InternalBG}). This is typically carried out by exposing the crystals to $\beta$- or $\alpha$-emitters at different locations on the surface of the detector. Also in this case, it is desired to perform a high statistics measurement of the background as it enters the background prediction and its uncertainty. 
For superheated liquid detectors, the thermodynamic conditions are adjusted such that the medium is not sensitive to $\gamma$-rays or electrons. Therefore, only the background from $\alpha$-decays need to be characterised (see section\,\ref{sec:SuperheatedLiquids}).

\begin{figure}[h]
  \begin{center}
    \includegraphics[angle=0,width=1.0\textwidth]{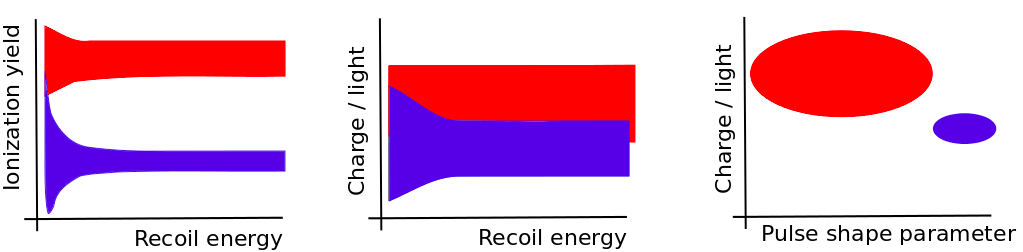}
   \caption[]{Schematic representation of signal (blue) and background (red) regions for a bolometer like a germanium detector (left), a liquid xenon TPC (middle) and a liquid argon TPC (right). \label{Fig:Det_SignalBG}}
  \end{center}	
\end{figure}

Figure\,\ref{Fig:Det_SignalBG} shows schematically how signal (in blue) and background (in red) events are distributed for some detector technologies. On the left the ionisation yield of a germanium bolometer is represented: the phonon signal is used to determine the energy scale and the normalised ratio of phonon to ionisation signal is used for signal discrimination. This type of detector achieves with this method a large separation of signal and background, e.g. a $10^6$ rejection of electronic recoils can be achieved\,\cite{Ahmed:2009zw}.

Only surface events with incomplete charge collection produce events leaking from the background region down to the signal region.
The middle panel of figure\,\ref{Fig:Det_SignalBG} shows the separation of signal and background for a liquid xenon detector using the ratio of charge to light signals. The signal discrimination is typically not as large as for bolometers. The highest $\gamma$-ray rejection factor achieved to date is at $5\times10^3$\,\cite{Lebedenko:2008gb}.
Finally,  the right panel shows, for a certain energy interval, the combination of two discrimination parameters as it can be done in liquid argon time-projection-chamber (TPC). In addition to the charge to light signal-ratio, the pulse shape of the scintillation can be used to separate signal and background. The WARP experiment, for example, made use of these two discrimination parameters in the data analysis\,\cite{Benetti:2007cd}. For each energy interval, the regions can be determined gaining a large signal acceptance and strong background suppression.

\section{Technologies and experimental results}
\label{Sec:Tech_Res}

In this section, the working principle of various technologies searching for dark matter is reviewed. As mentioned in section\,\ref{sec:Gene_results}, most experiments exploit either the 
phonon, charge or light signal, or a combination of some of those. In solid targets, phonon excitations of the crystal arise by the conversion of the kinetic energy from the scattering particle to lattice vibrations. The typical energy scale to create phonons in crystals is of the order of a few meV which is considerably lower than the energy of the quanta of light or charge.    
Charged particles moving through a medium ionise its atoms and the produced charges can be collected if an electric field is applied. 
To create an electron-hole pair in a semiconductor, a typical energy of a few eV is necessary\,\cite{Leo1987}, whereas for liquid nobel gases the ionisation energy is of the order of $(10-20)$\,eV\,\cite{Miyajima:1974zz}\cite{Takahashi:1975zz}. The photons emitted by scintillating materials are produced mainly by a relaxation of the excited medium. 
It is common to all scintillators (solid or liquids) that only a small fraction ($1-10$\,\%) of the total recoil energy is transferred to scintillation processes.

When describing the various existing detector technologies, the main challenges in direct detection will be considered:
a low energy threshold to detect the smallest recoil energies, a low background 
 to increase the signal significance and a large detector mass to increase the interaction probability inside the target. A forth and maybe underestimated goal is a stable detector performance
over time scales of a few years, where simpler detector configurations might be of advantage.    
In addition, the discrimination capabilities of different detectors will be discussed.
While this section describes the main technologies used in dark matter searches, including the respective main scientific results, overview figures summarising various experimental results are shown in section\,\ref{Sec:SumAndProsp}.

\subsection{Scintillator crystals at room temperature}\label{sec:Scin_crys}

Scintillators are some of the most used detection devices in particle physics. When radiation passes through a scintillating material, the atoms or molecules of the medium are excited and the subsequent de-excitation causes the emission of light. Among the various existing scintillators, mostly NaI(Tl) but also CsI(Tl) crystals are used in dark matter searches. 
In inorganic crystals, inhomogeneities are added to the crystalline structure as activators. These activators create crystal defects which act as additional luminescence centres\,\cite{Leo1987}. When adding the activator thallium  to NaI or CsI, the light emission of the crystals increases and the wavelength of the emitted light is shifted compared to the wavelength of the pure crystals to larger values (415\,nm and 580\,nm, respectively). At these wavelengths photosensors have a higher detection efficiency and the crystals show a better transparency.
The advantage of these inorganic crystals is the large stopping power arising from the high density (3.7 and 4.5\,g/cm$^3$ for NaI and CsI, respectively) and the large light output that results in a better energy resolution (around 8\,\% for 1\,MeV energy deposition) and lower energy threshold than other scintillators. Crystals can be grown with sizes of several cm$^3$. Therefore, in order to achieve larger target masses, the detectors are composed of several crystals.
An important advantage of this technology is its relative simplicity which allows to operate the detectors over large time periods of several years.
In these crystals only the scintillation signal is acquired, thus, no particle discrimination is possible, besides the rejection of multiple hits in different crystals. 
An event-by-event separation of signal and background is not possible but the annual modulation of the signal (see section\,\ref{sec:signatures}) can be used to identify dark matter interactions. 
Despite the absence of background rejection via discrimination, a high sensitivity can be achieved by keeping the overall background of the experiment sufficiently low. For this purpose, powders with low radioactive content on uranium, thorium and potassium are used to grow the crystals (see\,\cite{Shields:2015wka} as an example and section\,\ref{InternalBG}). In addition, most experiments use active vetoes operated in coincidence with the crystals to reduce further the background.

\begin{figure}[h]
  \begin{center}
   \includegraphics[angle=0,width=0.99\textwidth]{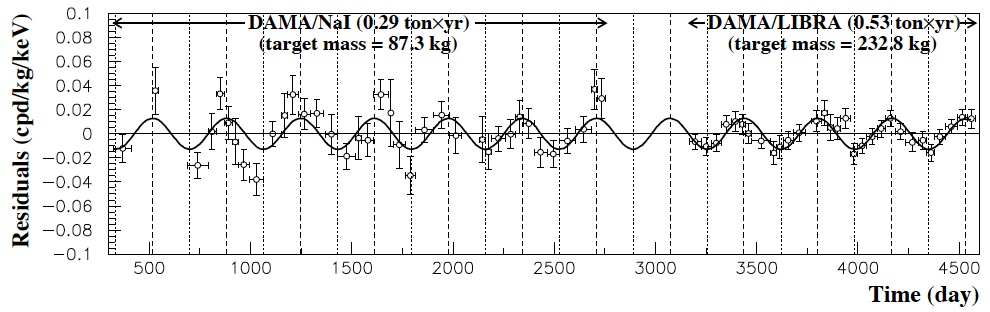}
   \caption[]{Annual modulation of the measured residual single-hit event rate by the DAMA experiment in the $(2-6)$\,keV energy range. The superimposed curve is a sinusoidal function with a period of one year and a phase equal to 152.5\,days (maximum on June 2\,nd). The figure covers the period between 1996 to 2007. Figure from \cite{Bernabei:2008yi}.\label{Fig:DAMAmod}}
  \end{center}
\end{figure}

The DAMA experiment at the LNGS underground laboratory\footnote{The complete names, location and shielding of the existing underground laboratories can be found in section\,\ref{sec:NeutronBG}.} is searching for dark matter using ultra low-radioactive NaI(Tl) crystals\,\cite{Bernabei:2008yh}.  A combined dataset of DAMA and its successor DAMA/LIBRA have collected 1.33\,ton\,$\times$\,y exposure showing an annual-modulated single-hit rate in the energy range $(2-6)$\,keV$_{ee}$ (keV electronic recoil equivalent, see section\,\ref{sec:cal_recoil_energy}). Its maximum is compatible with June 2\,nd within 2\,$\sigma$ which is the phenomenological expectation of the phase for dark matter interactions\,\cite{Drukier:1986tm}. Meanwhile, the significance of this signal is at 9.3\,$\sigma$ over a measurement of 14 annual cycles\,\cite{Bernabei:2013xsa}. The DAMA experiment has demonstrated, hereby, that this technology allows for a stable long-term  operation. Figure\,\ref{Fig:DAMAmod} shows the residual distribution of events of the DAMA experiment as function of time together with a fit to the data (black line). The residuals are calculated from the single-hit event rate after subtracting the constant background rate.
The DAMA experiment continues taking data and since 2010, the detector has been equipped with new photosensors which will allow for a lower energy-threshold of the experiment\,\cite{Bernabei:2014xoa}.

If the DAMA signals is interpreted as being caused by elastic WIMP-nucleus interactions, two favoured regions appear at $(10-15)$\,GeV/$c^2$ for scattering off sodium and $(60-100)$\,GeV/$c^2$ for scattering off iodine\,\cite{Savage:2008er}.  The calculation of the dark matter masses depend, as pointed out by the authors of~\cite{Xu:2015wha}, on the used scintillation efficiency. DAMA measured a scintillation efficiency of 0.3 independent of the recoil energy, however recent results shown in section\,\ref{sec:cal_recoil_energy}, favour significantly lower efficiencies. The new determination of the efficiency would result in larger reconstructed WIMP masses for the DAMA signal. 
Besides the standard spin-independent interpretation of the DAMA results, different interpretations of the annual modulation signal have been derived as, for example, the spin-dependent\,\cite{Bernabei:2001ve} and the inelastic scattering of WIMPs\cite{Bernabei:2002qr} (see section\,\ref{sec:OtherInterpr}). Since DAMA does not discriminate between electronic and nuclear recoil signals, dark matter candidates interacting dominantly with electrons are also considered\,\cite{Bernabei:2005ca}.

Results from other experiments that will be presented in the following are in tension with the various dark matter interpretations of this signal. Therefore, other non-dark matter related explanations of the DAMA signal are being discussed\,\cite{Schnee:2011ef}. The signal could be related to atmospheric muons, the rate of which is annually modulated due to temperature variations in the stratosphere\,\cite{Blum:2011jf}, or to combinations of muons and modulated neutrinos, caused by the varying Sun-Earth distance\,\cite{Davis:2014cja}. Furthermore, varying rates of background neutrons have been considered\,\cite{Ralston:2010bd}.
Some of those proposals have been refuted\,\cite{Barbeau:2014mla}\,\cite{Klinger:2015vga} but the signal and its interpretation remains controversial. A more detailed discussion about these explanations can be found in\,\cite{Davis:2015vla}.

To make independent cross-checks of the DAMA signal, a number of experiments are carried out using a similar technology at various underground locations. 
The SABRE collaboration has proposed\,\cite{Shields:2015wka} to test the DAMA signal at LNGS, in the same underground laboratory. The experiment will consist of  highly pure NaI(Tl) crystals in an active liquid scintillator veto to tag and reduce the $^{40}$K background from the crystals and the external background.
The Anais, DM-Ice and PICO-LON experiments use the same target as DAMA but are located at different locations.  While a signal confirmation at the same laboratory is desired to exclude experimental effects, a measurement at a different laboratory could give information on the possible origin of the modulation. 
The Anais experiment\,\cite{Amare:2014jta}\cite{Amare:2016rbf} is currently operating a 25\,kg NaI(Tl) detector at the LSC laboratory in Spain. First results showed the possibility to achieve a very low threshold at or below 2\,keV$_{ee}$ due to the excellent light yield of the crystals. On longer term, Anais aims to increase the total mass to 250\,kg improving the energy threshold and the internal radioactive contamination compared to the DAMA experiment. 
The DM-Ice experiment is currently operating 17\,kg of NaI crystals under the ice at the South pole at a depth of 2460\,m, and first annual modulation results have been released in 2016\,\cite{deSouza:2016fxg}. This data demonstrates the remote operation of the detector and stable environmental
conditions over 3.6 years and a measured background consistent with the expectation. Since the experiment is located in the southern hemisphere, any modulation related to seasonal effects (e.g. the modulation of atmospheric muons) would have a reverse phase to the northern hemisphere.
The PICO-LON project is developing low radioactive NaI crystals with the aim to construct  250\,kg setup in the Kamioka mine in Japan\,\cite{Fushimi:2015sew}.
Finally, the KIMs experiment located at the Yangyang laboratory in Korea used an array of 103\,kg CsI(Tl) crystals to test the DAMA signal caused by WIMP scattering off iodine. The energy region chosen for the analysis of the first data ranged from $(3-11)$\,keV$_{ee}$. This data, acquired between September 2009 and August 2010, showed no significant signal\,\cite{Kim:2012rza} and therefore, exclusion limits on the dark matter cross-section were derived assuming spin-independent interactions. These limits disfavours the WIMP-iodine nuclei interactions as the source of the DAMA signal. To test the interaction on sodium,  a program to develop ultra-low-background NaI(Tl) crystal detectors with lower background level and higher light yield than those of the DAMA experiment has been started\,\cite{Kim:2014toa}.

\subsection{Germanium detectors}\label{sec:germanium}

Germanium detectors combine a high radio-purity of the target material with a very low threshold down to $\sim 0.5$\,keV$_{ee}$ allowing to search for WIMPs down to masses of a few GeV/$c^2$. Such low energies are 
achieved when the detectors are operated in ionisation-mode, having no possibility to discriminate signal from background-like events. 
 To reduce the noise levels sufficiently,  the detectors are cooled down to the temperature of liquid nitrogen (77$\,$K), which is, in comparison to other technologies  (see section\,\ref{sec:bolometers}), relatively simple and does not require complex cooling systems.
The noise level scales up with larger crystal sizes due to an in general increased capacitance and dedicated optimisations in the detector layout are essential\,\cite{Radeka:1988}.    
The excellent energy resolution of these detectors (typically around 0.15\,\% at 1.3\,MeV) allows to identify and quantify background sources and, eventually, this knowledge 
can be used to reduce the background from radioactive contaminations. In contrast to n-type doped detectors, p-type semiconductors benefit from a dead-layer around the 
crystal which further shields external $\alpha$ and $\beta$ backgrounds. In addition, the rise-time of the signal can be used to discriminate surface background from bulk events. Still, a separation between electronic and nuclear recoils is not possible. 
At the energy threshold, the limiting feature for germanium detectors is, in general, noise from the detector itself as well as from the read-out electronics.

An ultra low background germanium detector was already used in 1987 to derive the first limits on dark matter interactions\,\cite{Ahlen:1987mn}. Nowadays, this technology is further improved, particularly to reduce the 
energy threshold and the background level.
For instance, the CoGeNT experiment\,\cite{Aalseth:2014eft} uses p-type point contact germanium detectors with a mass of 443\,g reaching an energy threshold of 500\,eV$_{ee}$. 
The CoGeNT detector has acquired, in total, 3.4\,years of dark matter data in the Soudan Underground Laboratory (see section\,\ref{sec:NeutronBG}) enabling a search for dark matter by an annual modulation 
of the measured event rate\,\cite{Aalseth:2012if}. An annual modulation of the rate was found in an energy interval of $(0.5\,-\,2)$\,keV$_{ee}$ with a phase corresponding to the phenomenological expectation for WIMPs 
at a level of 2.2\,$\sigma$. The amplitude of the signal is, however, a factor of $4-7$ larger than expected\,\cite{Aalseth:2014eft}. If this signal is interpreted as spin-independent interactions of WIMPs, a best fit value appears at a cross-section around $2.5\times10^{-41}$\,cm$^2$ for a 8\,GeV/$c^2$ WIMP mass. However, it is to mention that independent analyses on the released public data, with different assumptions for the background model \,\cite{Davis:2014bla}\cite{Aalseth:2014jpa}, did not find a significant signal.  
The CoGeNT data has also been investigated in order to search for signatures of axion-like particles\,\cite{Pospelov:2008jk} (see also section\,\ref{intro:DMcandidates}). These particles could interact in germanium via the axio-electric effect\,\cite{Derevianko:2010kz} (similar to photo-electric effect) producing an electronic recoil of an energy corresponding to the mass of the axion. The non-observation of such a peak in the spectrum has allowed to derive limits on the axion-electron coupling\,\cite{Aalseth:2008rx}. 
For the future, a larger experiment, named C-4, with 10 times more mass and lower background is currently designed\,\cite{Bonicalzi:2012bf}.

A similar detector technology is used at the MAJORANA low-background broad energy germanium detector,  MALBEK, operated at the Kimballton underground research facility\,\cite{Giovanetti:2014lxa}. The main motivation of this prototype detector is the demonstration of an ultra-low background level of approximately 3\,events/(ton$\cdot$y) in the neutrino-less double beta decay ($0\nu\beta\beta$) energy region of interest. A customised 465\,g  germanium crystal developed specifically for a low energy threshold of 600\,eV$_{ee}$ has been tested. So far, an exposure of 89.5\,kg\,$\cdot$\,d could be achieved, reaching a sensitivity down to $10^{-40}$\,cm$^2$. This excludes part of the CoGeNT signal region using a detector with the same target material.  

To avoid a high capacitance due to large crystal sizes which show generally a higher noise level, the CDEX-0\,\cite{Liu:2014juh} experiment uses an array of four smaller  (5\,g)  n-type germanium diodes, reaching a total exposure of 0.78\,kg$\cdot$d. This low-threshold development is based on the detectors used by the former TEXONO experiment\,\cite{Lin:2007ka}\cite{Li:2013fla} which was operated at a shallow site at the Khuo-Sheng reactor neutrino laboratory. The crystal array is surrounded by a NaI(Tl) crystal scintillator, serving as an anti-coincidence detector. The 
exceptional energy threshold of 177\,eV$_{ee}$ at 50\,\% signal efficiency allows dark matter searches for spin independent interactions down to masses of 2\,GeV/$c^2$. The measured spectrum agrees well with the background expectation allowing to place limits on the dark matter interactions. Earlier measurements with the CDEX-1 setup\,\cite{Yue:2014qdu} already disfavoured the CoGeNT result. The CDEX-10 prototype employs crystals in the order of 10 kg to study backgrounds and the energy threshold when the detector mass is scaled up. In the long-term a ton-scale project is aimed at\,\cite{Yue:2016epq}.

\subsection{Cryogenic bolometers}\label{sec:bolometers}

\begin{figure}[h]
  \begin{center}
   \includegraphics[angle=0,width=0.55\textwidth]{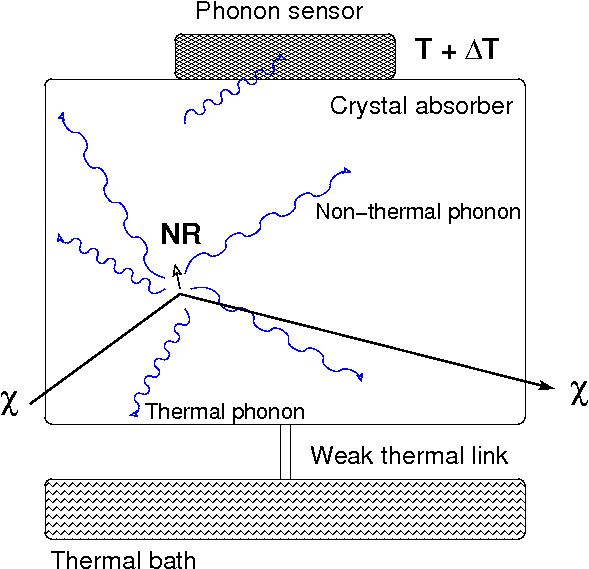}
   \caption[]{Schematic of a cryogenic phonon detector: an energy deposition $E$  from a nuclear recoil (NR) in an absorber of capacity $C(T)$ produces a temperature rise $\Delta T$ which is measured by a thermal sensor. \label{Fig:Bolometer}}
  \end{center}	
\end{figure}

Detectors collecting the phonon signal produced in a crystal are developed to reach very low thresholds and excellent energy resolution. If, in addition, the scintillation or charge signal is recorded, the
energy dependence of the signal quenching can be used to discriminate between nuclear and electronic recoils. 
This can be achieved in cryogenic bolometers, where an energy deposition by a nuclear or an electronic recoil is dissipated via collisions with the nuclei and electrons in the crystal lattice. 
A schematic representation of the phonon-detector working principle is shown in figure\,\ref{Fig:Bolometer}. 
The dissipated energy produces phonons which can be categorised in thermal and non-thermal phonons (also called athermal). Thermal phonons are related to the thermal equilibrium of the medium after an energy deposition and can be measured by the induced temperature rise. Athermal phonons describe a fraction of the initially produced phonons which are out of equilibrium and show a larger mean free path in the medium. These phonons carry not only the information about the energy deposition but can also be used to estimate, e.g., the location of the recoil. 
If an electric field is applied to the crystal, e.g. to read out the charge signal, the drifted electron-hole pairs dissipate further energy in the crystal lattice producing additional phonons (Neganov-Luke effect \cite{Luke:1990ir}\,\cite{Neganov:1985}). 
These Neganov-Luke phonons enhance the phonon signal which need to be accounted for in the estimation of the recoil energy. However, by a dedicated usage of this enhancement, the energy 
threshold might be significantly reduced. A general feature of cryogenic bolometers is, similar to germanium diodes, their limited crystal size ($\sim$\,1\,kg).  To achieve large exposures, these experiments generally use 
detector arrays which complicates not only the set up but also the analysis of the data.

 The crystal is weakly thermally coupled to a heat reservoir which is kept at a constant temperature of about $(10-100)$\,mK by a refrigerator cryogenic system. A temperature sensor measures a temperature evolution which 
can be expressed\,\cite{Booth:1997at} as
\begin{equation}\label{eq:thermalDet}
\Delta T = \frac{E} {C (T)}  \cdot \exp (-t/\tau),
\end{equation}
where $C(T)$ is the heat capacity of the absorber material and $\tau = C(T)/G(T)$ with $G(T)$ being the thermal conductance of the link between the crystal and the thermal bath. At cryogenic temperatures, $C(T)$ is very small for some materials due to their $T^3$ dependence of the heat capacity for a dielectric crystal.  This small heat capacity results in a relatively large temperature rise $\Delta T$. For example, germanium cooled down to 20\,mK temperature shows temperature rises of typically 1\,$\mu$K for nuclear recoils of a few keV.

To measure the temperature rise $\Delta T$, the most commonly used technologies are neutron-transmutation-doped germanium sensors (NTD) and transition-edge sensors (TES)\,\cite{Munster:2017lol}. To produce NTD sensors, small germanium crystals 
are exposed to thermal neutrons in order to produce a large amount of doping sites modifying the semiconductor properties of the crystal. The resistance of these thermistors changes strongly with temperature and can be constantly monitored by the voltage drop of the bias current running through them. A TES consists of a thin superconducting film, like tungsten for instance, operated at a temperature inside the phase transition between the conducting and the superconducting states. While this operation is demanding due to requirements on the temperature stability of the cooling system, in general a TES shows in comparison
to NTDs an increased sensitivity to measure small temperature changes and is sensitive to athermal phonons.  

Cryogenic bolometers exploit, in addition to the phonon signal, either the scintillation or the charge signal to provide particle discrimination. Independent of the read-out technology, the phonon
signal is unquenched, i.e. linear with deposited energy, and can be used to determine the recoil energy without dedicated measurements of the quenching factors (see section\,\ref{Sec:Calibr}).

The CDMS\,\cite{Ahmed:2008eu}  and its successor the CDMS II experiment\,\cite{Ahmed:2009zw} use germanium and silicon bolometers to search for dark matter. The experiments are located at the Soudan Underground Laboratory (see section\,\ref{Sec:BG}) and consist of up to 19\,Ge and 11\,Si detectors in the final configuration, with a mass of 230\,g and 100\,g each, respectively. These Z-sensitive Ionisation and Phonon-mediated (ZIP) detectors, exploit the phonon as well as the charge signal to allow for particle discrimination. In order to measure the athermal phonons TES are used. Information on the position of the interaction is obtained from the TES pulse arrival-times and the relative signal sizes in multiple sensors. This allows to select a fiducial volume to reduce the background. 
The dominant background of these detectors arises from events at the surface of the detectors
where a reduced ionisation yield is observed which leads to a misidentification of electronic recoils as nuclear recoils. However, phonon pulse-shape discrimination allows to identify surface events
with a misidentification rate of $1$ in $10^6$ electronic recoils\,\cite{Ahmed:2009zw}. 
A combined analysis of all CDMS II detectors yields an upper limit on the WIMP-nucleon spin-independent cross-section of 3.8$\times$10$^{-44}$\,cm$^2$ for a WIMP mass of 70\,GeV/$c^2$. Selecting the four germanium 
detectors with the best noise conditions 
and lowest energy thresholds allows to search for nuclear recoils in the energy range of $(3-14)$\,keV$_{nr}$. As a result, a WIMP mass of 7\,GeV/$c^2$ with a cross-section of 10$^{-41}$\,cm$^2$ is excluded\,\cite{Agnese:2014xye}. The analysis\,\cite{Agnese:2013rvf} using silicon crystals is based 
on an exposure of 23.4\,kg$\cdot$d for a nuclear recoil  energy range of $(7-100)$\,keV$_{nr}$. In this data set, an excess of events above the expected background is observed which corresponds to a WIMP mass of 8.6\,GeV/$c^2$ and a cross-section of $1.9\times10^{-41}$\,cm$^2$. 
The spin-dependent interpretation of the CDMS results can be found in\,\cite{Ahmed:2008eu}.  Although this signal indication created some excitement due to its cross section and mass reconstruction close to the signals by DAMA and CoGeNT, further measurements by the collaboration (see below) are in contradiction with the signal (see also figure\,\ref{Fig:SI_limits}). CDMS II performed additionally a study for an annual modulation of the event rate using data from October 2006 to September 2008\,\cite{Ahmed:2012vq}. No evidence for an annual modulation was 
found and this data disfavours the modulation claim of the CoGeNT experiment\,\cite{Aalseth:2014eft}  which also uses a germanium target.

The successor of the CDMS II experiment is the SuperCDMS detector which employs an improved interleaved ZIP technology (iZIP). These bolometers use an interleaved structure of the phonon and ionisation electrodes at the top and 
bottom faces of the crystals. This allows to improve the surface event rejection by using the asymmetry of the charge collection\,\cite{Agnese:2013ixa}. SuperCDMS uses 15 Ge crystals with masses of 0.6\,kg each and are sensitive to nuclear recoils between $(1.6\,-\,10)$\,keV$_{nr}$. A total of 577\,kg$\cdot$d science data 
was recorded focussing on dark matter masses below 30\,GeV/$c^2$, and a limit on the cross-section for a  8\,GeV/$c^2$ WIMP mass of $1.2\times10^{-42}$\,cm$^2$ was derived\,\cite{Agnese:2014aze}.
Another development of the CDMS collaboration is the CDMSlite (CDMS low ionisation threshold experiment) detector which uses a single crystal from the SuperCDMS detector for a 
dedicated WIMP search with a low energy threshold\,\cite{Agnese:2013jaa}\cite{Agnese:2015nto}.
In this operation mode, the bias voltage is increased in order to exploit the Neganov-Luke amplification of the phonon signal due to the drift from electron-hole pairs in the crystal lattice. 
Thus, the energy threshold could be significantly reduced to 56\,keV$_{ee}$ and an increase of the energy resolution is observed. However, in this operation mode 
the simultaneous measurement of the phonon and charge signal is not possible, thereby loosing the ability to discriminate between nuclear and electronic recoils. The results of SuperCDMS
 and CDMSlite set most sensitive exclusion limits at low WIMP masses\,\cite{Agnese:2014aze}\cite{Agnese:2015nto} (see also figure\,\ref{Fig:SI_limits}). SuperCDMS is also the first direct detection experiment which derives limits on more general WIMP interactions calculated with a 
non-relativistic effective field theory (see section\,\ref{sec:Gene_results}) \cite{Lin:2007ka}. 
The second generation of the SuperCDMS experiment will be located at SNOLAB using both, silicon and germanium detectors with two improved detector designs. The detectors will have a mass of 1.39\,g and 0.61\,g for Ge and Si, respectively. The goal is to detect WIMP masses down to $\sim0.5$\,GeV/$c^2$ lowering the sensitivity several orders of magnitude below current upper limits\,\cite{Agnese:2016cpb} by increasing the total mass and reducing further the current backgrounds.

A similar detector concept is used by the EDELWEISS collaboration which operates detectors at Laboratoire Souterrain de Modane (LSM). In contrast to the iZIP detectors, in EDELWEISS the signal is measured by thermalized phonons with NTDs. 
Since thermal phonons do not carry information about the spatial interaction inside the crystal, and surface events dominate the background for a WIMP search, an interleaved structure of the charge read-out is used. Already the EDELWEISS-II detectors used this technique to identify surface events with a reduced ionisation yield, enabling a rejection factor of more than $10^4$\,\cite{Broniatowski:2009qu}. 
The latest results from EDELWEISS-III stage of the experiment comprises a total of 582\,kg$\cdot$d fiducial exposure using an array of twenty-four $\sim 800$\,g germanium bolometers. The detectors are equipped with a set of fully inter-digitised electrodes which provide the possibility to select interactions in the bulk material allowing to reject events at the surface. 
This defines a fiducial volume with an average mass of 625\,g per detector.
Boosted decision tree algorithms are used in order to reject background. Searching in the energy interval $\sim (2.5-20)$\,keV$_{nr}$ in the detector's fiducial volume, results down to $4$\,GeV/$c^2$ WIMP mass have been derived. As no evidence for dark matter above the expected background is found, an upper limit is set on the spin-independent WIMP-nucleon scattering cross-section. For 5\,GeV/$c^2$ WIMP mass and a 90\% C.L. the derived limit is at $4.3\times10^{-40}$\,cm$^2$\,\cite{Armengaud:2016cvl}. A reanalysis of the data using a profile likelihood method achieves a higher sensitivity. As a result, an improvement of a factor of 7 for a 4\,GeV/$c^2$ WIMP mass is reached compared to the previous result while reproducing the results above 15\,GeV/$c^2$\,\cite{Hehn:2016nll}.  EDELWEISS is developing new improved detectors with the goal of reaching an exposure of 350\,kg$\cdot$d covering new regions of parameter space at low WIMP masses\,\cite{Arnaud:2016tpa}.
Both the CDMS and the EDELWEISS experiments have performed axion searches as described in section\,\ref{sec:germanium} for the CoGeNT experiment. The limits derived for the axion coupling to electrons are summarised in~\cite{Ahmed:2009ht} and\,\cite{Armengaud:2013rta} for CDMS and EDELWEISS, respectively.

The CRESST-II experiment at Laboratori Nazionali del Gran Sasso (LNGS) exploits, in addition to the phonon signal, also the scintillation light emitted by recoils in CaWO$_4$ crystals\,\cite{Angloher:2004tr}.
The phonon as well as the scintillation signal are read out by two optimised tungsten TES. Since particle discrimination
solely relies on the scintillation signal, it is necessary to achieve an effective collection of the generated photons. Thus, the housing of the crystals as well as the crystal surfaces
are optimised to avoid an absorption of the photons or inner total reflections. First limits on dark matter interactions have been derived already in 2004\,\cite{Angloher:2004tr}.
The second phase of CRESST-II had, in addition to a larger array structure for crystals, improvements of the neutron shield and an active muon veto. A total exposure of 730\,kg$\cdot$d with 8 
detectors was achieved, where each crystal weighs about 300\,g and shows an energy threshold in the range from 10.2\,keV$_{nr}$ to 19.0\,keV$_{nr}$.
An excess of events is observed, corresponding to a WIMP mass of 11.6\,GeV/$c^2$ (4.2\,$\sigma$) or 25.3\,GeV/$c^2$ (4.7\,$\sigma$) with a cross-section 
of 3.7$\times10^{-41}$\,cm$^2$ or 1.6$\times10^{-42}$\,cm$^2$, respectively\cite{Angloher:2011uu}. 
It is worth mentioning that the main background in this analysis is due to collisions of
lead nuclei with the crystal from $^{210}$Po $\alpha$-decays where the emitted $\alpha$ remains undetected. 
A further improvement of the detector layout increased the efficiency to measure the emitted $\alpha$ events from $^{210}$Po decays, leading to a strong suppression of the main background. In addition,
the detectors show an improved phonon and photon read-out efficiency, leading to a significant reduction of the energy threshold to 600\,eV$_{nr}$. With 29.35\,kg\,$\cdot$\,d of exposure
the previous signal claim could not be verified. A sensitivity to WIMP masses below 3\,GeV/$c^2$ was reached, while no background event in the signal region was observed\,\cite{Strauss:2014hog}.
With the same detector technology and exposure a dedicated low mass analysis was performed excluding WIMP interactions for a mass of 3\,GeV/$c^2$ at a cross-section of 8$\times10^{-40}$\,cm$^2$\,\cite{Angloher:2014myn}.  
Using the detector module with the lowest energy threshold of 307\,eV and an increased exposure of 52\,kg\,$\cdot$\,d, a cryogenic bolometer showed for the first time sensitivity to sub-GeV/$c^2$ dark matter masses at the cross-section level of  $10^{-37}$\,cm$^2$\,\cite{Angloher:2015ewa}.
In the future, the CRESST collaboration will focus on low mass WIMP detection by reducing the crystal size (24\,g) and lowering the energy threshold\,\cite{Angloher:2015eza} to less than 100\,eV. This would allow to
search for WIMP masses down to 1\,GeV/$c^2$. Another possible detector improvement is also considered to make use of the Neganov-Luke amplification of the phonon signal to lower the energy threshold\,\cite{Isaila:2011kp}.
Also the ROSEBUD experiment use scintillating bolometers to search for dark matter\,\cite{Cebrian:2000dz}. First results using sapphire crystals in the Canfranc Underground Laboratory showed promising results but due to the high background and the small exposure the results were not competitive.

A possible next generation experiment, EURECA (European Underground Rare Event Calorimeter Array), aims to build a facility to operate 1\,000\,kg of cryogenic detectors, both CaWO$_4$ and Ge detectors\,\cite{Angloher:2014bua}. This experiment is a joint effort mostly originating from the EDELWEISS, CRESST and ROSEBUD collaborations. The detectors would be located at the LSM laboratory, consisting of 150\,kg target material in a first phase, followed by a second phase with 850\,kg. The final goal is to reach a sensitivity of $3\times 10^{-46}$\,cm$^2$. In principle, a joint experiment between EURECA and SuperCDMS would be feasible, combining the various mentioned  technologies and exploiting their complementarity.

\subsection{Liquid noble-gas detectors}\label{LiquidNobleGas}

Liquid noble-gas detectors offer the advantage of large and homogeneous targets with high scintillation and ionisation yields. Currently, liquid argon (LAr) and liquid xenon (LXe) detectors are used as detector media. There are also some R\&D activities carried out for liquid neon\,\cite{Lippincott:2011zr} as well. The scintillation of both LAr and LXe is in the ultraviolet regime at 128\,nm and 175\,nm, respectively\,\cite{Cheshnovsky1972}. While for LAr it is common to use wavelength shifters and detect light in the blue wavelength region ($\sim 400$\,nm), in LXe the photons can be detected directly by using photosensors with windows made out of quartz which is transparent to the xenon scintillation light. 
After the passage of ionising radiation, ionisation or excitation of the medium takes place.
The excited or ionised atoms form excimers, $D_2^{\ast}$ or $D_2^{+}$ which de-excite emitting ultraviolet photons. The free electrons which appear in the ionisation can either recombine to produce further scintillation light or can be extracted with a drift field to be collected as an additional signal\,\cite{Bolozdynya:1999}.
Furthermore, liquid xenon has the advantage of containing almost 50\% of non zero spin isotopes, $^{129}$Xe and $^{131}$Xe, providing additional sensitivity to spin-dependent WIMP interactions\,\cite{Aprile:2013doa}. The high density of xenon (about 3\,g/$\ell$) provides excellent self-shielding such that a radio-clean innermost volume can be selected for analysis. 

In order to distinguish the main background due to $\gamma$ and $e^-$ interactions (electronic recoils, ER) from the interactions of WIMPs with nuclei (nuclear recoils, NR), two methods can be applied in liquid noble-gas detectors: pulse-shape discrimination and charge-to-light signal ratio. 
The short- (singlet) and long-lived (triplet) states that produce the luminescence in these media are populated at different levels for different types of particles. This results in a differentiation between ER and NR.  This technique gives large separation power in liquid argon due to the easily separable lifetimes (6\,ns and 1.6\,$\mu$s)\,\cite{Boulay:2006mb} of  the two components. However, pulse shape discrimination provides a good separation only for a large number of measured photons and therefore, a higher energy threshold has to be considered. In liquid xenon, the values for the decay constants are too close to each other, 4\,ns and about 22\,ns\,\cite{Hitachi:1983zz}, giving less rejection power. 

\begin{figure}[h]
  \begin{center}
   \includegraphics[angle=0,width=0.9\textwidth]{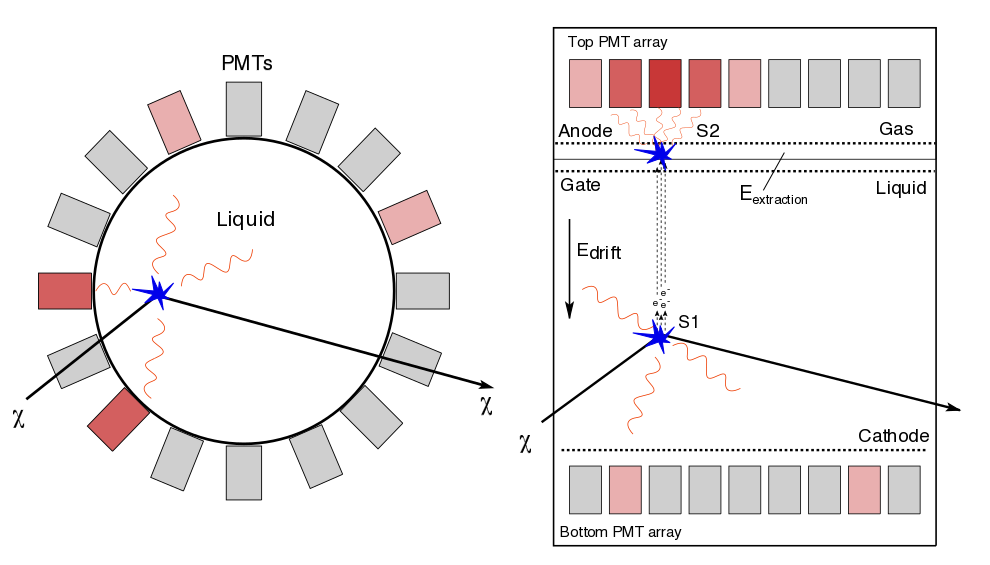}
   \caption[]{Schematic of single-phase (left) and double-phase (right) liquid noble-gas detectors.\label{Fig:DetConcepts_LiqNobl}}
  \end{center}	
\end{figure}

Single-phase (liquid) detectors consist typically of a spherical target, containing the liquid medium, which is surrounded by photo-detectors (see figure\,\ref{Fig:DetConcepts_LiqNobl} left). A main advantage is the $4\pi$-photosensor coverage which results in a larger light output compared to detectors which are only partially instrumented. The distribution and timing of the photons at the photosensors can be used to determine the position of the event typically with $\sim$\,cm resolution enabling the definition of a fiducial volume. Pulse shape is the main particle-discrimination parameter in single-phase detectors. 
DEAP\,\cite{Boulay:2012hq} and CLEAN\,\cite{Rielage:2014pfm} are examples of experiments using liquid argon in single phase. They are currently being commissioned at the SNO laboratory in Canada. Both detectors use light guides from the medium to the photosensor sphere in order to minimise the impact of the background from the radioactivity of the sensors. The DEAP-I prototype showed a discrimination power of $10^{-8}$ and an acceptance of 50\% for nuclear recoils above an energy of 25\,keV\,\cite{Boulay:2009zw}.  The background of this experiment  can be mainly explained by  radon daughters decaying on the surface of the active volume, misidentified electronic recoils due to inefficiencies in the pulse shape discrimination, and leakage of events from outside the fiducial volume\,\cite{Amaudruz:2012hr}. These backgrounds will be strongly reduced in the DEAP-3600 detector due to its higher light yield and simpler geometry compared to the first prototype. By the time of writing, DEAP-3600 is  taking data while already a larger detector with 50\,t fiducial mass (DEAP-50T) is being developed\,\cite{Amaudruz:2014nsa}.
The pulse-shape discrimination mechanism also works for liquid xenon\,\cite{Ueshima:2011ft} but not as efficient as in the argon case due to the similar decay components of the short- and long-lived states. The XMASS experiment\,\cite{Abe:2013tc} in Japan employs the single phase technology with about 800\,kg of liquid xenon. Ultra-low radioactive materials are used for construction to further reduce the experimental background. 
In the data acquired during 2012, an unexpected radioactive contamination originating from the photosensors appeared. Nevertheless, some results on low WIMP masses\,\cite{Abe:2012az} and on inelastic scattering off xenon\,\cite{Uchida:2014cnn} have been derived. 
The detector was refurbished to shield the contamination from the PMTs and new data has been analysed to search for a possible annual modulation of the background rate. In an exposure of 359\,live days $\times$ 832\,kg, the results show that the amplitude of a possible rate modulation derived is consistent with statistical fluctuations\,\cite{Abe:2015eos}.
In a next phase, XMASS1.5 plans to extend to 5\,ton LXe mass with about 1\,ton fiducial mass\,\cite{Liu:2014hda}. On larger time-scales, XMASS2 is proposed as multi-ton ($\sim\,24$\,ton) multi-purpose detector\,\cite{Hiraide:2015cba}.

A second method, the double-phase detectors (liquid and gas), enables to detect both the scintillation light and the charge signal from ionisation produced by an energy deposition\,\cite{Bolozdynya:1999}. The ratio of the two signals depends on the particle type and allows to separate signal-like events from background ones. Typically, two arrays of photosensors, on top and bottom of the detector, are employed to detect the prompt light signal.  Ionised electrons are drifted upwards to the liquid-gas surface and amplified via proportional scintillation in the gas phase\,\cite{Lansiart1976} which is also measured by the photosensor arrays. Therefore, double phase detectors are operated as a Time Projection Chamber (TPC, see figure\,\ref{Fig:DetConcepts_LiqNobl} right). 
Position reconstruction of events is performed obtaining the $z$ component from the time difference between the scintillation signal and the charge signal and by using the light pattern in the photosensors for the $(x,y)$-coordinates. The typical position resolution is in the order of millimetres.
WARP\,\cite{Benetti:2007cd}, operated during 2005\,-\,2006, was the first LAr detector which produced dark matter search results. It was located at the LNGS laboratory in Italy and consisted of 2.3\,$\ell$ liquid argon. Currently, the DarkSide experiment\,\cite{Agnes:2014bvk} is operating with about 50\,kg active mass and first results have shown a large light yield at $\sim 8$\,photoelectrons (PE) per keV energy which results in a very good separation of signal from background ($> 1.7\times 10^7$) using the information contained in the pulse shape. In the next run, Dark-Side employed underground argon which is depleted in $^{39}$Ar by a factor 1\,400\,\cite{Agnes:2015ftt}. In this run the threshold was placed at 13\,keV$_{nr}$. Combining  the null results of these two runs, an exclusion limit is placed which is at $2.0\times 10^{-44}$\,cm$^2$
 at 100\,GeV/$c^2$ WIMP mass.  
 On long-term, DarkSide plans on a multi-ton detector featuring 3.6\,tons\,\cite{Aalseth:2015mba} with an upgrade to 20\,tons LAr in the target volume\,\cite{Back:2016APS}.
 
Another ton-scale LAr TPC is the ArDM\,\cite{Badertscher:2013ygt} detector. This device was first tested at CERN and then was moved to the Canfranc underground laboratory in Spain where the commissioning took place in 2015. First data\,\cite{Calvo:2016hve} without drift field and using 850\,kg LAr has shown a successful detector performance and a light yield that allows for dark matter searches. Improvements on the system are currently on-going.

The liquid xenon TPCs used by the ZEPLIN\,\cite{Akimov:2006qw} and XENON10\,\cite{Aprile:2010bt} experiments showed already from 2006 to 2011 the potential of this technology to search for dark matter.  The exclusion limits on the coupling of dark matter particles to nuclei placed by these detectors were most constraining at that time\,\cite{Akimov:2011tj}\cite{Angle:2007uj}. The ZEPLIN detector, which operated at the Boulby underground laboratory, achieved a high separation between signal and background events by using a flat detector geometry allowing to increase the electric field in the liquid to a maximum of almost 9\,kV/cm\,\cite{Akimov:2006qw}. The $\gamma$-ray rejection factor was at $5\times10^3$ for the energy range $(2-16)$\,keV$_{ee}$\,\cite{Lebedenko:2008gb}. 
Besides the common spin-independent and dependent results, the XENON10 experiment performed a study using the charge signal (S2) alone. This allows to lower the detection threshold down to $\sim 1$\,keV$_{nr}$ but gives up the possibility to discriminate signal and background. In this mode, the liquid xenon technology obtained competitive sensitivities at WIMP masses as low as 5\,GeV/$c^2$\,\cite{Angle:2011th} (see figure\,\ref{Fig:SI_limits}).

The successor of XENON10, XENON100, started operation at the LNGS laboratory in 2009 and has been running until 2016.  Its total liquid xenon mass is 161\,kg,  where 62\,kg are contained inside the TPC and the rest is used for a LXe veto surrounding the TPC.  The latest XENON100 results combined the previous two science runs with a new one reaching a total exposure of  $1.75\times10^4$\,kg$\cdot$d\,\cite{Aprile:2016swn}.  When interpreting the data as spin-independent interactions of WIMP particles, a best sensitivity of $1.1\times10^{-45}$\,cm$^2$ for 50\,GeV/$c^2$ mass is derived at 90\% C.L. 
Natural xenon contains two nonzero nuclear-spin isotopes, $^{129}$Xe and $^{131}$Xe, with an abundance of 26.4\% and 21.2\%, respectively. 
Therefore, the absence of events above the background prediction also allows to exclude WIMP interactions which depend on the nuclear spin\,\cite{Aprile:2013doa}\cite{Aprile:2016swn}.  Both spin-independent and dependent exclusion limits are shown in section\,\ref{Sec:SumAndProsp}. Furthermore, the electron-recoil part of the data has been investigated in order to search for axion-induced signatures. Most sensitive upper limits on the coupling of axions to electrons are derived  ($g_{Ae}< 10^{-12}$ at 90\% C.L.) between 5 and 10\,keV/$c^2$ axion masses\,\cite{Aprile:2014eoa}. In addition, the 477\,live days XENON100 electron-recoil data has been used to study possible periodic variations 
of the event rate, allowing to exclude the DAMA annual modulation at $5.7\,\sigma$\,\cite{Aprile:2017yea}.Furthermore, exploiting the low background rate of the experiment, various leptophilic dark matter models have been excluded as explanations of the DAMA signal\,\cite{Aprile:2015ade}.
To further increase the sensitivity, a next generation detector, XENON1T\,\cite{Aprile:2012zx}, consisting of about 3\,tons of LXe has been commissioned and is taking data since end of 2016.  The goal is to reach two orders of magnitude improvement in sensitivity by also reducing the background by a factor of $\sim100$ compared to XENON100. 
XENON1T is built such that the main part of the infrastructure can host $\sim 7$\,tons of LXe and therefore, the planned upgrade to XENONnT can be performed with a moderate effort.

The LUX experiment, installed at the Sanford underground laboratory in the US,  operates a LXe TPC with an active mass of 250\,kg and realised their first results in 2013\,\cite{Akerib:2013tjd}. Since then, the experiment has continued taking data and the latest results from a combined analysis were released in 2016\,\cite{Akerib:2016vxi}. With a total exposure of $3.35\times10^4$\,kg$\cdot$day, the non-observation of a signal above the expected background results into an exclusion limit at $1.1\times10^{-46}$\,cm$^2$ for a WIMP mass of 50\,GeV/$c^2$. The analysis uses a low energy threshold at 1.1\,keV$_{nr}$ which results from an in-situ calibration of the nuclear recoil scale (see section\,\ref{sec:cal_recoil_energy}). 
Due to the larger light yield of the detector (at 8\,PE/keV at 662\,keV energy), 2.5 times higher than in XENON100, the experiment has set strong constraints at low WIMP masses. Indeed, the LUX result is currently the lowest limit of direct detection experiments for spin independent interactions for WIMP masses above  a few GeV/$c^2$. The LUX and ZEPLIN collaborations have joined to build the multi-ton LZ detector hosting about 7\,tons of liquid xenon in the target volume\,\cite{Malling:2011va}\cite{Akerib:2015cja} increasing, thereby, the sensitivity on WIMP-matter cross-sections.  
 
 The liquid xenon TPC technology is also used in the Chinese PandaX\,\cite{Cao:2014jsa} experiment which is operated at the Jin-Ping underground laboratory. In the first phase of the experiment, the target volume consisted of 120\,kg\,\cite{Xiao:2015psa}.  The detector was upgraded to host 500\,kg LXe and a first data was acquired in 2016. With a total exposure of $3.3\times10^4$\,kg$\cdot$day and having no dark matter signal identified above the background, upper limits on WIMP-nucleon cross section are derived with the lowest excluded value at $2.5\times10^{-46}$ for a 40\,GeV/$c^2$ WIMP mass\,\cite{Tan:2016zwf}.  The same dataset has been used to derive results on spin-dependent WIMP interactions with $^{129}$Xe and $^{131}$Xe\,\cite{Fu:2016ega}. The detector is currently running and new results are expected.
 In a final step, the detector will be upgraded to host a multi-ton target\,\cite{Cao:2014jsa}.

\subsection{Superheated fluids}\label{sec:SuperheatedLiquids}

Bubble chambers were often used in the last decades in accelerator experiments until new technologies as, for instance, gaseous detectors provided a better performance. Over the last years, the technology of using superheated
liquids has been revived in the context of dark matter searches\,\cite{Behnke:2008zza}. This branch of experiments can be divided into bubble chambers and droplet detectors. Both technologies use refrigerant targets
operated in a superheated state mildly below its boiling point. Interactions of particles with the target can be observed by the induced process of bubble nucleation. To create an observable bubble in the detector a phase transition of the medium is necessary. Therefore, the deposited energy by the particle must create a critically-sized bubble, requiring a minimum energy deposition per unit volume. This process can be described by the 'hot spike model'\,\cite{Seitz}. An event is then photographed with CCD cameras, and the position of the bubble can be determined with $\sim$\,mm resolution. This allows to define an innermost volume for the analysis, featuring lower background. 
After the formation of each bubble, the medium has to be reseted by a compression of the liquid phase followed by a decompression to a value below the vapour pressure.
In contrast to bubble chambers, droplet detectors make use of a water-based cross linked polymer to trap the bubbles resulting in a shorter dead time of the detector\,\cite{BarnabeHeider:2005ri}.   

The major advantage of this technology is that, being close below the temperature of the phase transition, bubble chambers are insensitive to minimum ionising backgrounds which generally dominate the backgrounds of other dark matter detectors.
In this way, most of the backgrounds created by $\gamma$-rays, X-rays and electrons from $\beta$-decaying isotopes are avoided. The remaining radiation which is able to produce nucleation are $\alpha$-particles, nuclear recoils from neutron interactions and WIMP-induced recoils. 
Due to the explosive character of the phase transition, acoustic signals can be used to discriminate $\alpha$-background events. 
For instance, the COUPP experiment has shown a $<99.3\,\%$ efficiency in rejecting $\alpha$-events, as they produce louder acoustic emission than nuclear recoils\,\cite{Behnke:2012ys}.   
Similarly, the rise-time and the frequency of the acoustic signal is used in the PICASSO experiment to mitigate the $\alpha$-background\,\cite{Archambault:2012pm}.
Although bubble chambers are threshold devices, i.e. counting events above a certain energy, by varying the temperature and/or pressure, the energy threshold can be changed. Existing detectors achieved energy 
thresholds of the order of a few keV nuclear recoil energy. 
Typically, the targets being used (CF$_3$I, C$_2$ClF$_5$, C$_3$ClF$_8$ and C$_4$F$_{10}$) contain fluorine which has an unpaired number of protons and is, thus, sensitive to spin-dependent interactions. 
Moreover, fluorine has a particular large expectation value for the proton spin content which enhances the sensitivity for spin dependent interactions to protons\,\cite{Menendez:2012tm}.  

Four different experiments have been operating during the last years using the bubble chamber (COUPP\,\cite{Behnke:2012ys} and PICO\,\cite{Amole:2015lsj}) and droplet 
detector (PICASSO\,\cite{Archambault:2012pm} and SIMPLE\,\cite{Felizardo:2011uw}) technologies. The used target masses reached only a few kg, hence they are not competitive in the spin-independent interpretation of the data. Nevertheless since their target contains fluorine, these detectors are sensitive to proton-coupling spin-dependent interactions. 
Experiments using germanium or liquid xenon have unpaired neutrons and consequently lower sensitivity to proton-coupling. As a result, the results of bubble chamber detectors have best sensitivities within this interpretation 
of the data (see figure\,\ref{Fig:SD_limits}, right). 
One of the first bounds on the dark matter cross-section from a detector using superheated fluids was achieved by the SIMPLE experiment which is operated at LSBB in France (see section\,\ref{Sec:BG}). It used 215$\,$g of $\rm{C_2ClF}_5$ as a  target and reached exposures up to 13.7$\,$kg$\,$days. With an energy threshold of 9$\,$keV, a sensitivity to the spin-dependent WIMP-proton cross-section of $5.7\times\,10^{-39}\,\rm{cm}^2$ at 35$\,$GeV/$c^2$ was achieved\,\cite{Felizardo:2011uw}.

Among the above mentioned detectors, PICO (formed from the PICASSO and COUPP experiments) shows the strongest exclusion limits on the spin-dependent WIMP-proton cross-section  (see figure\,\ref{Fig:SD_limits}). The experiment operates at the SNOLAB underground laboratory (see section\,\ref{Sec:BG}) both the PICO-60\,\cite{Amole:2015pla} and the PICO-2L\,\cite{Amole:2016pye} bubble chambers.
The 2-liter PICO-2L C$_3$F$_8$ bubble chamber has been operated with a threshold at 3.3\,keV and a total exposure of 129\,kg$\cdot$days\,\cite{Amole:2016pye}. The data obtained is consistent with the predicted background rate from neutrons and provided at the time of publication the most constraining limit to proton coupling spin-dependent below 50\,GeV/$c^2$.
In the latest results of 2017, the collaboration used a bubble chamber filled with 52\,kg C$_3$F$_8$. With an exposure of 1167\,kg$\cdot$day and with an energy threshold at 3.3\,keV, no nuclear recoil candidates are found and the most stringent limit on the WIMP-proton spin-dependent cross section is placed at  $3.4\times\,10^{-41}\,\rm{cm}^2$ for a 30$\,$GeV/$c^2$ WIMP mass\,\cite{Amole:2017dex}.
It is to mention that the PICO detector is able, for the first time, to discriminate efficiently alpha events by the acoustic signal due to the efficient bubble nucleation processes in $\rm{C_3F}_8$\,\cite{Amole:2015lsj}.   For spin-independent interactions, the PICO experiment demonstrates that it is possible to increase the target mass of bubble chambers to be competitive, for instance, to the LAr technology. 
For the future, a ton-scale detector has been proposed which would feature a spin-dependent sensitivity on proton-coupling at a similar level as future liquid xenon detectors (e.g. XENONnT, LZ) to neutron-couplings.

\subsection{Directional detectors}\label{sec:DirectionalDet}

In the previous sections, dark matter signatures based on either the annual modulation of the recoil rate or on the spectral shape of the signal are exploited. An additional possibility, introduced in section~\ref{sec:signatures}, is to measure directly the recoil track produced by the dark matter interaction. This would provide information on the ionisation density (d$E$/d$x$) as function of position, on the range and eventually on the direction of the recoiling nuclei. In the reference frame of the Earth, the WIMPs of the Milky Way halo are expected to originate from a preferred direction, approximately the Cygnus constellation.  Therefore, an asymmetry in the number of events scattering forwards and backwards is expected\,\cite{Spergel:1987kx}. 

The range of dark matter-induced nuclear recoils is below 100\,nm for energies $<200$\,keV in liquids and solids making the track reconstruction very challenging. The most promising strategy for directional searches is, instead, the use of low pressure gases such that the track length of the induced recoiling nucleus is large enough to be resolved. 
For a pressure $< 100$\,Torr ($< 130$\,mbar), the range of a WIMP-induced recoil with a mass of 100\,GeV/$c^2$ and a speed of 220\,km/s is, for instance, $(1-2)$\,mm. Note that this range varies significantly with the gas pressure. Such low pressures result in a low target mass and, consequently, very large detectors are necessary to achieve sensitivities comparable to the experiments mentioned before.  However, the measurement of the nuclear recoil direction would constitute the ultimate confirmation of dark matter detection.

\begin{figure}[h]
  \begin{center}
   \includegraphics[angle=0,width=0.7\textwidth]{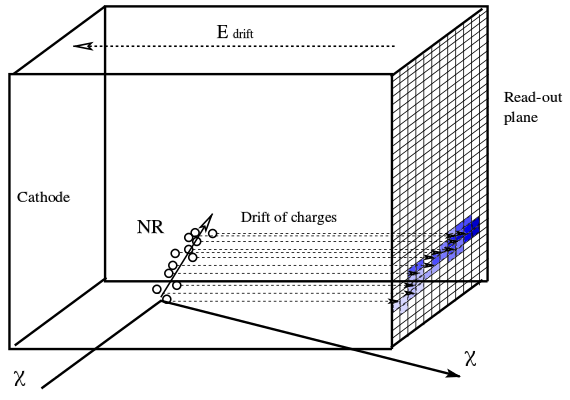}
   \caption[]{Schematic of a track reconstruction in a directional low-pressure gaseous time-projection chamber (TPC).\label{Fig:DetDirect}}
  \end{center}	
\end{figure}

Current directional developments use the gaseous time projection chamber technology for directional dark matter searches\,\cite{Ahlen:2009ev}\cite{Mayet:2016zxu}\cite{Battat:2016pap}. The drift gas serves as target material and detector simultaneously. Commonly used 
gases are CS$_2$, CF$_4$ and $^3$He, where the last two are favoured due to the unpaired nucleons that give sensitivity to spin-dependent interactions. The ionisation charge produced after a nuclear recoil is drifted by a homogeneous field to the read-out plane.   Using the ionisation pattern, the $(x,y)$-projection of the recoil can be reconstructed. The extraction of the $z$-projection is dependent on the detector and is discussed for each detector below.  Angular resolution, i.e. the precision of the reconstructed angle, are around 30$^{\circ}$. Figure\,\ref{Fig:DetDirect} shows a schematic view of a gaseous time projection chamber (TPC).
Several read-out technologies are being developed including multi-wire proportional chambers (MWPC), CCD cameras or gas electron multipliers (GEMs\,\cite{Sauli:1999ya}). The amplification of the drifted charges before the read-out plane allows to lower the detection energy threshold of the experiments.

In general, all charged particles produce tracks in the gaseous target. However, gamma- or beta-induced electronic recoils can be distinguished from nuclear recoils by determining the track length. The length for electronic recoils is at all energies about factor of 10 larger than for the equivalent nuclear-recoil energy. In contrast, alpha-induced tracks are not as well separable from nuclear recoil tracks\,\cite{Bertone:1900zza}. Therefore, a low alpha contamination is necessary. The energy threshold of these detectors is coupled to the energy to create an electron-ion pair (called W-value). These W-values are in the range of tens of eV which should, in principle, enable to reach sub-keV energy thresholds.

The DRIFT-II experiment\,\cite{Battat:2014van}, currently the largest directional detector, is operated at the Boulby underground laboratory in the UK (see also section\,\ref{sec:NeutronBG}). The detector is a TPC read-out by a MWPC. Its volume of 0.8\,m$^3$ is filled with a low-pressure mixture 30:10:1\,Torr of CS$_{2}$:CF$_4$:O$_{2}$ gas. 
This mixture provides 33.2\,g of fluorine in the active volume as target mass for spin-dependent WIMP interactions. The DRIFT detector uses the drift of negative ions (in particular CS$_{2}^{-}$) which travel at different velocities to determine the $z$-component (drift direction) of the event. This enables to fiducialize the volume by choosing the cleanest region for the analysis. Another advantage of using CS$_{2}^{-}$ is the reduction of diffusion while drifting. 
Using this technique, the DRIFT collaboration produced at the end of 2014 a background-free directional result for spin-dependent (proton coupling) interactions. This result is the most sensitive for directional searches but it is not yet competitive with the leading results from bubble chambers\,\cite{Amole:2015lsj}.
The MIMAC experiment\,\cite{Billard:2013cxa} also aims to measure the nuclear recoil energy and its angular distribution. Since 2012, a prototype-chamber  is being operated at the Modane underground laboratory (LSM) in France with 50\,mbar of a mixture with CF$_{4}$, 28\,\% CHF$_{3}$ and 2\,\% C$_{4}$H$_{10}$. The read-out of the ionisation electrons consists of pixelated micromegas\,\cite{Giomataris:1995fq}. The initial acquired data showed that a main background is due to the decays of radon. This data has been used to show, for the first time, the observation of a low energy nuclear recoil originating from the $\alpha$-decay of radon\,\cite{Riffard:2015rga}. In addition, a parameter based on the diffusion size of the electron cloud is defined to allow for a $z$-component determination which in turn allows to define a fiducial volume. The DMTPC experiment\,\cite{Battat:2014mka} is a planned m$^3$-scale TPC using CF$_4$ at 50\,Torr. The detector is a TPC with an charge amplification region. The primary ionisation is drifted to this region where an avalanche with a gain of 50\,000 takes place to amplify the signal. Scintillation photons from ion-recombination in the amplification region is acquired with CCD cameras. The image allows to determine the track geometry and the direction of the recoil. First prototypes are tested to measure both the energy and the direction of nuclear recoils\,\cite{Deaconu:2015ama}.  Finally, there are two further directional gas-TPCs where the readout is based on the GEM technology:  the NEWAGE\,\cite{Miuchi:2010hn} 
and the D$^3$\,\cite{Vahsen:2011qx} experiments. While the former, located at the Kamioka underground laboratory in Japan, has performed two science runs in 2010 and 2013\,\cite{Nakamura:2015iza}, the latter is in the prototyping phase\,\cite{Ross:2014tha}. 
The goal of the experiments mentioned above is to achieve volumes of $\sim$\,m$^3$ with the required signal topology. The increase of mass would then be realised by a multi-module detector.

Beside the low-pressure gaseous detectors mentioned above, nuclear emulsions can be used to reconstruct sub-micrometer particle tracks. Fine grained emulsions using silver-halide crystals of several tens of nm have been produced\,\cite{Naka:2013mwa} and their tracking capabilities have been shown using nuclear recoils from a neutron source. The read-out is performed via optical and X-ray microscopes and an angular resolution of about 20$^{\circ}$ has been achieved. The read-out efficiency for 120\,nm long tracks is larger than 80\,\%. After an initial R\&D phase,  a path in order to perform a first measurement with a target mass of 1\,kg on a time scale of six years has been proposed\,\cite{Aleksandrov:2016fyr}.

\subsection{Novel detectors}\label{sec:NovelDetectors}

R\&D activities are on-going in the development of detectors made out of solid xenon\,\cite{Yoo:2015sya}. Besides some of the advantages of xenon mentioned in section\,\ref{LiquidNobleGas}, solid xenon shows an increased amount of light collection and a faster electron drift compared to the detectors using liquid phase. 
In addition, if it would be possible to read a phonon signal, the energy resolution and the energy threshold would be greatly improved. 
Moreover, it would open the possibility to measure all three signals (ionisation, scintillation and heat) which eventually results in an increased particle discrimination power. It has been noted, however, that the development of crystals at sub-Kelvin temperatures might be challenging. So far, the scalability of transparent solid-xenon detectors to masses in the kg-scale has been demonstrated by a 2\,kg crystal which was grown at a temperature of 157\,K.

DAMIC\,\cite{Barreto:2011zu} is a detector using silicon charge-coupled devices (CCDs) to search for light WIMPs with $(1-10)$\,GeV/$c^2$ masses. Due to their low electronic noise, these devices can be operated with thresholds as low as 40\,eV$_{ee}$. First studies of the radioactive contamination of these devices show  that the levels are sufficiently low to reach a competitive sensitivity at low WIMP masses. DAMIC detectors with a total mass of 5.8\,g have been used to derive exclusion limits at low WIMP masses\,\cite{Aguilar-Arevalo:2016ndq}  (see figure\,\ref{Fig:SI_limits})  and  competitive constrains in searches for eV-scale hidden photons\cite{Aguilar-Arevalo:2016zop}. Currently, a low-radioactivity 36\,g detector is running at SNOLAB. The future goal of the experiment is to reach a kg-scale target mass while reducing the energy threshold to $0.5$\,keV$_{ee}$.

Also low-background gaseous detectors are being developed to gain sensitivity at low-WIMP masses. Due to the amplification capabilities, the sensors allow to reach sub-keV thresholds. For certain configurations, sensitivity to the ionisation of single electrons can be reached. 
The NEWS\,\cite{Gerbier:2014} spherical gaseous detector is developed to exploit low-WIMP masses well below 10\,GeV/$c^2$. The detectors consist of cylindrical vessels with diameters from 0.15\,m to 1.3\,m. Ionisation charges are drifted to a central metallic ball located at the centre. One of the prototypes, running at the Modane underground laboratory, has shown a background rate of 100 events/(keV$\cdot$kg$\cdot$d) for an energy threshold of 200\,eV. 
Another experiment focussing on the detection of low-WIMP masses is TREX-DM\,\cite{Iguaz:2015uza}. The detector is a  gaseous Micromegas-based TPC with a pressure up to 10\,bar. The aim is to operate an active mass of $\sim0.3$\,kg having an energy threshold of 0.4\,keV$_{ee}$ or lower. A first R\&D prototype has been operated with 1.2\,bar of Ar+2\%iC$_{4}$H$_{10}$. On longer term, a low-radioactivity version filled with neon or helium is developed to be operated at the Canfranc underground laboratory.

An improvement of detectors using superheated fluids as the detection medium is the 'geyser technique' (or condensation chamber) exploited by the MOSCAB experiment\,\cite{Bertoni:2013tua}. While
showing the advantages of the droplet and bubble chamber technologies (see section\,\ref{sec:SuperheatedLiquids}), the geyser technique allows to reset the detector within a few seconds.
This technology not only simplifies the overall detector set up significantly, allowing to increase the target mass, but also reduces the dead time of the detector.
To date, a prototype detector of 0.5\,kg (C$_{3}$F$_{8}$) is operated and a 40\,kg detector is being developed.  Already the 40\,kg detector is expected to surpass the current best sensitivity by the PICO experiment by several orders of magnitude. For the future, extensions are planned up to 400\,kg allowing to probe spin-dependent proton interactions down to $10^{-43}$\,cm$^2$. 

An interesting proposed idea is to use detectors made out of DNA or RNA to search for dark matter\,\cite{Drukier:2012hj}. A possible realisation would consist of thin gold films with strings of nucleic acids hanging from it. A gold nuclear-recoil produced by the interaction of a WIMP, would create a break in the sequence of these strings. The location and geometry of the break can be reconstructed by techniques common for biologists. By using the track reconstruction, the directionality of the signal can be employed to reduce the background. A threshold at $\sim$\,0.5\,keV would allow to focus on the low-WIMP mass region. The aimed target mass would consist of about 1\,kg of gold.  Note that this technology has, so far, not been used in any astroparticle physics-related experiment and the feasibility of such a detector has to be experimentally shown.


\section{Summary and prospects for the next decade}
\label{Sec:SumAndProsp}

In order to proof the existence of weakly interaction massive particles (WIMPs), experimental efforts can be 
categorised into indirect detection, i.e. via secondary particles created by dark matter self-annihilation, production of dark matter in particle colliders and direct detection of dark matter scattering off a target.
This article summarises the main concepts of direct detection experiments, namely dark matter detection signatures, methods for background reduction, detector calibrations, the statistical treatment of data and the 
interpretation of results. The focus lies on section\,\ref{Sec:Tech_Res} where various technologies aiming to directly detect dark-matter interactions are discussed together with their current status and plans. In the following, some of the possible interpretations of results are presented and prospects for the next years are discussed. 

\begin{figure}[h]
  \begin{center}
   \includegraphics[angle=0,width=0.496\textwidth]{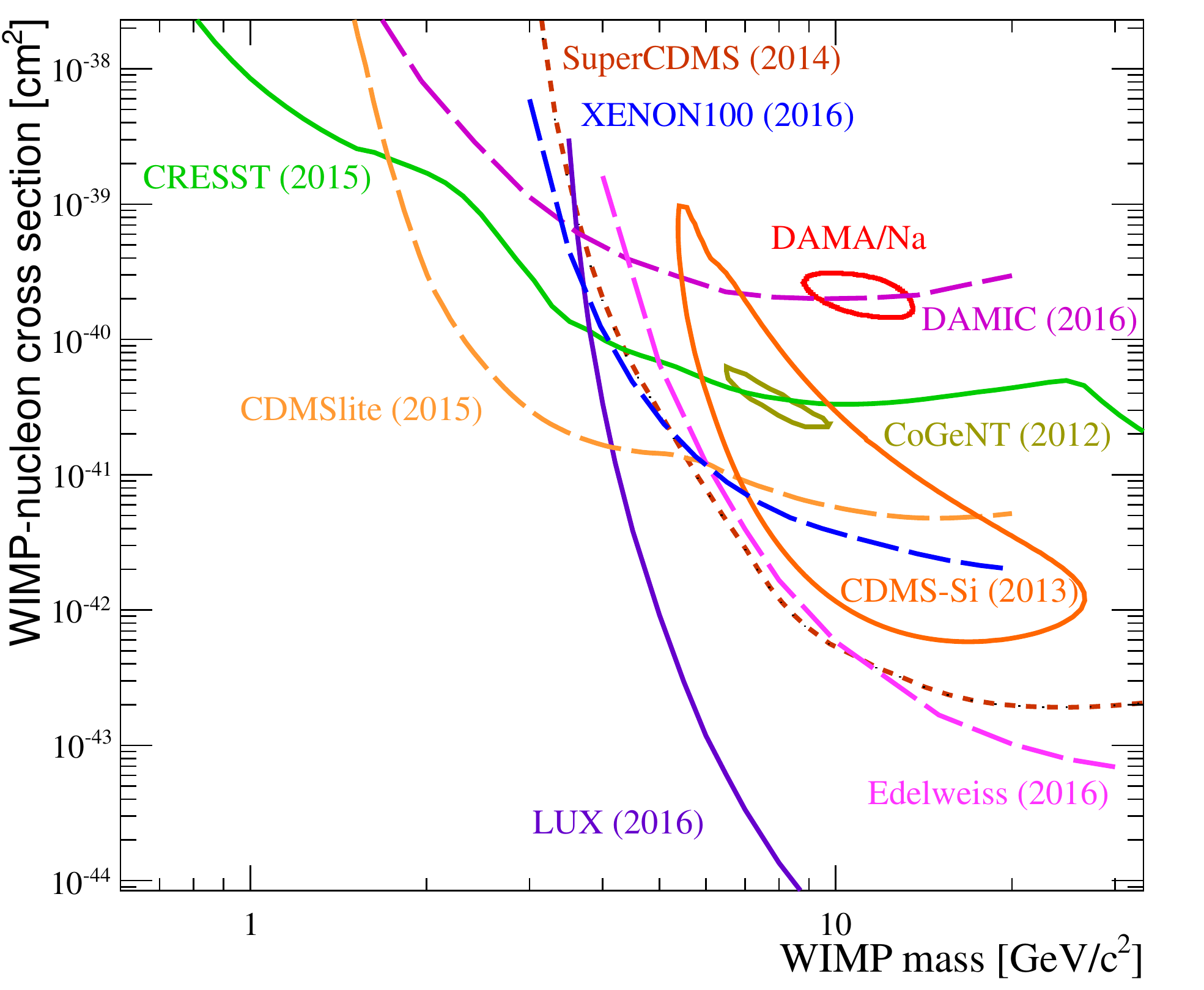}
   \includegraphics[angle=0,width=0.496\textwidth]{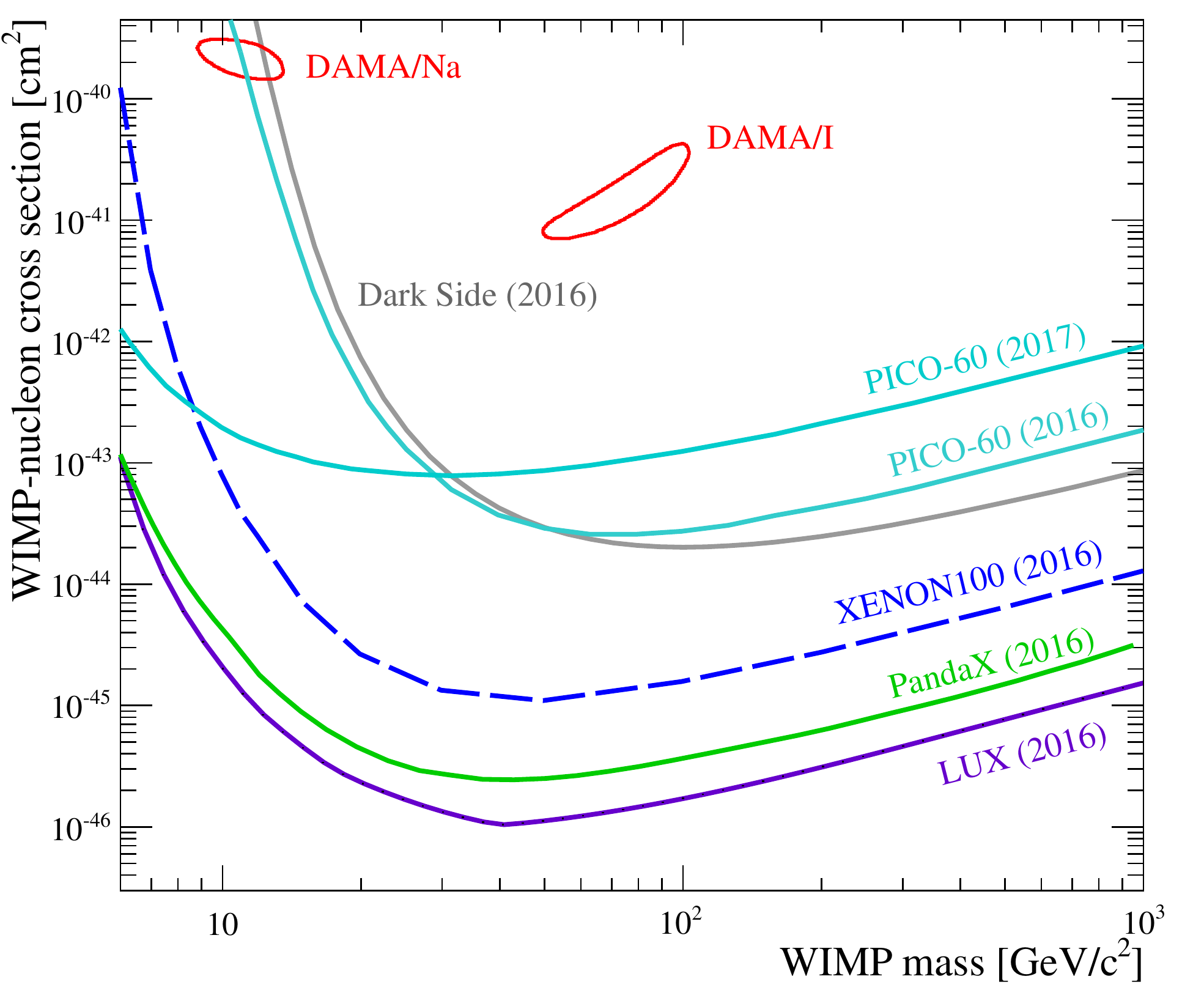}
   \caption[]{ Overview of signal indications and exclusion limits from various experiments for spin-independent WIMP-nucleon cross-section for low WIMP masses (left) and high WIMP masses (left)  as of early 2017. Data from dmtools\,\cite{dmtools} or private communications.\label{Fig:SI_limits}}
  \end{center}	
\end{figure}

\begin{figure}[h]
  \begin{center}
   \includegraphics[angle=0,width=0.496\textwidth]{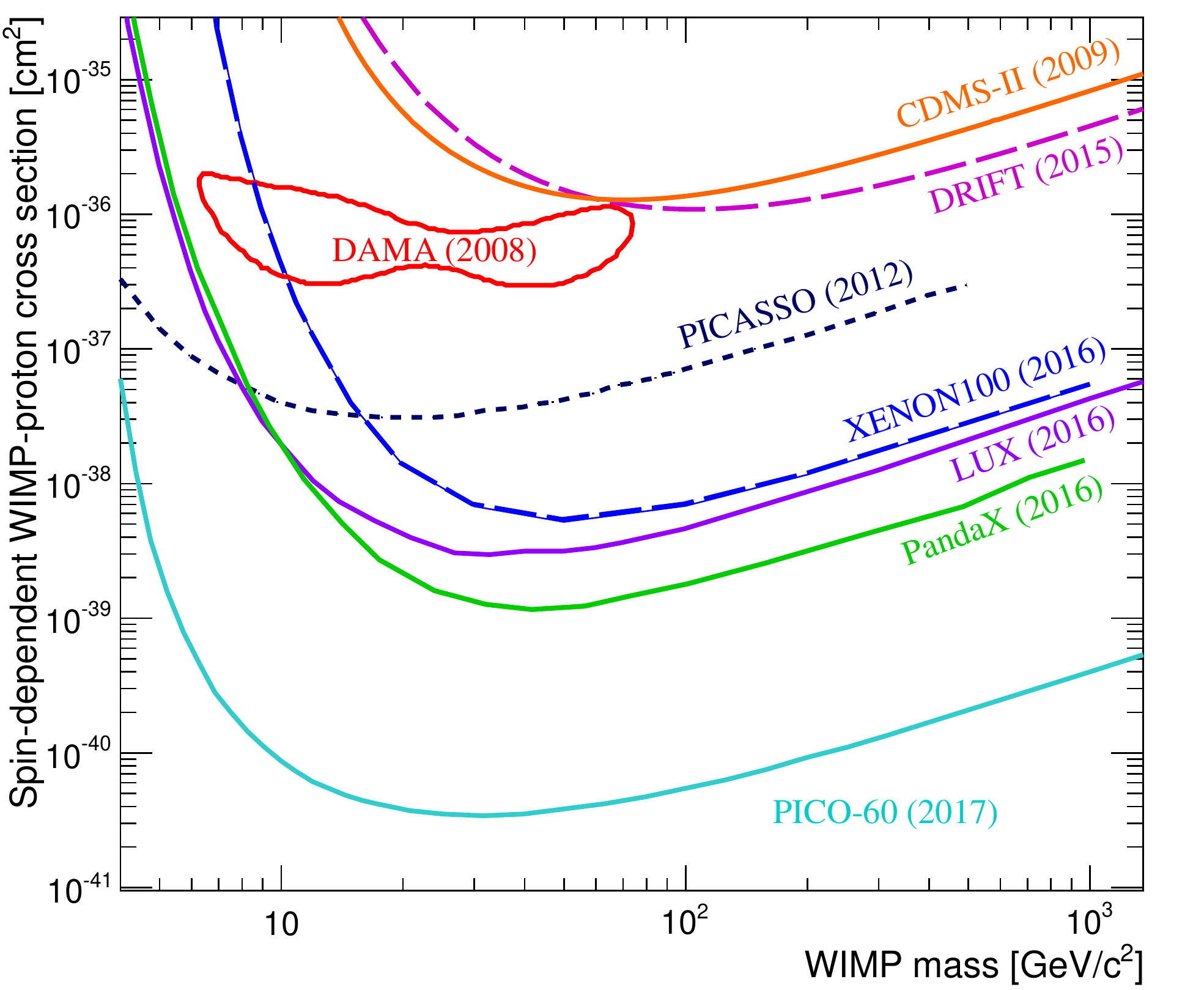}
   \includegraphics[angle=0,width=0.496\textwidth]{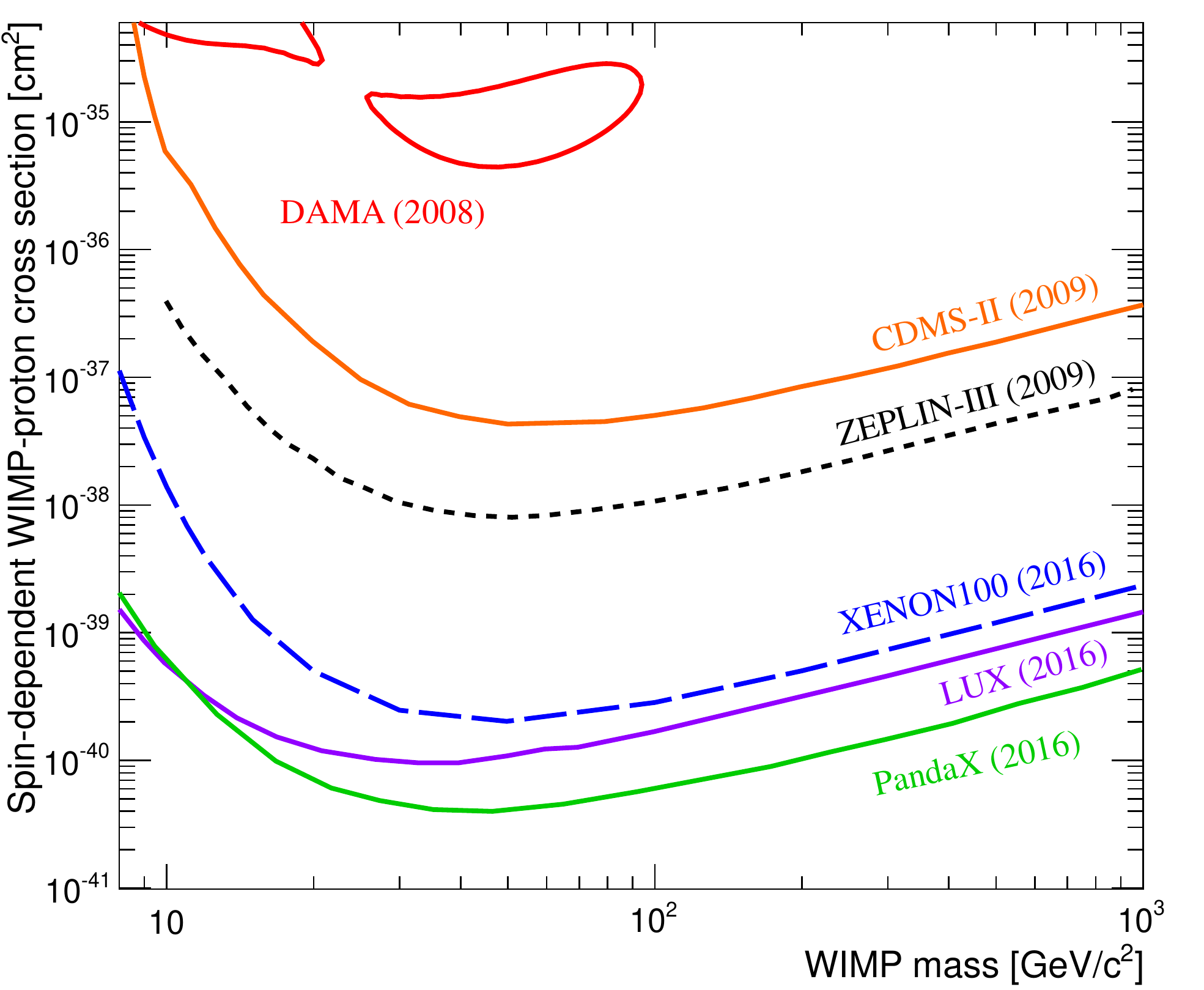}
   \caption[]{Exclusion upper-limits for spin-dependent WIMP-nucleon cross-section assuming pure proton coupling (left) and pure neutron coupling (right) as of early 2017. Data from dmtools\,\cite{dmtools} or private communications.\label{Fig:SD_limits}}
  \end{center}	
\end{figure}

WIMP interactions with the target of an experiment can be detected by a characteristic energy spectrum, an annual modulation of the measured event rate or by a directional dependence of 
interaction tracks (see section\,\ref{sec:intro_prin_dir_det}). 
Figure\,\ref{Fig:SI_limits} compiles signal indications and exclusion limits for both, low WIMP masses (left) and high WIMP masses (right). Signal indications stated by several 
experiments are shown as closed contours, whereas limits are represented by curves excluding the parameter space above.  
The separation of the tested parameter space in the two WIMP-mass ranges became more important in recent years, since various experiments have started to focus on a particular mass scale to exploit the specific advantages of the individual technology (see section\,\ref{Sec:Tech_Res}).  
For experiments showing sensitivity to low WIMP masses, the determination of the energy threshold becomes a crucial aspect. For this purpose, dedicated measurements of the target energy scale are performed. The systematic uncertainties  in the determination of these scales can affect, indeed, the results shown in figures\,\ref{Fig:SI_limits} and\,\ref{Fig:SD_limits}. In section\,\ref{Sec:Calibr}, the calibration strategies for various detector types are summarised.

Only a few experiments analysed the data for an annual modulation of the event rate, mainly due to the requirement to achieve a long-term stability of the detector. 
The annual modulation of the rate measured by the DAMA experiment has a significance 
of 9.3\,$\sigma$\,\cite{Bernabei:2013xsa} and, depending on the analysed target atom (Na or I), the derived signal regions (solid red) are shown in figure\,\ref{Fig:SI_limits}. The origin of the signal, however, remains
controversial, especially since results of other experiments are in strong tension with the DAMA claim. In section\,\ref{sec:Scin_crys}, possible explanations for the DAMA signal are discussed including both dark matter and non dark-matter related origins. Using a germanium detector as the target, an annual modulation of the signal at the level of 2.2\,$\sigma$ (solid dark green) was also claimed by the CoGeNT collaboration\,\cite{Aalseth:2012if}. A reanalysis of the data with different background assumptions shows, however, even lower significances\,\cite{Davis:2014bla}\cite{Aalseth:2014jpa}. In addition, the data from the 
CDMS II detector could not verify a modulation of the measured event rate by evaluating the germanium data\,\cite{Ahmed:2012vq}. 

Commonly the spectral shape of signal candidate-events is used to constrain dark matter interactions with the assumption of SI (and isospin-conserving) elastic scattering off WIMPs. 
In 2013, the presence of 3 observed events in the CDMS silicon detectors were above the expected background\,\cite{Agnese:2013rvf} (solid orange). Although no further data using silicon has
been released so far, the SuperCDMS collaboration 
performed a science run in 2014 with improved low-threshold germanium detectors\,\cite{Agnese:2014aze} which can not verify the previous signal (dashed brown).
Shortly before, an event excess measured by the CRESST experiment in 2012\,\cite{Angloher:2011uu} could be interpreted by WIMP interactions with masses of 11.6\,GeV/$c^2$ (4.2\,$\sigma$) or 25.3\,GeV/$c^2$ (4.7\,$\sigma$). 
However, new results derived by the same collaboration using an upgraded detector with improved background conditions and the same target element could not reproduce this excess\,\cite{Strauss:2014hog} (solid green).
Therefore, the initial signal claim is not shown in figure\,\ref{Fig:SI_limits}.  
In addition, the results from the liquid xenon detectors XENON100\,\cite{Aprile:2016swn} (dashed blue), LUX\,\cite{Akerib:2013tjd} (solid violet) and PandaX\,\cite{Xiao:2015psa} (solid green) disfavour the signal indications described above.
The presence of various dark-matter indications in the region around $\sim10$\,GeV/$c^2$ created some excitement, however, meanwhile improved results from several experiments indicate 
that probably in most cases, background was responsible for the observed events. This emphasises the relevance of the background prediction and the quantification of its uncertainty.
The presented experimental results are, in addition, derived with different statistical frameworks which consider systematic and statistical uncertainties to different degrees. Section\,\ref{sec:StatTreatm} discusses briefly the implicit assumptions made in the various statistical frameworks used by the experiments. 
In conclusion, these figures show the strength of detector technologies with a low energy threshold (e.g. cryogenic bolometers) since they are most constraining for WIMP interactions 
with masses below 5\,GeV/$c^2$. In contrast, liquid xenon TPCs have the highest sensitivity for larger dark matter masses due to their large target masses.  
Another signal indication of dark matter interactions is given by a directional dependence of the interaction tracks (see section\,\ref{sec:intro_prin_dir_det}). 
Low pressure gaseous detectors aim to measure the direction of the recoil atoms (see section\,\ref{Sec:Tech_Res}), however, their exposures are currently not competitive with the sensitivities of other technologies.

As discussed in section\,\ref{CrosSec_NuclPhys}, the results of a dark matter experiment can be also interpreted by spin-dependent interactions if the isotopes in the target material contain an 
unpaired number of nucleons resulting in an non-zero spin expectation value. It is
common to derive results separately for spin couplings to neutrons and protons. 
Figure\,\ref{Fig:SD_limits} shows the spin-dependent results from various experiments for pure neutron-coupling (left) and pure proton-coupling (right). 
To date, PandaX\,\cite{Fu:2016ega} shows the strongest limit for spin-dependent interactions on neutrons due its large exposure. In general liquid xenon detectors are most competitive in the neutron coupling channel because $^{129}$Xe and $^{131}$Xe have a high neutron spin expectation-value. In contrast, for spin-dependent WIMP interactions with protons, experiments using $^{19}$F have the highest sensitivities also because of the large spin expectation-value for this isotope. 
Currently, the most constraining limits are derived
from technologies using superheated liquids containing $^{19}$F as exploited by the PICO experiment\,\cite{Amole:2017dex} (solid cyan in figure\,\ref{Fig:SD_limits}), despite the lower exposures.  
Measuring the directionality of the recoil tracks with low-pressure gaseous detectors containing $^{19}$F enables to search also for spin-dependent interactions. 
The DRIFT experiment sets one of the first limits from this technology on spin-dependent proton interactions\,\cite{Battat:2014van} (dashed magenta).

Note that the choice of present experimental results interpreted by  spin dependent and independent interactions with matter is given by their relative strength in comparison to general coupling terms but are not the only possibilities. A more generalised interpretation of dark matter interactions containing, for instance, also velocity suppressed operators in the context of a non relativistic effective field theory is summarised in section\,\ref{CrosSec_NuclPhys}. Although this general approach is not yet widely used, in 2015 first experimental results have been displayed in this framework\,\cite{Schneck:2015eqa}. 

Systematic uncertainties in astrophysical parameters of the dark matter halo distribution are entirely neglected in figures\,\ref{Fig:SI_limits} and\,\ref{Fig:SD_limits}. Even though
the results are usually derived by a common choice of astrophysical parameters using the standard halo model (see section\,\ref{sec:DMdistrib}), a comparison of different detector targets is demanding. This is caused by the varying kinematics of WIMP interactions on different target elements and the various energy thresholds. 
A method to display results independently of astrophysical assumptions to avoid a possible bias in the comparison among results is presented in section\,\ref{sec:GenericResults}. It has to be noted that, even using this method, the discrepancy between the DAMA signal and the null results from other experiments remains.

Several experiments with a significant discovery potential for WIMP dark matter will be operated in the next years. Such discovery becomes possible in large liquid noble gas detectors with several tons of target material\,\cite{Aalseth:2015mba}\cite{Amaudruz:2014nsa}\cite{Akerib:2015cja}\cite{Hiraide:2015cba}\cite{Aprile:2015uzo} or in experiments with extremely low energy thresholds\,\cite{Agnese:2016cpb}\cite{Angloher:2015eza}\cite{Cushman:2013zza}. If a positive signal is observed by a detector, a confirmation by an independent one would be mandatory. Although a confirmation using the same technology would  strengthen its validity, a measurement with a different target would be required. A reconstruction of the mass and cross section is viable, however, this is improved by the combination of signals measured with differing target elements due to the degeneracy of the parameters  $\sigma/m_\chi$ (see equation\,\ref{eq:diff_rate1}). Furthermore, this degeneracy can only be broken if the dark matter mass is at the same order as the target nucleus\,\cite{Pato:2010zk}.
Different interaction channels such as spin dependent  and spin independent interactions could be exploited  additionally to constrain further the properties of dark matter\,\cite{Cerdeno:2013gqa}. Furthermore, if an annual modulation of the event rate is detected, information on the dark matter mass might be extracted from the reconstruction of a phase reversal towards lower energies\,\cite{Freese:2012xd}.

Although this review focusses on direct detection experiments, a comparison to collider searches and indirect detection experiments is beneficial due to the complementary of the approaches. This complementarity will be especially relevant in presence of a signal. 
Collider searches exploit mostly "mono-signatures" (e.g. mono-jets, mono-photons) accompanied by missing transverse-energy to constrain dark matter masses. Since, no indication of dark matter particle-production at colliders appeared so far, these results, can in principle be compared to the one of direct detection experiments. 
However, a direct comparison between the both is not possible and first dark matter limits have to be mapped to a common parameter space.  Instead of comparing individual dark matter models, collider searches can be interpreted in an effective field theory (EFT) approach (see e.g.\,\cite{Goodman:2010ku}\cite{Beltran:2010ww}\cite{Fox:2011pm}) 
or by considering minimal simplified dark-matter models (e.g.\,\cite{DiFranzo:2013vra}\cite{Buchmueller:2014yoa}\cite{Abercrombie:2015wmb}). 
It has to be remarked that an EFT approach can only be used for certain dark matter masses and coupling strengths as pointed out in e.g.\,\cite{Buchmueller:2013dya}\cite{Busoni:2013lha}. 
 Tevatron and LHC Run1 searches employed mostly EFT to compare the dark matter results with direct searches.
Meanwhile it has been realised that kinematics of the mono-signatures searched for at LHC occur via TeV mediators were the validity ofEFT is questioned at certain energies. Therefore, LHC Run2 results are interpreted in the framework of simplified models\,\cite{Abercrombie:2015wmb}.
In general, due to the limited center-of-mass energy in colliders, direct detection experiments show a higher sensitivity at heavy WIMP masses. Collider searches are, in turn, most constraining below the energy threshold of dark matter experiments and, hence, at low WIMP masses. 
Moreover, for spin-dependent interactions, direct detection signatures loose the A$^2$ enhancement of the event rate (see section\,\ref{CrosSec_NuclPhys}) and colliders, i.e. results from the LHC, have a higher sensitivity at most WIMP masses (see for example~\cite{Sirunyan:2017hci}\cite{Aaboud:2016tnv}).
Note, however, that collider searches can neither directly measure the dark-matter particle nor tests its lifetime. Therefore, the definitive confirmation of such detection would only occur in combination with direct detection results.

Also for indirect detection, a comparison with direct searches is, in general, demanding since the former approach is only sensitive to the thermally averaged self-annihilation cross-section of dark matter.
Hence, these processes do not allow to constrain elastic scattering of dark matter particles to baryons. 
However, it is possible to gravitationally capture dark matter particles inside the Sun via elastic scattering. The strength of the elastic scattering cross-section
determines the dark matter density inside the Sun, which is in turn proportional to the dark matter pair-annihilation rate. From all possible dark matter self-annihilation products, the only detectable particles that would reach the Earth are neutrinos\,\cite{Press:1985ug}\cite{Silk:1985ax}. Therefore, spin-independent interactions can be constrained by the elastic scattering on solar hydrogen 
and helium, whereas spin-dependent interactions can only be probed by scattering off hydrogen (protons). Few experiments as Super-Kamiokande\,\cite{Choi:2015ara}, Ice-Cube\,\cite{Aartsen:2012kia} and Baksan\,\cite{Boliev:2013ai} have exploited these channels to constrain the dark matter cross-section to baryons by searching for high energy neutrinos from the Sun. Similar to the collider searches, the constraints from these experiments are not competitive with direct detection experiments for spin-independent interactions, except at very 
low WIMP masses. Spin-dependent proton scattering can be, instead, very well constrained exceeding the sensitivities from direct detection experiments but not the limits from LHC.   

\begin{figure}[h]
  \begin{center}
   \includegraphics[angle=0,width=0.99\textwidth]{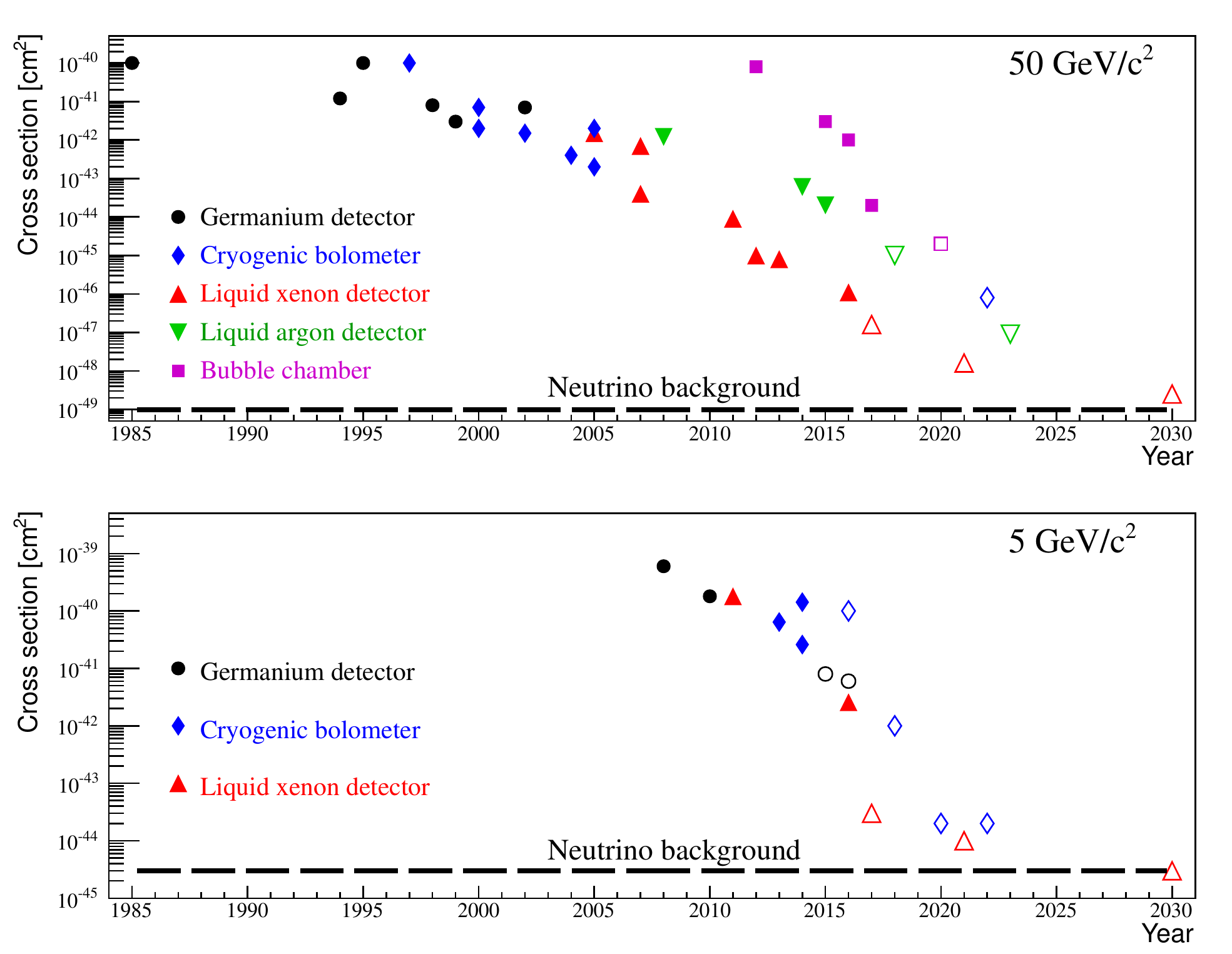}
   \caption[]{Evolution of sensitivity for spin-independent WIMP-nucleon cross-section for 5\,GeV/$c^2$ (top) and 50\,GeV/$c^2$ WIMP mass (bottom) as of early 2017. Data from germanium and silicon detectors is shown by black circles, cryogenic bolometers by blue diamonds, liquid xenon detector by red triangles (upwards), liquid argon by green triangles (downwards) and bubble chambers in purple squares. Empty markers represent the planed sensitivity for each technology. Below the horizontal line, the sensitivity to discover dark matter is limited by coherent neutrino scattering.\label{fig:SensitEvol}}
  \end{center}	
\end{figure}

During the last decades, although no definitive evidence for dark matter has appeared, great progress has been achieved in direct dark-matter searches. 
Figure\,\ref{fig:SensitEvol} summarises the time evolution of the spin-independent cross-section sensitivity since the first results of a germanium detector in 1985\,\cite{Ahlen:1987mn}.  
The top panel of figure\,\ref{fig:SensitEvol} shows upper limits for a 50\,GeV/$c^2$ WIMP mass and the bottom panel for a 5\,GeV/$c^2$ dark matter particle. 
The data is separated in low and high WIMP-mass regions due to the recent development towards an increase of sensitivity at low WIMP masses.
After the first results at high WIMP masses from germanium detectors (black circles), cryogenic bolometers (blue diamonds) showed most competitive exclusion limits. From $\sim$\,2005, the development of liquid noble-gas  detectors (red and green triangles) made it possible to significantly increase the target masses by keeping the background sufficiently low.  
Bubble chambers feature, in general, best sensitivities for spin-dependent proton-coupling but the fast progress of this technology in the last few years is also visible in terms to spin-independent interactions (purple squares).
The black dashed line represents the level at which coherent neutrino scattering limits the WIMP sensitivity. While for 5\,GeV/$c^2$ solar $^8$B-neutrinos would be the first to undergo coherent 
neutrino scattering, for 50\,GeV/$c^2$ atmospheric and the diffuse background of supernova neutrinos contribute.

The next generation of experiments (empty markers) will aim at enlarged target masses to achieve an even higher sensitivity. At the same time, it is essential to simultaneously reduce backgrounds from natural radioactivity and of cosmogenic origin (see section\,\ref{Sec:BG}). For this purpose, experiments are placed deep underground inside efficient shields and use techniques as careful screening of the materials, cleaning procedures, coating or etching the surfaces, etc., to keep the background at a minimum level. As, in general, an increase of target mass is easier for liquid noble-gas detectors, we expect that this technology will continue leading the sensitivity at large WIMP masses (above $\sim10$\,GeV/$c^2$). There is, indeed, a number of proposed detectors aiming to reach cross-section sensitivities down to $\sim10^{-46}$\,cm$^2$. 
The DEAP3600\,\cite{Boulay:2012hq}\cite{Fatemighomi:2016ree} and XENON1T\,\cite{Aprile:2015uzo} are ton-scale experiments which are expected to  release results in 2017. As a next step, detectors with target masses of several tons as DarkSide\,\cite{Aalseth:2015mba}, DEAP-50T\,\cite{Amaudruz:2014nsa}, LZ\,\cite{Akerib:2015cja}, XMASS2\,\cite{Hiraide:2015cba} and XENONnT\,\cite{Aprile:2014zvw} are planned. The success 
of the liquid noble gas TPC technology has motivated proposals for even larger detectors like the DARWIN  (dark matter wimp search with liquid xenon) facility\,\cite{Aalbers:2016jon} in Europe, consisting of a large liquid xenon detector. In the case of an evidence of a dark matter signal DARWIN, with a total mass of about 50\,t (40\,t target), could make a high statistics measurement of the dark matter particle properties, i.e. its mass and cross-section.
For an exposure of 200\,t\,$\cdot$\,y, spin-independent cross sections as low as $2.5\times10^{-49}$\,cm$^2$ can be tested for WIMP masses around 40\,GeV/$c^2$\,\cite{Schumann:2015cpa}.
However, at this sensitivity the neutrino background becomes a significant background.

For WIMP masses below $10$\,GeV/$c^2$, cryogenic bolometers and new developments as CCD cameras\,\cite{Barreto:2011zu} feature best sensitivities. Therefore, instead of focussing on increasing the mass to the ton scale, these technologies have started to develop ideas in order to reach lower energy thresholds and improved background levels. The SuperCDMS\,\cite{Cushman:2013zza} and the EURECA\,\cite{Angloher:2014bua} experiments aim to operate few hundreds of kg of target material to cover a significant part of the parameter space at low WIMP masses. As shown in figure\,\ref{Fig:SI_limits}, the DAMIC experiment is, as well, sensitive to the currently lowest
 detectable WIMP masses. A future run using 100\,g target material expects to improve the current sensitivity by more than two orders of magnitude\,\cite{Chavarria:2014ika}.

The future projects mention above will be challenged by the requirement of reducing the external and internal backgrounds to lowest levels. However, the experiments are entering into a cross-section region in which  the background from neutrinos can not be neglected anymore. Neutrinos can produce both electronic recoils from their interactions with electrons and nuclear recoils from coherent neutrino scattering. Solar neutrinos interacting coherently with nuclei will start limiting the sensitivity of dark matter experiments for low WIMP masses (few GeV/$c^2$) for cross-sections around $\sim10^{-45}$\,cm$^{2}$. For experiments with larger energy thresholds, the coherent scattering of atmospheric neutrinos will limit the sensitivity for dark matter searches at cross-sections of $\sim 10^{-49}$\,cm$^{2}$\,\cite{Strigari:2009bq}\cite{Ruppin:2014bra}. Although there are strategies to overcome the neutrino background at these cross-sections\,\cite{O'Hare:2015mda}, ideally, a dark matter discovery would appear before neutrino become a challenging background. 
Such a measurement would provide information on one of the most important topics of modern physics.

\section*{Acknowledgments}

 We gratefully acknowledge the support by the Max-Planck society and the DFG research training group 'Particle physics beyond the standard model'. 
We thank our colleagues Jan Conrad, Franz von Feilitzsch, Steffen Hagstotz, Jacob Lamblin, Thomas Schwetz-Mangold, Hardy Simgen, Quirin Weitzel and Michael Willers for useful comments to this document. We would also like to thank Francis Froborg,  Dongming Mei and D'Ann Barker for providing numerical data for figures in chapter 6.


\section*{References}
\bibliography{DarkMatterReview_A2}


\end{document}